\documentclass[revtex]{emulateapj}

\usepackage{etoolbox}
\usepackage[backref,breaklinks,colorlinks,citecolor=blue]{hyperref}
\usepackage{wasysym}
\usepackage{bmpsize}
\usepackage{graphicx} % needed for including graphics e.g. EPS, PS
\usepackage{anysize}
\usepackage{multirow}
\usepackage{float}
\usepackage{afterpage}

\newtoggle{emulateapj}
\toggletrue{emulateapj}
\newcommand{\nth}{\textsuperscript{th}}
\newcommand{\twaste}{\hbox{\ensuremath{T_{\mbox{\rm \scriptsize waste}}}}}

\newcommand{\wise}{\mbox {\it WISE}}

\newcommand{\hii}{\mbox{\ion{H}{2}~}}

\newcommand{\Msun}{M$_{\odot}$}
\newcommand{\Lsun}{L$_{\odot}$}

\newcommand{\kone}{\hbox{{\rm Type }\kern 0.1em{\sc i}}}
\newcommand{\ktwo}{\hbox{{\rm Type }\kern 0.1em{\sc ii}}}
\newcommand{\kthree}{\hbox{{\rm Type }\kern 0.1em{\sc iii}}}

\iftoggle{emulateapj}{\newcommand{\Sigurdsson}{Sigur{\scriptsize \DH}sson}}{\newcommand{\Sigurdsson}{Sigur\dh sson}}
\def\spose#1{\hbox to 0pt{#1\hss}}
\def\simlt{\mathrel{\spose{\lower 3pt\hbox{$\mathchar"218$}}
     \raise 2.0pt\hbox{$\mathchar"13C$}}}
\def\simgt{\mathrel{\spose{\lower 3pt\hbox{$\mathchar"218$}}
     \raise 2.0pt\hbox{$\mathchar"13E$}}}

\begin{document}

\title{The \^G Infrared Search for Extraterrestrial
  Civilizations with Large Energy Supplies. III. The Reddest Extended
  Sources in \wise}

\author{Roger L. Griffith\altaffilmark{1,2},
  Jason T. Wright\altaffilmark{1,2},
  Jessica Maldonado\altaffilmark{3}, Matthew
  S. Povich\altaffilmark{3}, Steinn \Sigurdsson\altaffilmark{1,2}, Brendan Mullan\altaffilmark{4}
}

\altaffiltext{1}{Department of Astronomy \& Astrophysics, 525 Davey
  Lab, The Pennsylvania State University, University Park, PA, 16802, USA}

\altaffiltext{2}{Center for Exoplanets and Habitable Worlds, 525 Davey
  Lab, The Pennsylvania State University, University Park, PA 16802, USA}

\altaffiltext{3}{Department of Physics and Astronomy, California State
  Polytechnic University, Pomona, 3801 West Temple Ave, Pomona, CA
  91768, USA}

\altaffiltext{4}{Carnegie Science Center, 1 Allegheny Ave.\ Pittsburgh, PA 15212, USA}


\begin{abstract}

%In this third installment of the \^G series of papers, w
Nearby Type {\sc{iii}} (galaxy-spanning) Kardashev supercivilizations
would have high mid-infrared (MIR) luminosities. We have used the
Wide-field Infrared Survey Explorer (\wise)
to survey $\sim 1 \times 10^5$ galaxies for 
extreme MIR emission, $10^3$ times more galaxies than the only
previous such search.  We have calibrated the \wise\ All-sky Catalog pipeline
products to improve its photometry for extended sources.
We present 563 extended sources with $|b| \ge 10$
and red MIR colors, having visually vetted them to remove artifacts.
No galaxies in our sample host an alien
civilization reprocessing more than 85\% of its starlight into the
MIR, and only 50 galaxies, including Arp 220, have MIR
luminosities consistent with $>50$\% reprocessing.  Ninety of these (likely) extragalactic sources have little literature
presence; in most cases they are likely barely resolved 
galaxies or pairs of galaxies undergoing large amounts of star
formation.  Five are new to science and deserve
further study. The Be star 48 Librae sits within a MIR nebula, and we suggest that it may be creating dust.  \wise,
2MASS, and {\it Spitzer} imagery shows that IRAS 04287+6444 is
consistent with a previously unnoticed, heavily extinguished cluster
of young stellar objects. We identify five ``passive'' (i.e.\ red) spiral
galaxies with unusually high MIR and low NUV luminosity.  We search a
set of optically ``dark'' \ion{H}{1} galaxies for MIR emission, and
find none. These 90 poorly understood sources and five anomalous passive
spirals deserve follow-up via both SETI and conventional
astrophysics. 

\end{abstract}

\section{Introduction}

This is the third paper in a series describing the \^G infrared
search for extraterrestrial civilizations with large energy supplies.
 The first two papers \citep{WrightDyson1,WrightDyson2} provide the justification 
 and framework for the search.  Here we give a brief summary of those
 works, and the purpose of this paper.

\subsection{Justification}

\citet{hart75} argued that the failure of SETI to date was because
humanity is alone in the Milky Way, based on a comparison of likely
colonization timescales for the Milky Way and its age.  Hart's
argument also implies that any galaxy with a spacefaring species will
become thoroughly colonized in a time short compared to the galaxy's
age, suggesting that most galaxies should either contain no spacefaring species or
be filled with them.

\citet{kardashev64} parameterized potential alien civilizations by their energy
supply compared to the starlight available to it, with a Type {\sc{i}}
civilizations (K1 in our notation) commanding its planet's entire
stellar insolation, a Type {\sc{ii}} civilization commanding an entire
star's luminosity (i.e.\ a Dyson sphere, K2), and a Type {\sc{iii}}
civilization (K3) commanding most of the stellar luminosity in a
galaxy.  Expressed in these terms, Hart's argument is that the
timescale for the appearance of the first K2 to its growth into a
K3 is very short, implying that we should expect many K3
civilizations in the Universe if spacefaring life is common.

Indeed, the technological sophistication required to construct a Dyson sphere
seems far greater than that required for achieving interstellar
travel: while humanity's solar panels currently fall short of complete
coverage of the Sun by a factor of $\sim 10^{17}$, our deepest space
probes today these fall short of the distance to the nearest star by a
factor of only a few thousand.  

If Hart's reasoning is sound, then we should expect that, unless
intelligent, spacefaring life
is unique to Earth {\it in the local universe}, other galaxies should
have galaxy-spanning supercivilizations, and a search for K3's may be
fruitful.  If there is a flaw in it, then intelligent, spacefaring
life may be endemic to the Milky Way in the form of many K2's, in
which case a search within the Milky Way would be more likely to
succeed.  It is prudent, therefore, to pursue both routes. 

\subsection{Prior Searches and the Promise of WISE}

\citet{dyson60} and \citet{slysh85} demonstrated that waste heat would
be an inevitable signature of extraterrestrial civilization, and that
such signatures might be detectable to mid-infrared (MIR) instrumentation
for civilizations with energy supplies comparable to the luminosity of
their host star.  The first effective all-sky search sensitive to such
namesake ``Dyson spheres'' was performed by {\it IRAS}, but the
infrared cirrus and the poor angular resolution of {\it IRAS} limited
its sensitivity to only the brightest sources.

\citet{Carrigan09a} and \citet{Carrigan09b} used the {\it IRAS} low
resolution spectrometer (LRS) to determine whether candidate Dyson
spheres' SEDs were consistent with blackbodies with $T$ = 100--600K.
\citeauthor{Carrigan09a} concluded from infrared colors
and low resolution spectra that the best of these of the most Dyson sphere candidates were
typically reddened and dusty objects such as heavily 
extinguished stars, protostars, Mira variables, AGB stars, and
planetary nebulae (PNs).  Nonetheless, of the 11,000 sources he studied,
\citeauthor{Carrigan09a} identified a few weak Dyson 
sphere candidates with spectra consistent with carbon stars. One
candidate, IRAS 20369+5131, showed a nearly featureless blackbody
spectrum with $T$ = 376K, but Carrigan concluded it is likely a
distant red giant with no detectable SiC emission.

\citeauthor{jugaku91} performed a series of follow-up searches of sources with
anomalously red $(K-12\mu)$ colors
\citep{jugaku91,jugaku95,jugaku97,jugaku00}, and found no highly
complete Dyson spheres around any of the 365 solar-type stars within 
25 pc studied, or another 180 stars within the same distance
\citep{jugaku04}.
 
To date, the only search for the
waste heat of a K3 civilizations in the peer-reviewed literature has
been that of \citet{annis99a}, who searched for outliers to the
Tully-Fischer relation to identify K3s intercepting a significant
fraction of their starlight. \citet{carrigan12} also suggested
searching for the morphological signatures of K3s, especially in
elliptical galaxies.  

The advent of large solid angle, sensitive
MIR surveys makes a waste-heat based K3 search more feasible
today.  The {\it Wide-field Infrared Survey Explorer} \citep[{\it WISE},][]{WISE}
performed an all-sky MIR survey at
3.4, 4.62 12 and 22 $\mu$ (the $W1$,$W2$,$W3$, and $W4$ bands) with
superior angular resolution (by a factor of 5) and sensitivity (by a
factor of 1000) than {\it IRAS}.  {\it WISE} is thus the first
sensitive survey for both K2's in the Milky Way and K3's among the
approximately $1\times 10^5$ galaxies it resolved.  The {\it Spitzer Space
  Telescope} is another powerful tool for waste heat searches, having
superior sensitivity and angular resolution.  Its survey of the
Galactic plane will be a powerful tool in the search for K2's in the
Milky Way.  Since its large area surveys are generally restricted to
star-forming regions and the Galactic Plane, where sensitivity to K3's
is more limited, we have restricted our efforts in this paper to \wise.

\subsection{The AGENT Formalism}

 In \citet{WrightDyson2}, we developed the AGENT formalism for quantifying
the expected MIR spectra from galaxies hosting K3s in terms
of the energy supply of an alien civilization, and outlined the
methodology of the {\it Glimpsing Heat from Alien Technology} (\^G) search for such civilizations in the local 
Universe using the results of the \wise\ All-sky MIR
survey.  In particular, we argued in that work that extended sources
have the lowest false positive rate because many of the confounding
sources, primarily dusty and extinguished stars and cosmological
sources, would not be present in that sample.

The AGENT formalism parameterizes the power used by an alien
civilization in terms of starlight absorbed (represented by the parameter
$\alpha$), energy generated by other means ($\epsilon$), thermal waste
heat emitted ($\gamma$), and other energy disposal ($\nu$).  

Most relevant to the present paper are the parameters $\gamma$ (waste
heat luminosity, expressed as a
fraction of the starlight available to the civilization) and
$T_{\rm  waste}$, the characteristic temperature of the waste heat
(which dictates its infrared colors).  For values of $T_{\rm waste}$
  in the 100--600 K range, values of $\gamma$ near 1 would
imply that most of the luminosity of a galaxy is in the MIR (in the
form of the waste heat from alien engines), while
values near 0 would imply that the alien waste heat was very small
compared to the output of the stars in the galaxy.  For dust-free elliptical
galaxies with little of their luminosity in the MIR, values of $\gamma$ of a few percent
would be detectable as an anomalous MIR excess.

\subsection{Scope and Purpose of this Paper}

As an essential step in our waste heat search, we have produced a
clean catalog of the reddest sources {\em resolved} by \wise.  The
purpose of our focus, in this work, on resolved sources is twofold:
resolved sources present their own challenges of interpretation and
photometry, necessitating this separate effort; and we wish to first deal with
a relatively small and clean sample of nearby galaxies, consistent
with a search for nearby K3's.  \wise\ resolves approximately $1\times
10^5$
galaxies (see Section~\ref{number}).  

Sources unresolved by \wise\ include a wide variety of potential false
positives, such as
false detections, data artifacts, cosmologically redshifted objects,
dusty objects at cosmological distances, dusty stars, and heavily extinguished
stars. The study of these sources requires a different sorts of
analysis from those needed for nearby galaxies, which we will describe in
later papers.  By contrast, there are many fewer, and more easily
excluded, sources of false positives among the extended sources in the
all sky catalog.  

Our primary objective in this paper is to ``map the landscape'' among
galaxies resolved by \wise\ by identifying the nature of the very
reddest of these extended sources in the \wise\ All-sky survey, using
several metrics for ``red-ness'', including the AGENT parameter
$\gamma$.  A byproduct of this effort is a clean catalog of the
reddest extended sources in \wise, which we present here.

Our secondary objective is to identify and explain the most extreme
objects in this catalog, which by their superlative nature are
inherently scientifically interesting,
regardless of the origin of their MIR luminosity.  In most
cases, these are well-known objects; in many of the remaining cases
their nature seems clear.  A few cases, however, are new to the
scientific literature and their nature is uncertain.  These sources of uncertain
nature will be natural candidates for followup, and those that appear
consistent with Kardashev civilizations warrant followup by
communication SETI efforts, in particular.   
 
As a tertiary objective, we place a zeroth-order upper limit on the energy
supplies of nearby K3s by identifying the most MIR-luminous galaxies in
our sample.  This upper limit can be pushed down to the
degree that these galaxies' MIR emission can be shown have purely
natural origins.  A rigorous upper limit will require a more detailed
analysis of these galaxies' SEDs, and a more precise calculation of
the number of galaxies considered in our sample.  We save this
exercise for another paper in this series.

Finally, we also illustrate the waste heat approach by performing a quick check of two classes of anomalous
galaxies to confirm that they are not hosts to MIR-bright K3's.

\subsection{Plan}

In Section~\ref{Sample} we describe how we have analyzed 
the \wise\ All-sky catalog, and how we performed a series of cuts to
select a sample of only real, red, extended sources (our
``Extended Gold Sample'' of 30,808 sources).  Sections~\ref{calibrate} describes how we
calibrated the \wise\ photometry, and our estimates of our photometric precision.

Section~\ref{classification} describes our efforts to classify the
sources in the Extended Gold Sample, mostly via SIMBAD object types,
so that we could identify previously unstudied sources and reject
many Galactic sources.  

Section~\ref{visualgrading} describes our more detailed efforts to
understand the reddest sources in the Extended Gold Sample using
careful visual inspection and literature searches.

Section~\ref{platinum} describes how we performed a second round of
vetting on the Extended Gold Sample, using our calibrated photometry
from Section~\ref{calibrate}, our classifications from
Section~\ref{classification}, and our visual inspections and literature searches
from Section~\ref{visualgrading} to confidently and carefully identify
the reddest sources and determine the best photometry for
them on a case-by-case basis.  The result is the 563 source ``Platinum
Sample,'' which we present in our catalog, and whose fields are described
in Table~\ref{FITS}.

Section~\ref{sec:extreme} describes the reddest objects in the
Platinum Sample, for several definitions of ``red'': all six
combinations of the \wise\ bandpasses and the AGENT parameter
$\gamma$.  Section~\ref{new} describes five sources that are
effectively new to science, having little or no literature presence
(beyond having been detected with {\it IRAS}).  

We present our conclusions in Section~\ref{conclusions}, and in the
appendices we use the \wise\ imagery to examine two categories 
of anomalous galaxies, \ion{H}{1} dark galaxies and so-called
``passive spirals,'' and show that they do not exhibit sufficient
MIR emission to have their anomalous natures explained by the presence
of a K3.

\section{Sample Selection}

\label{Sample}
The \wise\ mission began scientific operations on 2010 January 7 at wavelengths of 3.4, 4.6, 12, and 22 $\mu$m, 
hereafter referred to as {\it W1, W2, W3}, and {\it W4}, respectively. The All-sky Data Release was subsequently issued on 2012 March 14 and 
reached 5$\sigma$ point source sensitivities in unconfused regions to better than 0.08, 0.11, 1 and 6 mJy, 
respectively \citep{WISE}. It should be noted that a more recent data release by the \wise\ collaboration, dubbed
ALLWISE\footnote{\url{http://wise2.ipac.caltech.edu/docs/release/allwise/}} was issued on 2013 November 13. The added sensitivity and depth, proper motion measurements and 
improved flux variability information in the ALLWISE data products means that they supersede the earlier All-sky Data Release 
Catalog and Atlas for {\it{most}} uses. The \wise\ team suggests that the All-sky Release Catalog may have better photometric 
information for objects brighter than saturation limits in {\it W1}
and {\it W2} ({\it W1}$<$ 8.1 mag and {\it W2}$<$6.7 mag). Given that our project 
was well underway before the ALLWISE data release was issued, and the
fact that ALLWISE does not add many ($<1\%$) new, bright and
extended sources in W3, we rely on the All-sky Data Release
measurements for this analysis. 

\subsection{12$\mu$m Extended Sample}
\label{rchi2}

The following methodology was used to select the full sample of 12$\mu$m-selected sources with extended photometric profiles from the \wise\ All-sky catalog. 
The instrumental profile-fit reduced $\chi^2_\nu$
(W3RCHI2)\footnote{We refer to fields/nomenclature in the All-sky catalog in all
  caps} was used to distinguish sources with extended profiles
(W3RCHI2 $\ge$ 3) from sources with point source profiles (W3RCHI2 $<$3). Given that the majority of sources in the WISE All-sky catalog are
unresolved in W3, i.e. angular sizes $< 6.5^{\prime\prime}$, we adopt a conservative $\chi^2_\nu$ threshold of 3, which excludes only a small number of marginally
resolved sources while admitting a large but manageable number of
point sources with anomalously poor point-spread function (PSF) fits. We also required that the uncertainty in the 
W3 ``standard'' aperture magnitude to be measured (i.e.\ not null). A null result means that the W3 ``standard'' aperture magnitude is a limit, or that no aperture measurement was possible. Finally, we applied a cut in the Galactic latitude ($|b| \ge 10$) to remove contamination from nebular emission in the Galactic Plane:

\begin{displaymath}
{\rm W3RCHI2} \ge 3\ {\rm and}\ |b| \ge 10 \ {\rm and}\
{\rm W3SIGM}\ \mbox{ is not null}
\end{displaymath}

\noindent
These search criteria yield a total of 202,851 sources, which composes our parent sample. 

\subsection{Photometry}

The \wise\ All-sky database provides photometry measurements using a variety of methods: PSF profile fitting, variable aperture photometry (eight circular apertures), curve of growth (COG) aperture photometry, and
elliptical aperture photometry (for sources matched to the 2MASS
Extended Source Catalog (XSC)).

The profile fitting photometry is referred to as W\#MPRO and is defined as the magnitude measured with profile-fitting photometry. In addition to magnitudes this procedure also derives the 
signal-to-noise ratio (SNR) and WRCHI2 (i.e., the goodness of fit to
the PSF model of the source) or the goodness of fit to a PSF model of the source.

The COG or ``standard'' aperture photometry is referred to as W\#MAG
. According to the \wise\ explanatory supplement ``This is the curve-of-growth 
corrected source brightness measured within an 8.25\arcsec radius
circular aperture centered on the source position. The background sky
reference level is measured in an annular region with inner radius of
50\arcsec\ and outer radius of 70\arcsec '' \footnotemark{\ref{wexp}}. 

The \wise\ pipeline performed nested circular aperture
photometry. They used eight apertures from 5.5\arcsec\ to
24.75\arcsec\ in W1-W3 and 11\arcsec\ to 49.5'\arcsec\ in
W4\footnote{\url{http://wise2.ipac.caltech.edu/docs/release/allsky/expsup/sec2_2a.html} \label{wexp}}. They provide these in the form of parameters named W\#MAG\_1 for the first aperture of 5.5$''$ in W1 through W3 and 11$''$ in W4, W\#MAG\_2 for the 2nd aperture, W\#MAG\_3 for the 3rd aperture and so on.

The elliptical aperture measurements, referred to as W\#GMAG were based on matched
sources between \wise\ and the 2MASS Extended Source Catalog (XSC). The shape of the elliptical aperture was determined by utilizing the shape information of the source provided by the 2MASS XSC. We
discuss the uncertainties in our best photometry from calibrating
aperture magnitudes to careful extended source photometry in Section~\ref{calibrate}.

Since our analysis is primarily interested in identifying galaxy-scale extraterrestrial civilizations K3s, we required that photometric measurements of extended 
sources in \wise\ be as reliable as possible. The PSF profile
fitting photometry is best used when extracting photometry from
point-like sources in the \wise\ survey. For the majority of extended
sources in \wise, the COG photometry generally provides more reliable measurements, though the strict 8.25$''$ radius circular aperture fails for large and extended galaxies. Measuring reliable 
photometry for extended galaxies has historically proven to be a non-trivial and difficult task, the reason being that galaxies come in all shapes and sizes, and a single ``standard'' aperture fails to encompass
the full range of galaxy sizes and structures.   

Recently, \citet{Jarrett12} have attempted to extract reliable galaxy photometry for extended sources in the \wise\ catalog. They have developed a complex algorithm in order to construct the 
\wise\ High Resolution Galaxy Atlas and present their initial results in \citet{Jarrett13}. Even more recently, Tom Jarrett has measured reliable photometry for $\sim 67,000$ 
extended sources in the \wise\ catalog contained within the South Galactic Cap (SGC: $b < -60$). We use this analysis to calibrate the aperture magnitudes as presented by the \wise\ All-sky database, as we describe in Section~\ref{calibrate}

\subsection{Quality Flags}

\begin{figure*}[htp]
\begin{center}
\begin{tabular}{l}
\includegraphics[width=7in,trim=0 2in 0 1in,clip]{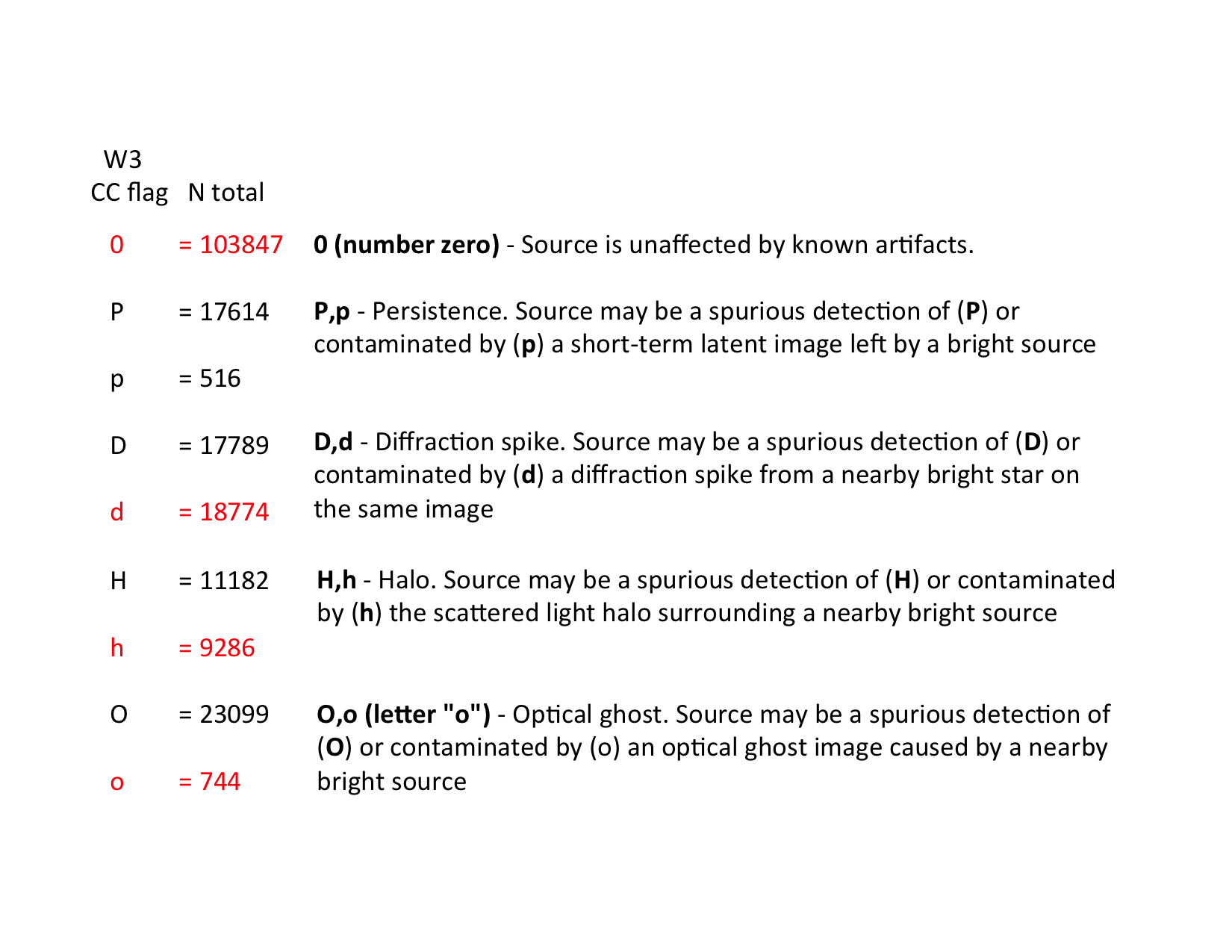}
\end{tabular}
\end{center}
\caption[CAPTION]{\label{flow} Description of cc flags, copied verbatim from the \wise\ supplementary catalog.\footnote{Available online at
\url{http://wise2.ipac.caltech.edu/docs/release/allsky/expsup/sec2_2a.html\#cc_flags}}}
\end{figure*}

In order to remove spurious sources we use the contamination and confusion flags (CC\_FLAG) given for W3. In Figure~\ref{flow}, we describe 
the definitions and provide the number counts of objects having the
various CC\_FLAGs. We retain sources with the following 
CC\_FLAG: `0',`d',`h', and `o'. We consider sources with CC\_FLAG `P',`p',`D', `H' and `O' to be contaminated and we removed them from
further analysis. Our decision for choosing these particular flags are motivated on empirical examination of a significant fraction of the flagged
entries (i.e.\ we found that the rejected flags have very good
reliability in flagging false sources, but the ones we retain often
appear for real sources). Using these quality flags reduces the
12$\mu$m sample to 132,651 sources.

\subsection{Coordinate and Color Cuts}

In addition to removing the Galactic Plane, we also applied three
additional coordinate cuts to remove the most obvious high density
regions of foreground contamination from objects in star-forming
regions.  Region 1 is composed of the Galactic Bulge, (i.e., -12 $<$ $l$ $<$ 10 and 20 $<$ $b$ $<$ 10); the total area covered by this region is 
$212.23$\ deg$^2$ and contained 9,323 sources. We removed all sources in Region 1 from our sample. Region 2 comprises sources associated with the Large Magellanic Cloud (LMC); this region 
contains a total of 13,104 sources. Region 3 is composed of sources
associated with the Orion Nebula and contains a total of 12,733
sources. Instead of imposing a blanket
coordinate cut for these regions of patchy contamination, we found that we could rather reliably identify
foreground sources by rejecting highly clustered sources and retaining
relatively isolated sources using a surface density algorithm.  This
reduces our sensitivity to K3s in regions 2 and 3 without rendering
us completely blind to them.  After we removed the sources in these
three regions, the 12$\mu$m sample was reduced to 97,491 sources. 

We use two simple color cuts to eliminate the most obvious stellar contaminants (it should be noted that these cuts also eliminate a large number of elliptical galaxies, since these generally tend to be blue in the infrared). 
We identify the stellar locus in the lower left (blue) corner of the
diagram (since stellar photospheres have neutral MIR colors) and use the following criteria 

\begin{displaymath}
{\rm  W2MPRO} - {\rm W3MPRO} < 2\ {\rm and}\ {\rm W3MPRO} - {\rm W4MPRO} \le 1 
\end{displaymath}

\noindent
to remove them from further analysis.
We are motivated to use the profile fit photometry because these sources are considered to be stellar-like and thus
should have reliable profile fitting photometry. 

% FIGURE 3
\begin{figure}[htp]
\begin{center}
\begin{tabular}{l}
\includegraphics[width=3.5in,trim=0 2.5in 0 0,clip]{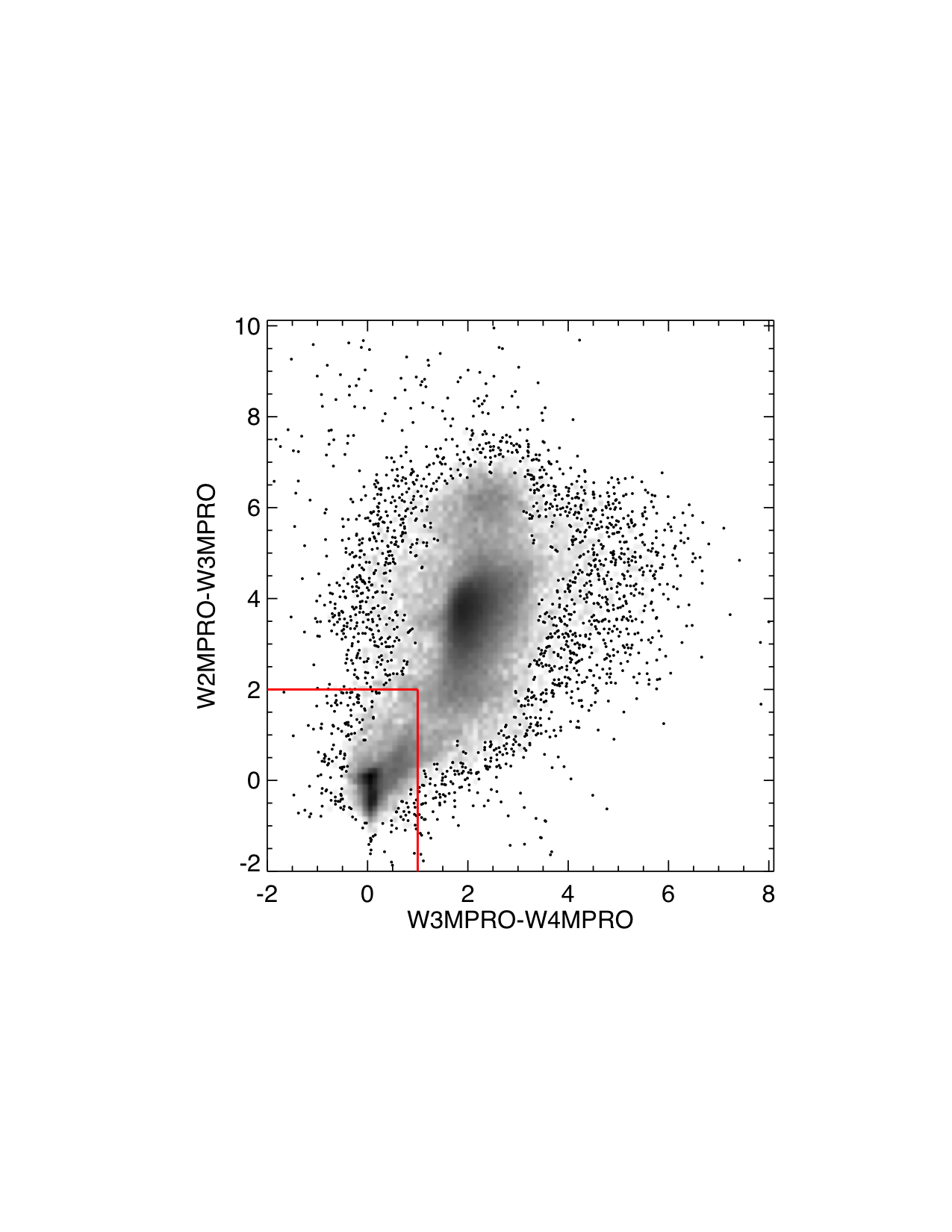}
\end{tabular}
\end{center}
\caption[CAPTION]{\label{fig:3} W3MPRO - W4MPRO versus W2MPRO - W3MPRO for a sample of $\sim$ 70,000 sources. We use this diagram 
to identify the stellar locus and remove these sources from further analysis.}
\end{figure}

\begin{figure}[htp]
\begin{center}
\begin{tabular}{l}
\includegraphics[width=3.5in,trim=1.5in 2in 0 0,clip]{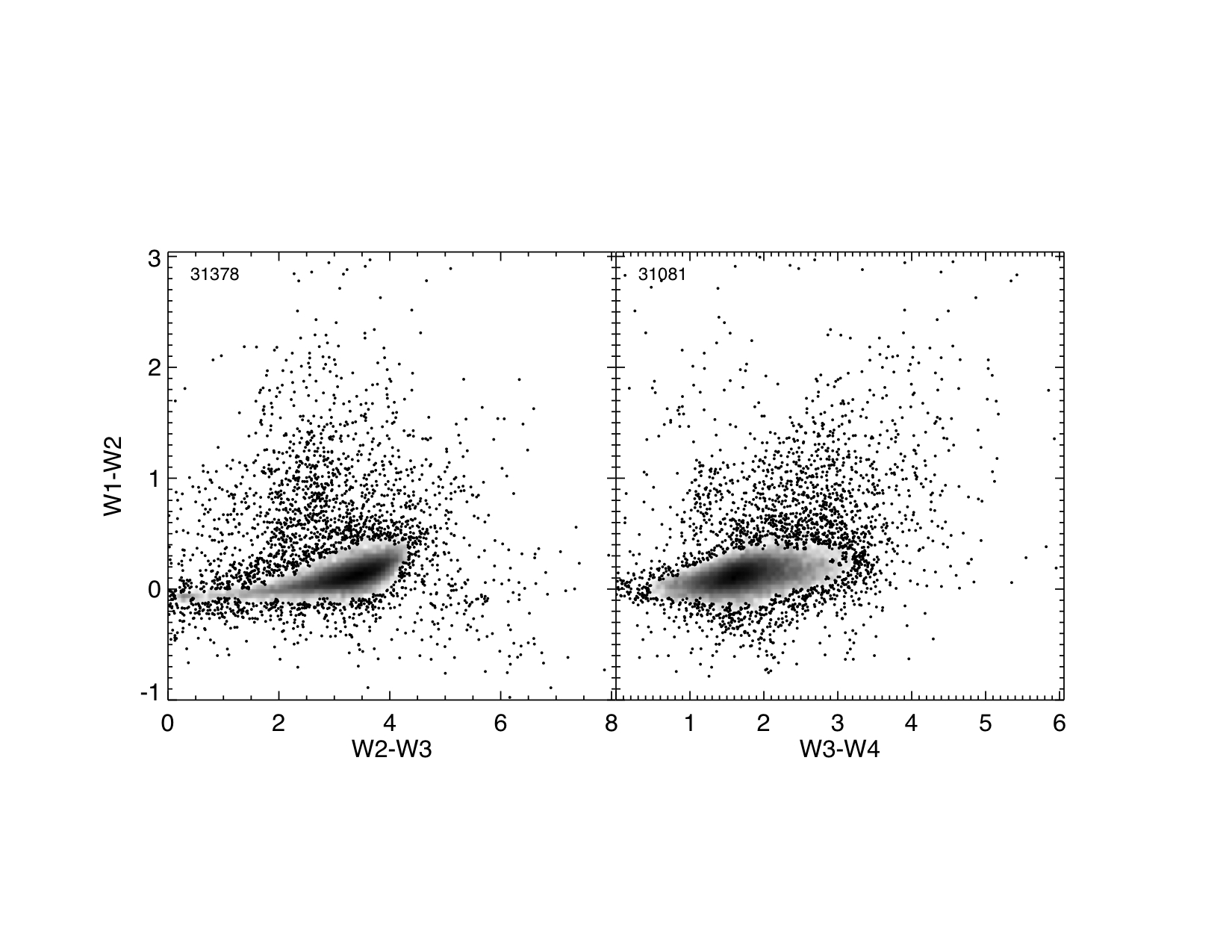}
\end{tabular}
\end{center}
\caption[CAPTION]{\label{ExtendedCC} Corrected W1CMAG - W2CMAG versus W2CMAG - W3CMAG
  {\it (left)} and W3CMAG - W4CMAG {\it (right)} for the W3 Extended Red
  Sample (i.e.\ our sample of All-sky entries that appear to be
  extended in W3 and with red MIR colors). High density regions are
  represented in a logarithmic 
  greyscale; the total number of sources in each plot is listed in the
  plot's corner.}
\end{figure}

We visually examined a representative sample of sources in this region
and concluded that they are indeed stellar-like objects with
anomalously high W3RCHI2 values and since we are primarily concerned 
with red extended objects, this cut is compatible with our overall search. 
These criteria identified a total of 21,645 stellar-like sources,
leaving 75,846 sources in the 12$\mu$m extended sample for further
inspection.  We present a color-color map of these sources in Figure~\ref{ExtendedCC}.

\subsection{Visual Classification}
\label{visual}
\begin{figure*}[htp]
\begin{center}
\begin{tabular}{l}
\includegraphics[width=7in,trim=0 1in 0 0,clip]{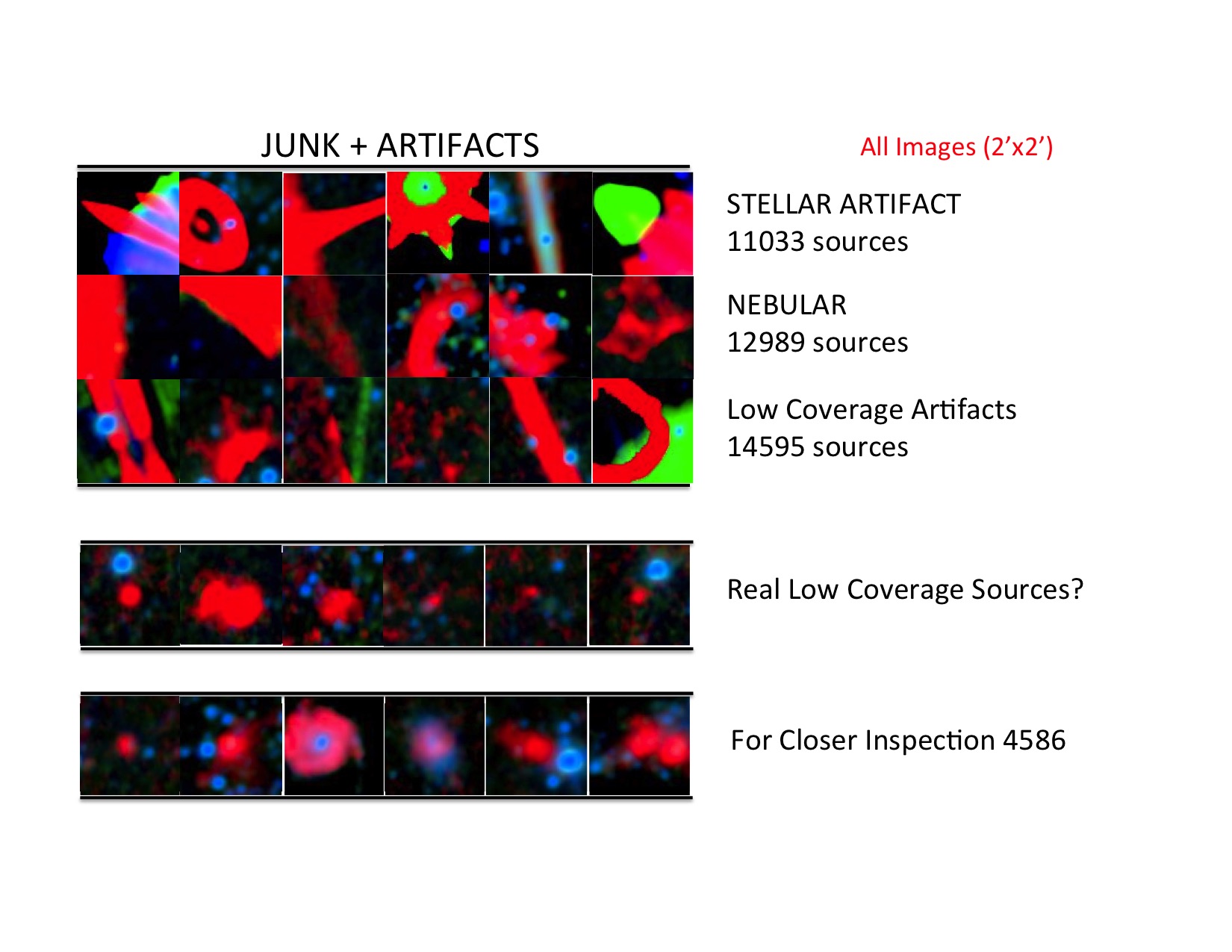}
\end{tabular}
\end{center}
\caption[CAPTION]{\label{artifact_examples} We present examples of the color images for the visually classified sources. The top 3 rows show cases where the source is considered to be the most obvious examples
of astrophysical and instrumental artifacts. The fourth row shows
sources with low coverage W3M $\le 5$ but which could potentially be real
astrophysical sources. The fifth row shows sources with nominal
coverage but lacking 2MASS associations.  Some such objects are real
sources (often blends of multiple sources), and we separated these from artifacts by visual inspection.}
\end{figure*}

We constructed $2^\prime \times 2^\prime$ \wise\ color images ((W1+W2)/2 = Blue, W3 = Green, and W4 = red) for the remaining 75,846 sources and have visually classified them 
into five primary groups. The five groups are: {\bf{stellar artifacts}}, {\bf{low coverage artifacts}}, {\bf{nebular}}, {\bf{needs closer inspection}}, and 
{\bf{high quality}}. Representative examples are presented in
Figure~\ref{artifact_examples}.

The {\bf{stellar artifacts}} comprise 11,033 sources which were caused by bright saturated stars (i.e., halos, streaks, and latents). Most of these did not have any CC\_FLAGs 
indicating a problem in the All-sky release.

The {\bf{low coverage artifacts}} are sources which are likely an artifact because they were observed
fewer than 6 times by \wise\  under nominally good conditions(i.e.\
W3M $<$ 6, where W3M gives the number of individual 8.8s W3 exposures
on which a profile-fit measurement of the source was possible). To put
this number in perspective, the median W3M for all 202,851 sources was
13. There were a total of 14,595 low coverage sources. 
We visually examined this sample and recovered 215 sources, which,
though having low coverage, appear to be real astrophysical sources
and are considered for further analysis.

The {\bf{nebular}} sources comprise 12,989 sources and we determined them to be locally bright regions of large, nebular 
networks of Galactic dust, and so not discrete objects. These sources required images with a much larger FOV (20$'\times20'$) to be classifiable. Since these are highly extended 
in nature we first identified sources within a 10$'$ radius and constructed $20'\times 20'$ color images. The brightest W3 source within a cluster was used as the center of a single color image for the field. 
We recovered 192 sources that appeared to be `nebular' in the 2$'\times 2'$ image, but we reclassified as being discrete objects
after inspecting the $20'\times 20'$ images. 

There were 4,727 sources laking a 2MASS association the we labeled
{\bf{for closer inspection}}, since they appear to be legitimate sources
with good coverage. The majority of these sources were duplicates of
sources already included in the {\bf{high quality}} sample. 
The sources which were unique were rematched to the 2MASS catalog using a two-fold process: The first matching used a 30$''$ radius and recovered
photometry for 509 sources. The second matching used a 60$''$ search radius and recovered photometry for 44 sources. 

\begin{figure*}[htp]
\begin{center}
\begin{tabular}{rl}
\includegraphics[scale=0.27,trim=3.5in 4in 1in 2in,clip]{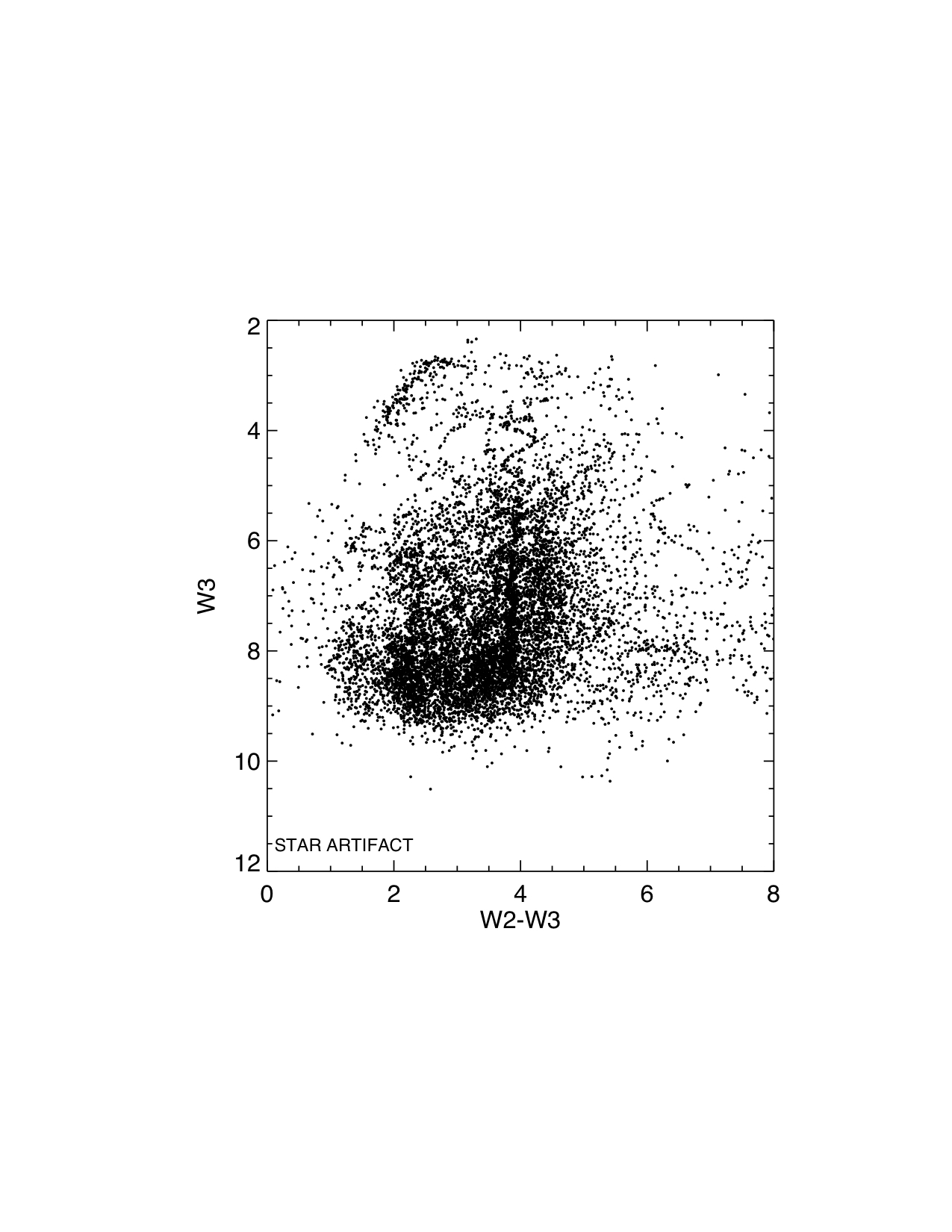}
\includegraphics[scale=0.27,trim=3.5in 4in 0 2in,clip]{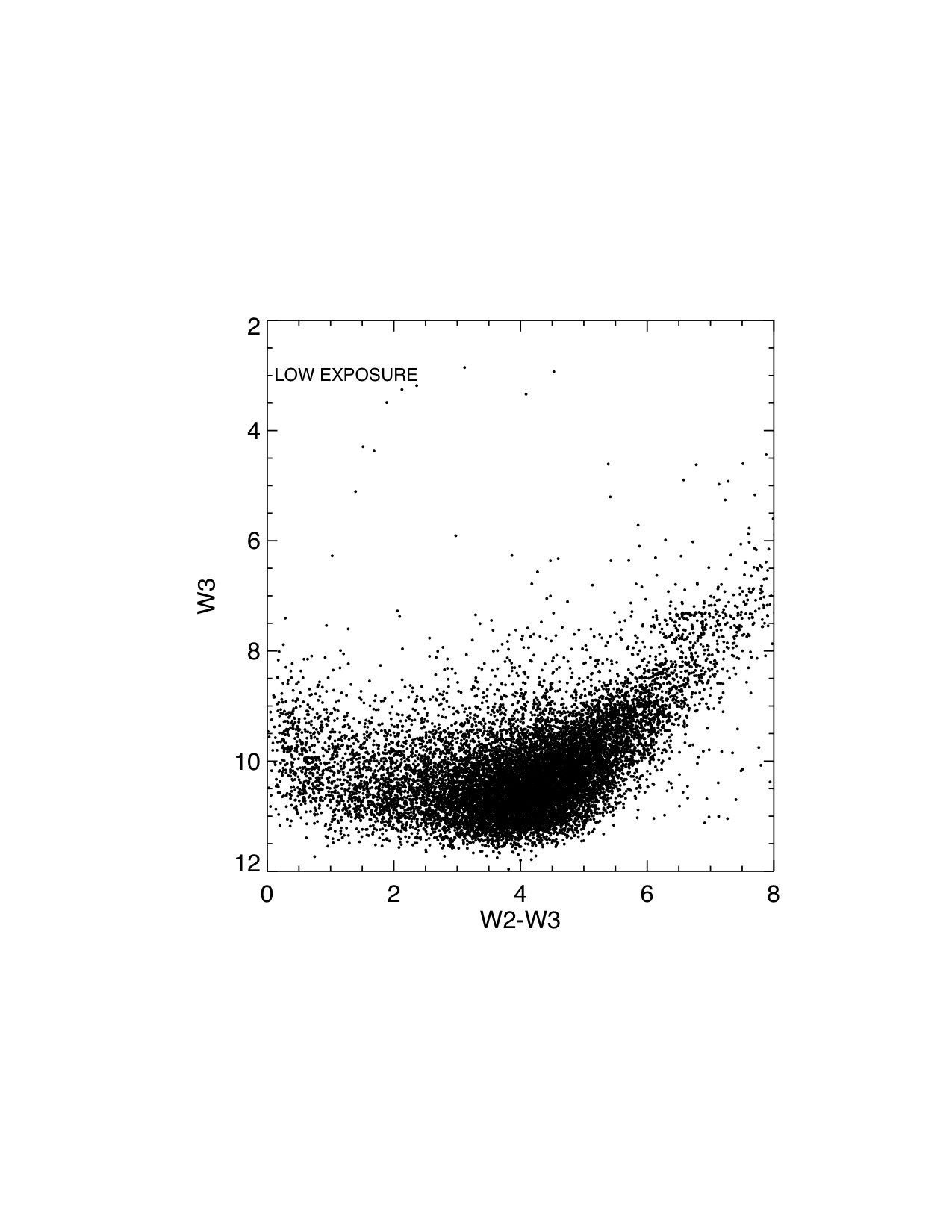}\\
\includegraphics[scale=0.27,trim=3.5in 2in 1in 4in,clip]{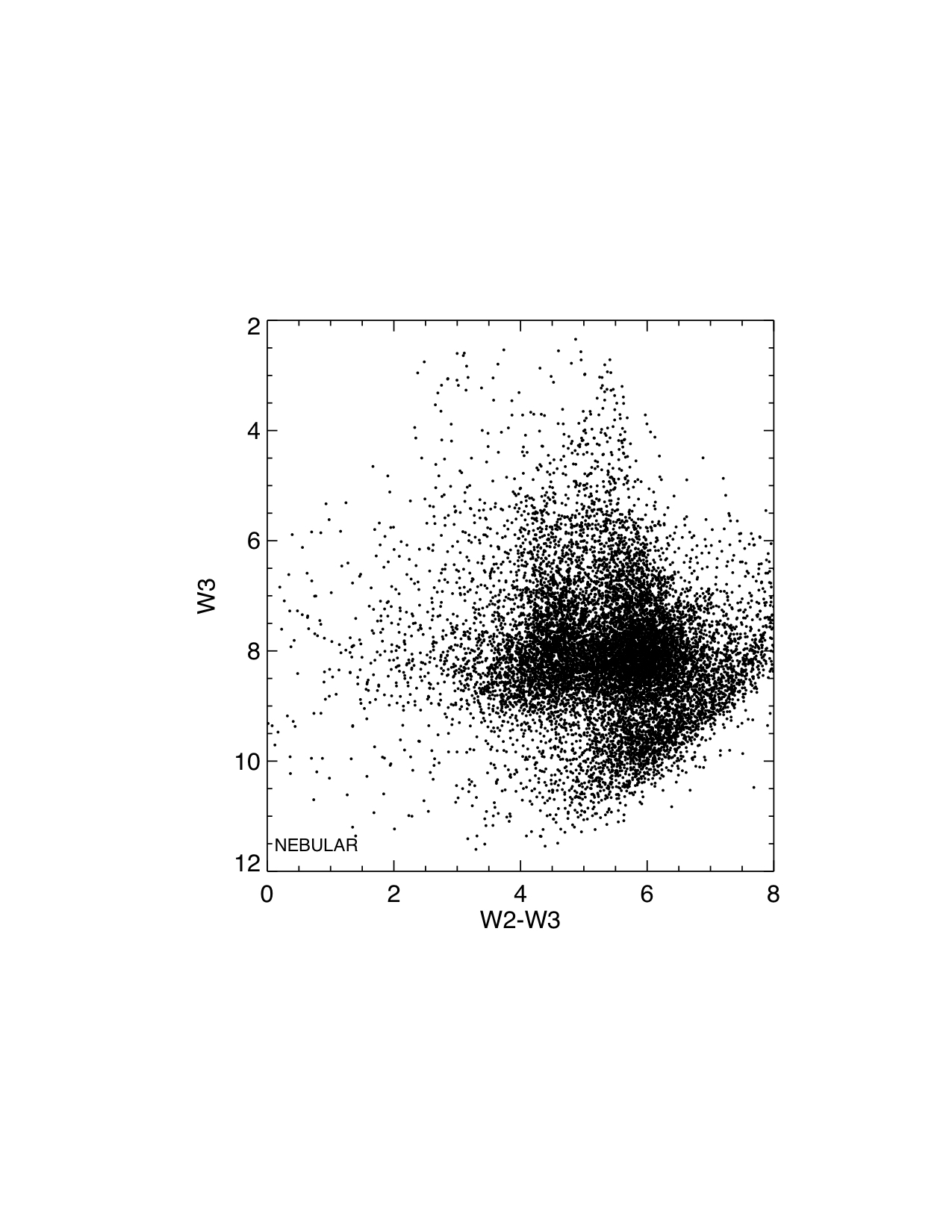}
\includegraphics[scale=0.27,trim=3.5in 2in 0 4in,clip]{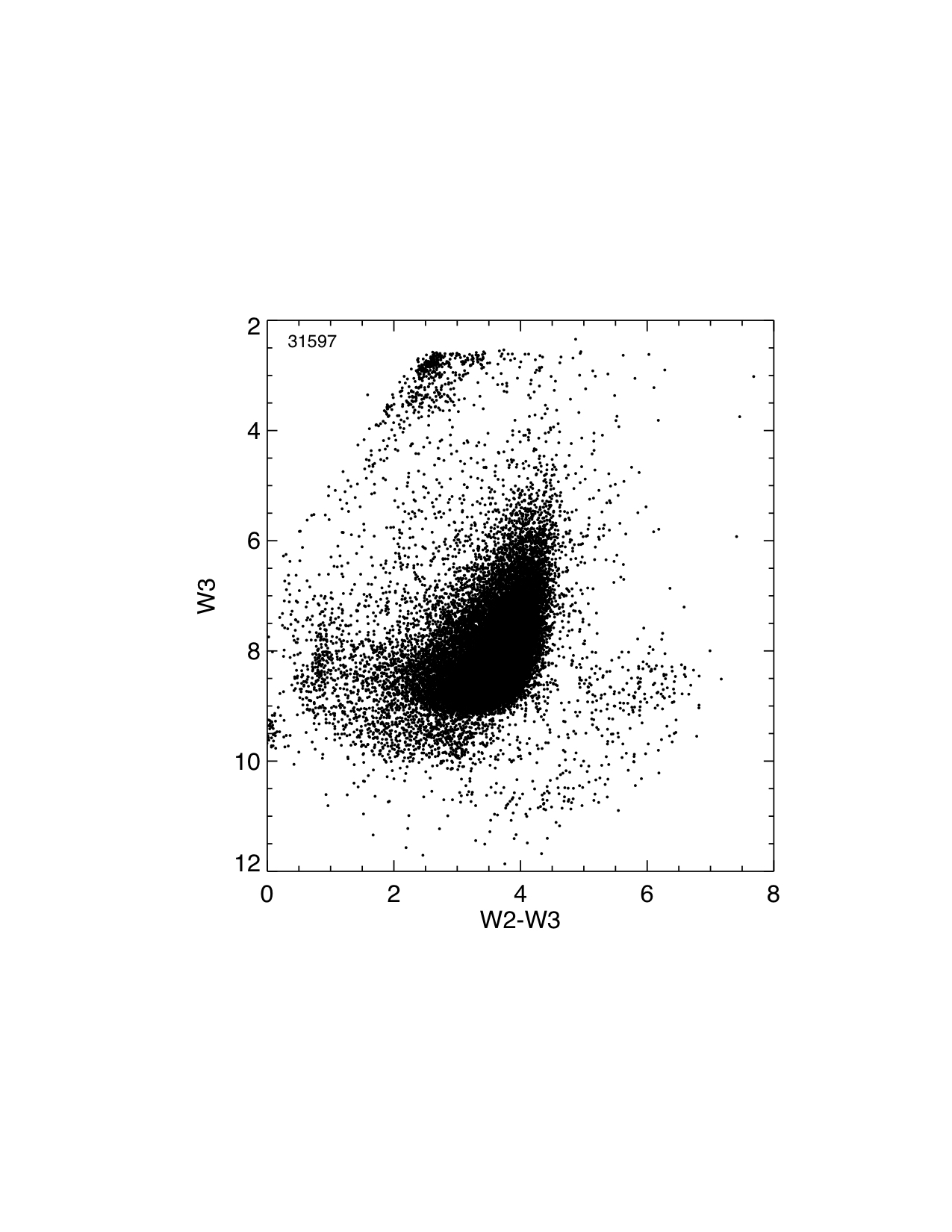}
\end{tabular}
\end{center}
\caption[CAPTION]{\label{fig:4}We show W3 magnitude versus W2 - W3 color for classified sources. Top Left: Stellar artifacts. Top Right: Low exposure sources.
Bottom Left: Nebular type sources. Bottom Right: W3 Extended Gold Sample.}
\end{figure*}

Figure~\ref{fig:4} illustrates the typical colors and magnitudes of these categories
of sources, and of the W3 Extended Gold Sample (i.e.\ our sample of
{\it real} sources extended in W3 with red MIR colors).

% FIGURE 4

We identified 32,502 sources which had 2MASS associations and appear to be real astrophysical objects based on their color images as {\bf{high quality}} sources.
Some extended objects appear in the catalog multiple times. We removed the duplicate entires (those within 60$''$) of the brightest entry in the catalog.
We identified a total of 1,694 duplicated sources which we removed
from the sample.

The number of galaxies we have in our final sample here makes sense.  Most of the high latitude extended
sources in \wise\ are galaxies, and as we show in Section~\ref{number},
there are approximately 100,000 galaxies that would be extended in the W3
\wise\ band if they had significant W3 emission.  We have thus
selected approximately the reddest 1/3 of the galaxies on the sky
larger than the angular resolution of the \wise\ W3 band.

We have thus constructed a sample of mostly real,
extended, and discrete sources in the \wise\ All-sky release which we
refer to as the {\it W3 Extended Gold Sample}.
\label{GoldBuild}

\begin{figure*}[htp]
\begin{center}
\begin{tabular}{l}
\includegraphics[width=7in,trim=0 2in 0 1in,clip]{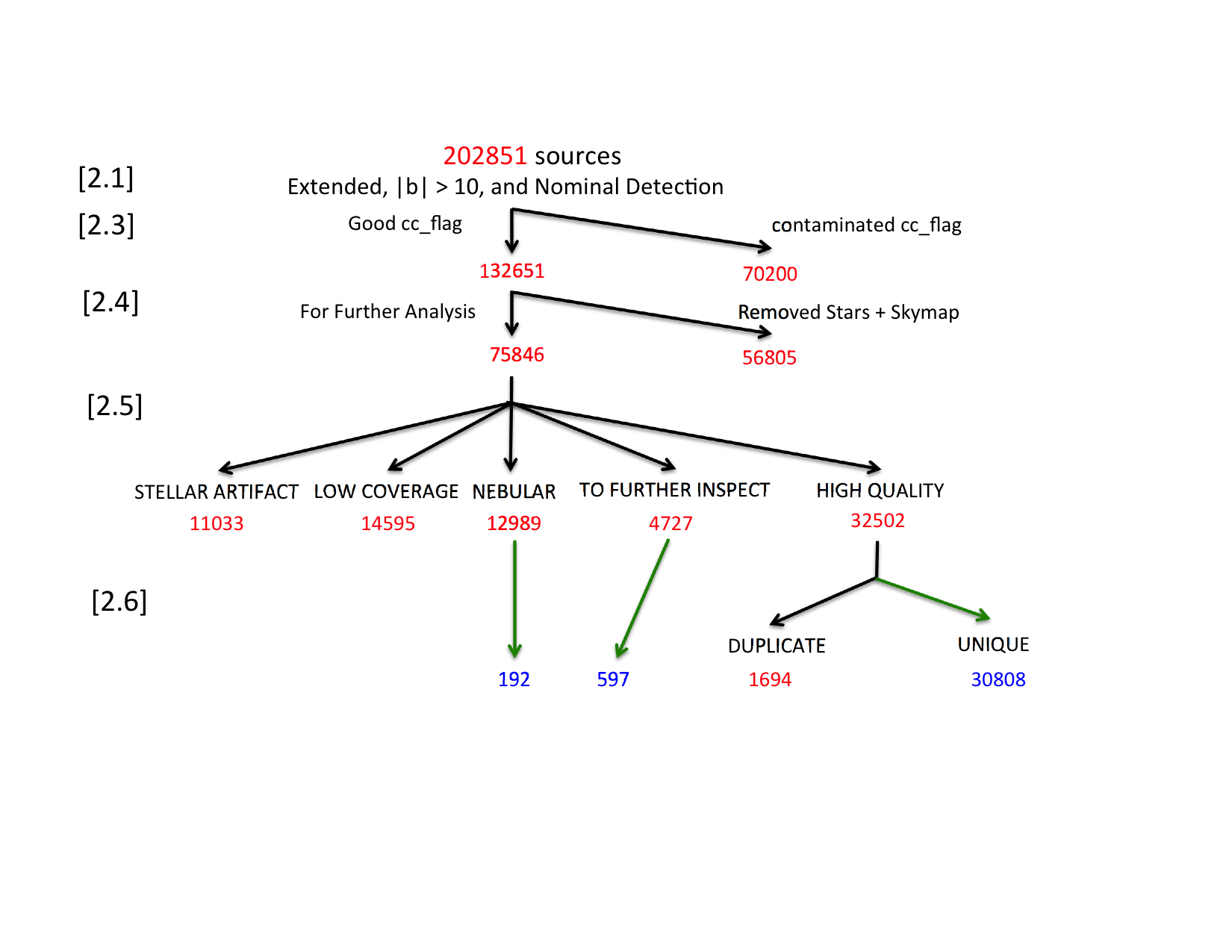}
\end{tabular}
\end{center}
\caption[CAPTION]{\label{fig:2} Schematic diagram of W3 extended
  source analysis. Numbers to the left of Figure ~\ref{fig:2} refer to the section numbers in this manuscript.
Extended sources are identified as those with W3RCHI2 $\ge 3$ and nominal detection are those where W3SIGM is not null.}
\end{figure*}

\begin{figure*}[htp]
\begin{center}
\begin{tabular}{l}
\includegraphics[scale=0.35,trim=1.5in 3in 0 2in,clip]{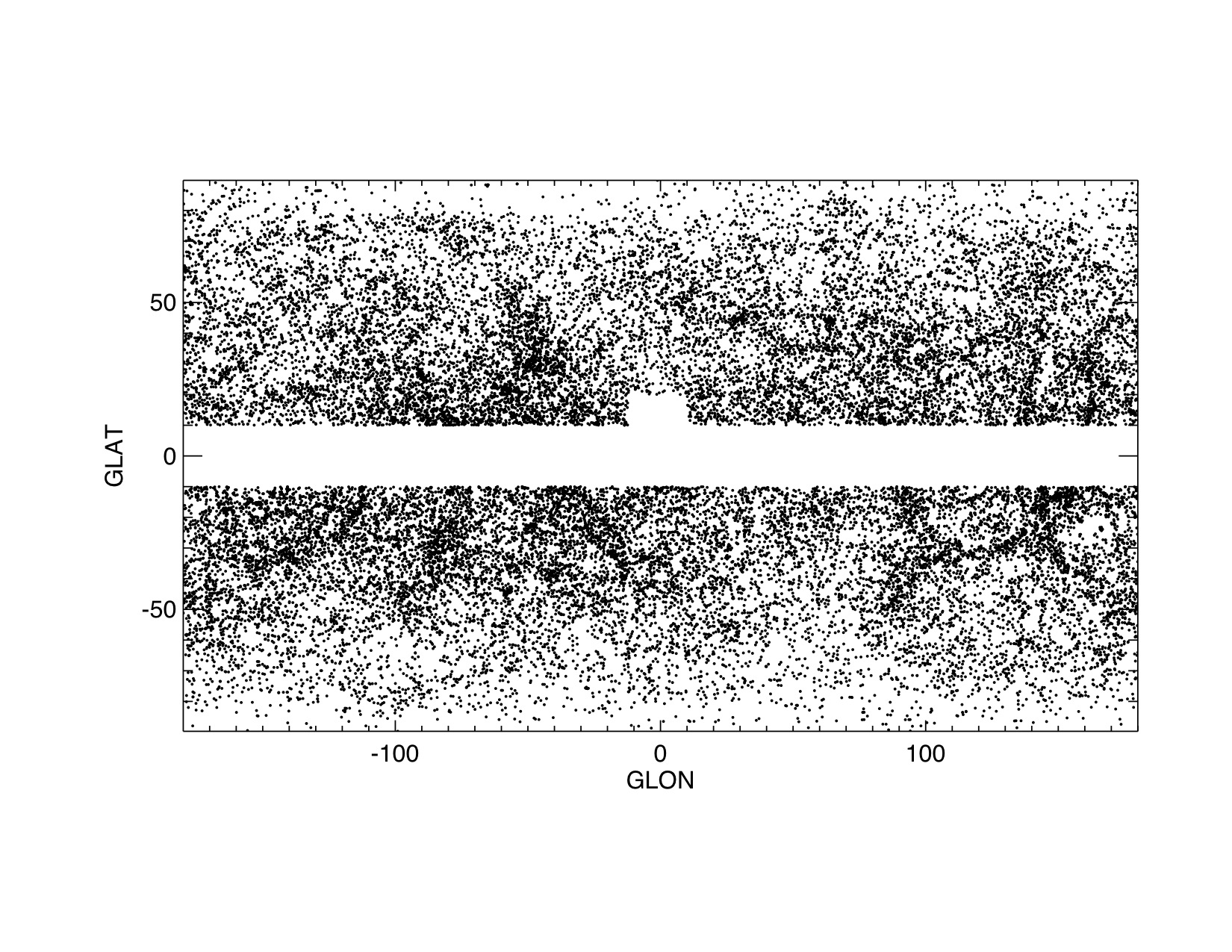}
\end{tabular}
\end{center}
\caption[CAPTION]{\label{fig:5} We present the sky distribution for the W3 Extended Gold Sample.}
\end{figure*}

We summarize the vetting procedure that produced the W3 Extended Gold
Sample in Figure~\ref{fig:2} and present the sky distribution for this sample in Figure~\ref{fig:5}

% FIGURE 2

% FIGURE 5

\subsection{W3 Extended Gold Sample Photometric Calibration}

\label{calibrate}

The W3 Extended Gold Sample comprises 30,808 apparently extended
sources from the {\bf{high quality}} sample, 597 sources from the
{\bf{for closer inspection}} sample and 192 sources from the
{\bf{nebular}} sample, bringing the final sample to 31,597 sources
(shown in the lower right plot of Figure~\ref{fig:4}). As
discussed in Section~\ref{GoldBuild} , the \wise\ COG photometry is generally
reliable for most sources in this sample, but some care must be taken
before using these measurements.

\begin{figure*}[htp]
\begin{center}
\begin{tabular}{l}
\includegraphics[width=8in,trim=1.5in 3in 0 2in,clip]{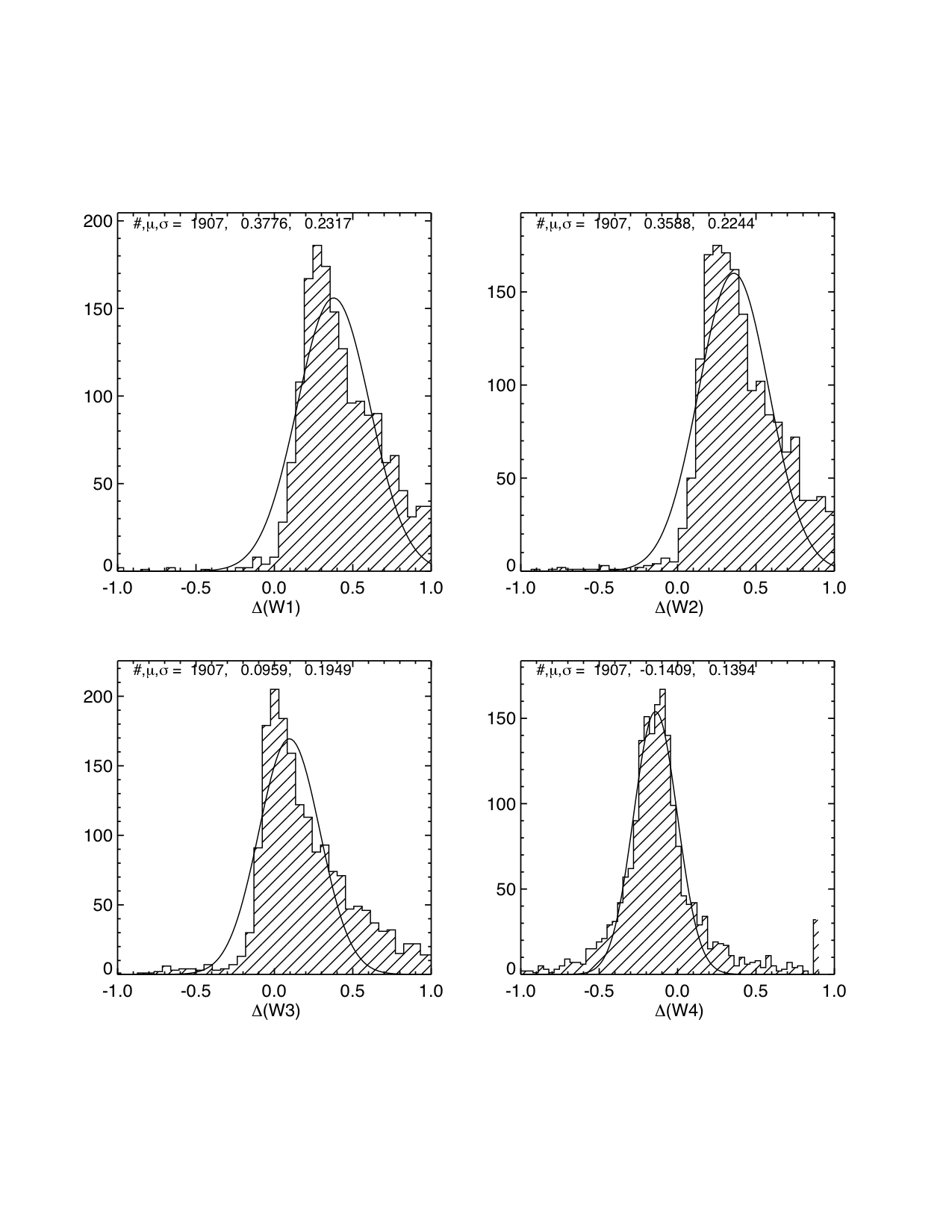}
\end{tabular}
\end{center}
\caption[CAPTION]{\label{AllSkyvsTJOriginal} \wise\ WMAG
  (curve-of-growth corrected) All-sky photometry
  vs.\ aperture photometry for extended sources in the South Galactic Cap carefully extracted by Tom Jarrett.  Numbers in
  the plots describe the  number, mean ($\mu$), and standard deviation ($\sigma$) of Gaussian
  fits to the data.  The pipeline photometry has offsets and
  color-dependent systematic errors for these extended sources.} 
\end{figure*}

Tom Jarrett has kindly provided us with a preliminary version of his
extended source catalog for the south Galactic Cap ($b<-60$) in
advance of this catalog's publication.  To further calibrate the
extended source photometry and determine the degree of systematic
errors in the \wise\ pipeline extraction for extended sources, we have cross-matched
1,907 \wise\ All-sky sources with sources in Jarrett's preliminary
catalog (210 sources did not match, usually because the source was
only marginally extended in one of the catalogs). We found that for our W3 Extended Gold Sample, the distribution of
differences between the All-sky COG magnitudes and Jarrett's magnitudes
were significantly offset from zero (by 0.4, 0.4, 0.1, and -0.15
magnitudes in W1, W2, W3, and W4) and skewed with long tails past 1
magnitude, with the Jarrett magnitudes typically brighter than the COG
magnitudes.  
%The systematic errors removed by Jarrett's improved analysis
%(measured as the ``robust sigma" of these distributions, see below), are XX, XX, XX, and XX in the 4 bands. 
In Figure~\ref{AllSkyvsTJOriginal} we present the comparison between the COG magnitudes
presented in the All-sky release to those measurements presented in Jarrett's preliminary catalog.

We estimated the size of the sources using the difference between the
profile, 4\nth, and 8\nth\ aperture magnitudes, W$_\mathrm{profile}$,
W\#\_4, and W\#\_8.  For unblended point sources these magnitudes are
typically identical, but for extended sources the larger aperture
magnitudes are more faithful, because they measure flux further up the
curve of growth.   We formed ``size'' parameters from the quantities
(W\#\_8-W\#\_4) and (W\#\_8-W$_\mathrm{profile}$), and found that by
using the W\#\_8 magnitudes, with quadratic corrections in the size
parameters and constant offsets, we could reproduce the Jarrett
magnitudes to within 0.06 mag in most cases and colors to within 0.08
mag.

The transformation we used to generate accurate extended source
magnitudes W\# from the \wise\ pipeline photometry is:

\begin{eqnarray}
{\rm S1} &=& {\rm W\#}\_8-{\rm W\#}\_4 \\
{\rm S2} &=& {\rm W\#}\_8-W_\mathrm{profile} \\
{\rm W\#} &= & a_0 + a_8 {\rm W\#}\_8 + a_{s1,1} {\rm S1} + a_{S2,1}\\
&&{\rm S2} + a_{S1,2} {\rm S1}^2 + a_{S2,2} {\rm S2}^2
\end{eqnarray}

We performed a singular value decomposition to determine the best
values for the coefficients, and report our coefficients in
Table~\ref{correctioncoefficients}:

\begin{center}
\begin{deluxetable}{rcccccc}
\tablewidth{0pt}
\tablecaption{Correction coefficients to produce corrected magnitudes
for extended sources.}
\tablehead{\colhead{band} & $a_0$ & $a_8$ & $a_{s1,1}$ & $a_{S2,1}$ &
$a_{S1,2}$ & $a_{S2,2}$}
\startdata
W1 &-1.083 & 1.093 & 0.107 & -0.591 & 0.021 &-1.114 \\
W2 &-1.211 & 1.103 & 0.246 & -1.104 & 0.062 &-1.540 \\
W3 & -0.419 & 1.050 & 0.030 &-0.449 & 0.049 &-1.130 \\
W4 & -0.588 & 1.106 & 0.006 &-1.021 &-0.228 & 0.676
\label{correctioncoefficients}
\enddata
\end{deluxetable}
\end{center}

\begin{figure*}[htp]
\begin{center}
\begin{tabular}{l}
\includegraphics[width=7.5in,trim=1.5in 3in 0 2in,clip]{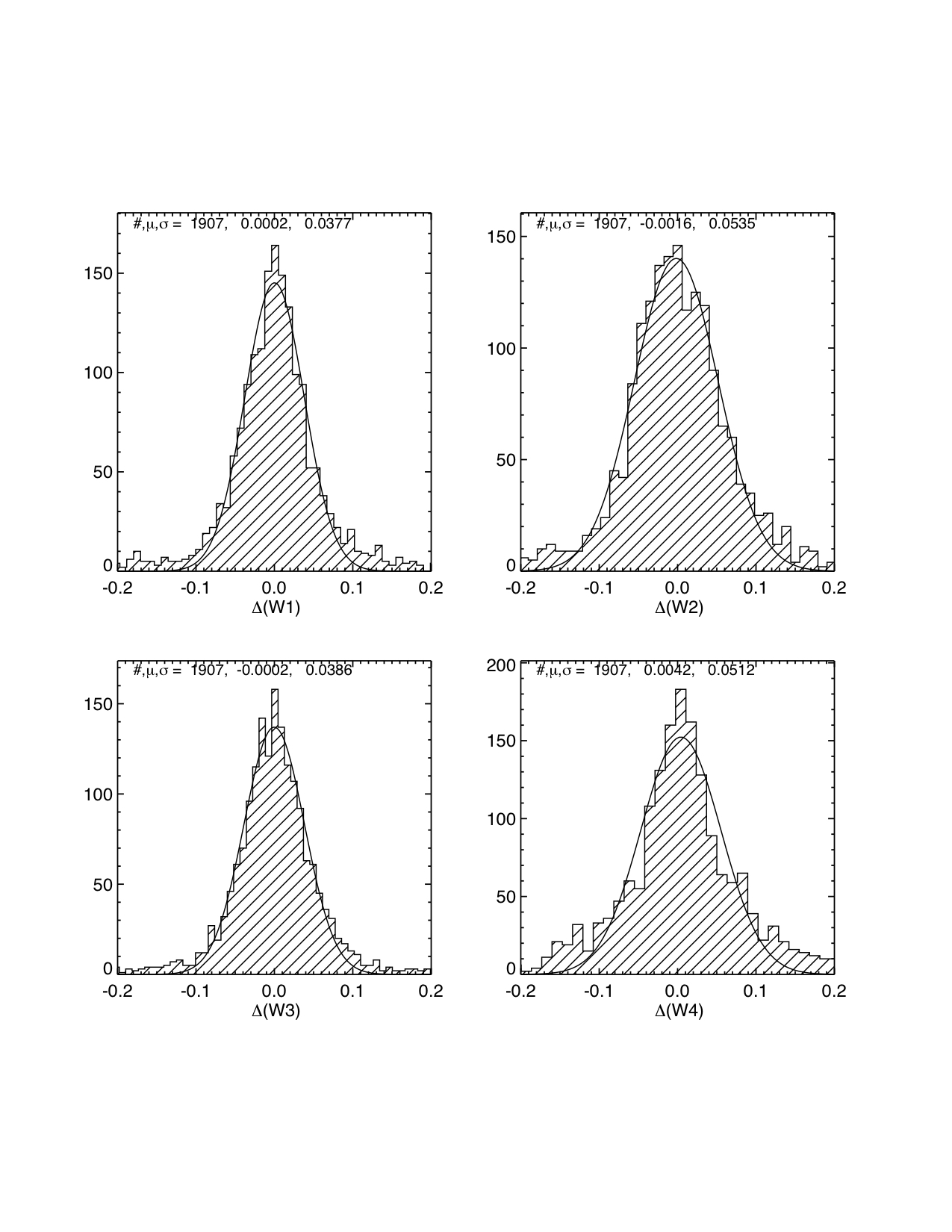}
\end{tabular}
\end{center}
\caption[CAPTION]{\label{AllSkyvsTJ} Results of our calibration of
  \wise\ photometry to careful extended source photometry of sources
  in the south Galactic Cap ($b < -60$) provided by Tom Jarrett.
  We present the distribution in differences in our corrected \wise\
  magnitudes in the four \wise\ bands to the Jarrett photometry for common sources.  Numbers in
  the plots describe the  number, mean ($\mu$), and standard deviation ($\sigma$) of Gaussian
  fits to the data.  These ``corrected'' magnitudes have roughly
  Gaussian errors and little or no systematics with color or size for
  most sources.} 
\end{figure*}

\begin{figure*}[htp]
\begin{center}
\begin{tabular}{l}
\includegraphics[width=8in,trim=1in 1.5in 0 1in,clip]{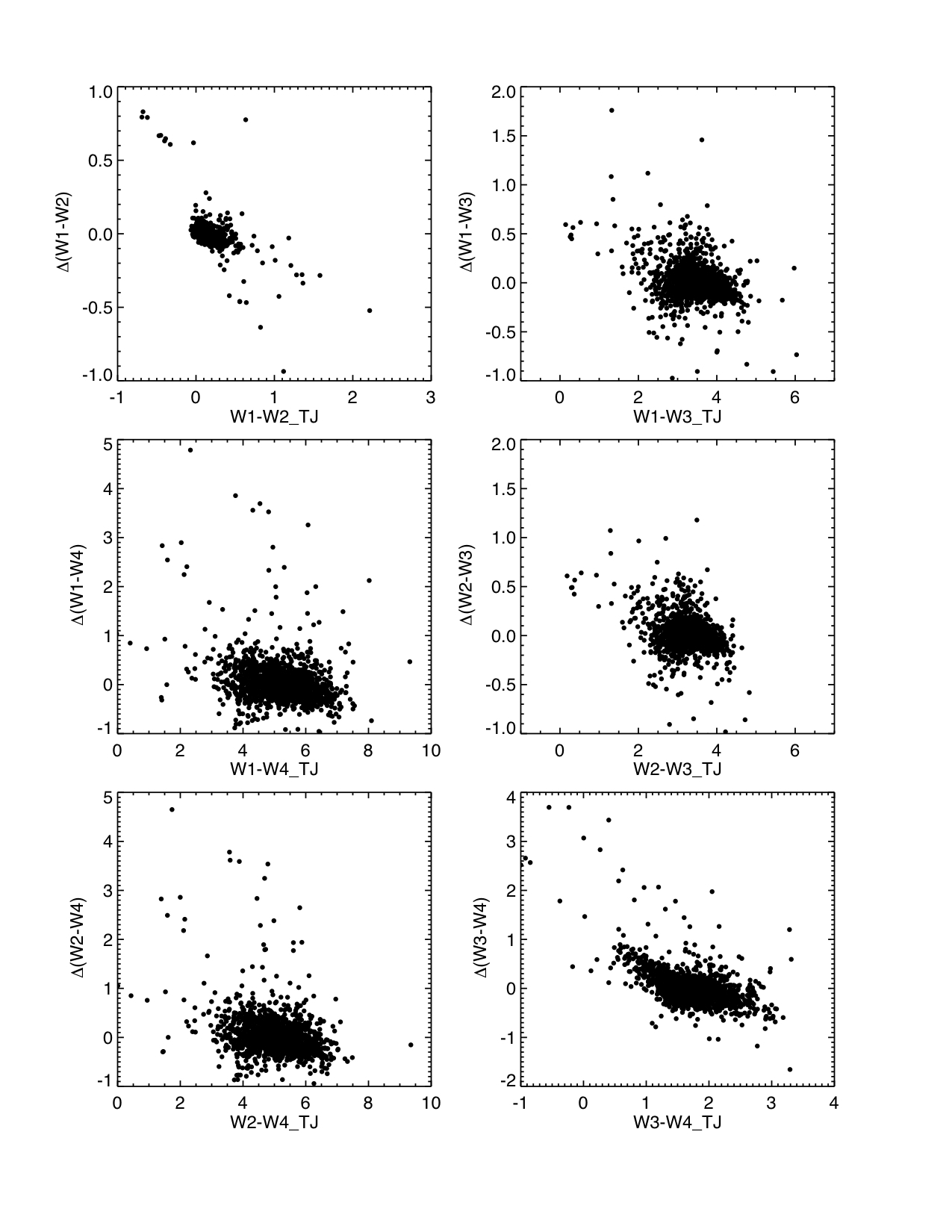}
\end{tabular}
\end{center}
\caption[CAPTION]{\label{AllSkyvsTJColor} Results of our calibration of
  \wise\ photometry to careful extended source photometry of sources
  in the south Galactic Cap ($b < -60$) provided by Tom Jarrett.  For
  all six color combinations of the four \wise\ bands (using the
  Jarrett magnitudes),
  we present the distribution in differences in our corrected
  magnitudes to the Jarrett photometry for common sources.  Numbers in
  the plots describe the  number, mean ($\mu$), and standard deviation ($\sigma$) of Gaussian
  fits to the data.  These ``corrected'' colors have roughly
  Gaussian errors and little or no systematics with magnitude or size for
  most sources.}
\end{figure*}

We present a comparison of our corrected photometry to that of the
preliminary Jarrett photometry in Figure~\ref{AllSkyvsTJ} \&
\ref{AllSkyvsTJColor}.  Because these distributions are roughly
Gaussian but with extended wings, we report the precision of our
photometry in three ways.  In our figures we show the best-fit
Gaussians to each of the residual distributions, and report the width
of these Gaussians as $\sigma$; this represents the typical systematic
error due to limitations of the \wise\ pipeline for most of our
extended sources.  The long wings of the residual distribution
typically represent blended and highly structured sources for which our
simple calibration scheme failed to match Jarrett's more careful
analysis, and a small number of extreme outliers inflate the standard
deviation of these distributions beyond utility.  To quantify the
systematic photometric errors including the bulk of the non-Gaussian
wings in a more robust manner, we calculated a ``robust sigma''
including outlier rejection\footnote{Using the IDL routine {\tt
ROBUST\_SIGMA} written by H. Freudenreich, who cites ``Understanding
Robust and Exploratory Data Analysis'' \citep{Hoaglin}}, and
also the value of the 68\nth\ percentile absolute deviation from the
median for each magnitude and color combination.  We present the
results of all three error estimates in Table~\ref{Errors}.

\begin{deluxetable}{cccc}
\tablecaption{Three measures of systematic errors in our extended
source photometry.}
\tablehead{\colhead{Quantity} & \colhead{$\sigma$} & \colhead{Robust
st. dev. (\# rejected)} & \colhead{68\%-ile}}
\tablewidth{0pt}
\startdata
W1 & 38 & 54  (148) & 54 \\
W2 & 54 & 70  (115) & 69 \\
W3 & 39 & 49 (121) & 50 \\
W4 & 51 & 81 (142) & 84 \\
W1-W2 & 30 & 38 (41) & 36 \\
W1-W3 & 42 & 56 (132) & 56 \\
W1-W4 & 73 & 100 (132) & 100 \\
W2-W3 & 50 & 67 (103) & 65 \\
W2-W4 & 79 & 110 (114)& 100 \\
W3-W4 & 74 & 100 (73) & 90
\enddata
\tablecomments{Values in mmag, measured as the width of the
distribution of differences between our calibrated magnitudes and
those of a preliminary version of Jarrett's unpublished extended
source photometry.  We present three measures:  the width of the
best-fit Gaussian ($\sigma$), the ``robust sigma'' calculated as a
standard deviation with outlier rejection (number of rejected sources
in parentheses, out of 1907 total sources), and the 68\nth percentile
absolute deviation from the median.  For Gaussian distributions these
three quantities should be nearly identical.\label{Errors}}
\end{deluxetable}

While a detailed study of individual targets would benefit from the
more precise photometry of Jarrett's final catalog, this precision is
sufficient for the initial exploration of the \wise\ data set and
identification of superlative objects.

\section{Characteristics Of The Extended Gold Sample}

\label{classification}

To help us characterize sources in the Extended Gold Sample, we
employed the source classifications presented in the SIMBAD database,
where available, as a starting point. SIMBAD provides over 130 different types of
astrophysical classifications ranging from the stellar to the
cosmological. We recovered source classifications for $\sim$ 87\% of the
Extended Gold Sample. The median distance between the \wise\ and SIMBAD source is
of the order $\sim 0\farcs5$. We find that $\sim$93\% of these sources
have been categorized as being external galaxies vs. Galactic
foreground objects  We find that the 7\% that are ``stellar" are
typically not point sources but are resolved in the IR for some reason.  
These are encouraging results because they validate our selection and vetting procedures.  

In the following sections we discuss
the two most commonly identified types of astrophysical sources ---
those of Galactic origin and those of extragalactic origin, the latter
being of primary interest in this study. 

\subsection{Extragalactic Sources}

We have grouped the SIMBAD classifications for extragalactic sources into five broad categories:

\begin{itemize}

\item {\bf normal} Galaxy: Includes the following SIMBAD types, Galaxy (G), Galaxy in Pair (GiP), Galaxy in Group (GiG), Galaxy in Cluster (GiC), Cluster of Galaxies (GlC), Group of Galaxies (GrG), Brightest galaxy in a Cluster (BiC), and Compact Group of Galaxies (CGG)

\item {\bf active} Galaxy: LINERS (LIN), QSO, Radio Galaxy (rG), Active Galactic Nuclei (AGN), Seyfert Galaxies (Sy1 + Sy2), BL Lac - type ob ect (BLL), and Blazar (Bla))

\item {\bf star-forming} Galaxy: Emission Galaxy (EmG), HII Galaxy (H2G), Starburst Galaxy (SBG), and Blue Compact Galaxy (bcG)

\item {\bf interacting} Galaxy: Interacting Galaxy (IG), and Pair of Galaxies (PaG)

\item  {\bf low surface brightness} Galaxy: low surface brightness galaxy (LSB)

\end{itemize}

\begin{figure*}[htp]
\begin{center}
\begin{tabular}{l}
\includegraphics[width=7in,trim=0 2in 0 0,clip]{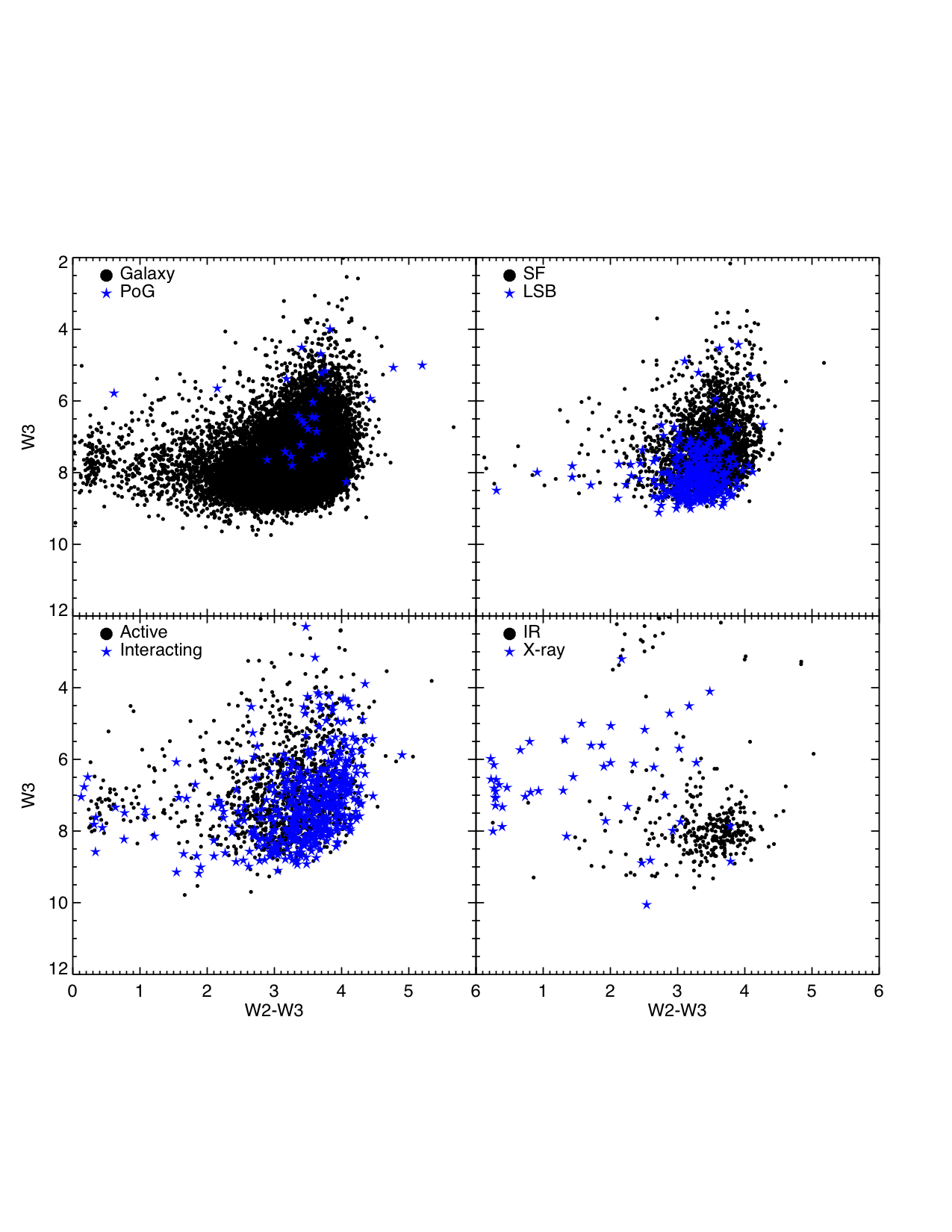}
\end{tabular}
\end{center}
\caption[CAPTION]{\label{GoldCMD} Color-magnitude diagrams for a variety
  of extragalactic sources in the Extended Gold Sample.}
\end{figure*}

To characterize the galaxy population we construct a simple
color-magnitude diagram (W2-W3 versus W3). Figure~\ref{GoldCMD} present W2-W3
versus W3 for a variety of (mostly extragalactic) sources. The two
most striking features in this figure are the almost complete absence of
galaxies with W3 $> 10.0$, and W2-W3 $\ge 4.5$. 

The faint limit is significantly above the \wise\ detection limit in W3
of $\sim 11.2$ (for point sources), and so requires explanation.  This
limit originates in our requirement that a galaxy be both detected
{\it and extended} in \wise\ in the W3 band.  We measure extended-ness with W3RCHI2,
and it appears that this parameter rarely takes values above
3 for sources with W3 $\gtrsim$ 9, no matter how extended
the object is.  This lower limit is thus an artifact of the \wise\ data
pipeline and our Extended Gold Sample selection criteria.

The upper left plot shows where the {\it normal} galaxies and the SIMBAD classification
Part of a Galaxy (PoG) reside in this color-magnitude space. These sources
compose $\sim$ 82\% of the total galaxy population. In the upper right
we show the color-magnitude diagram for star forming and LSB galaxies,
which compose $\sim$ 12\% of the population. The lower left shows the
active and interacting galaxies, composing $\sim$ 6\% of the
population. The lower right plot shows the generic SIMBAD
classifications of (IR) and X-ray (X) sources.  These last two are
somewhat ambiguous classifications, and we include them
only for completeness.  

These results are quite consistent with our expectations validating
the SIMBAD types. The {\it normal} galaxies appear to occupy the overall
parameter space seen in the other populations suggesting this to be a
mixture of a variety of galaxy types. The {\it star-forming} galaxies
appear to congregate at redder W2-W3 colors (W2-W3 $\ge 3$), as 
expected. Interestingly, the LSB galaxies appear to tightly congregate on the fainter
end of the {\it{star-forming}} population. The
{\it{interacting}} galaxies appear to populate the same parameter
space as seen by the {\it{star-forming}} galaxies, which is an
expected result as a majority of  {\it{interacting}} galaxies are
experiencing high star formation rates. The {\it active} galaxies populate
similar parameter space as seen in the {\it normal} galaxy population. The
IR sources appear to be a mixture of both extragalactic sources and
Galactic sources, though we do see a tight clustering occupying
similar parameter space as the LSB galaxies.  

\subsection{Galactic Sources}

\label{galactic}

Though {\it Galactic sources} (i.e.\ sources within the Milky Way
Galaxy) compose only $\sim$ 7\% of our sample it
is important to understand how they might produce contamination in our
investigation. SIMBAD provides classifications for over 70 different
types of these Galactic sources, which we have reduced to six primary types:
$normal$ stars, Young Stellar Objects (YSO), Variables, Evolved, Evolved +
IR, and stars in clusters. Our grouping methodology is as follows: 

\begin{itemize}

\item {\bf normal} Star:  Includes the following types, Star (*), Emission-line Star (Em*), Peculiar Star (Pe*), High proper-motion Star (PM*),  	Star in double system (*i*),  Be Star (Be*), and Eclipsing binary (EB*)

\item {\bf YSO}: T Tau-type Star (TT*), Variable Star of FU Ori type (FU*), Herbig-Haro Object (HH), Pre-main sequence Star (pr*), and Young Stellar Object (Y*O)

\item {\bf variable} Star: Variable Star (V*), Variable Star of beta Cep type (bC*), Variable Star of Orion Type (Or*), Semi-regular pulsating Star (sr*), Variable Star of W Vir type (WV*), Long-period variable star (LP*),
CV DQ Her type (intermediate polar) (DQ*), Variable Star of delta Sct type (dS*), Variable Star of alpha2 CVn type (a2*), Cepheid variable Star (Ce*), Variable Star with rapid variations (RI*), Variable Star of RV Tau type (RV*),
Pulsating variable Star (Pu*), Variable Star of Mira Cet type (Mi*), Variable Star of R CrB type (RC*), Eruptive variable Star (Er*), Variable Star of irregular type (Ir*), and Cataclysmic Variable Star (CV*)

\item {\bf evolved} Star: Asymptotic Giant Branch Star (He-burning) (AB*), Red Giant Branch star (RG*), Wolf-Rayet Star (WR*), and S Star (S*)

\item {\bf evolved+IR} Star: Post-AGB Star (proto-PN) (pA*), Star with envelope of OH/IR type (OH*), and Carbon Star (C*)

\item {\bf Star in Cluster}: Star in Cluster (*iC), Cluster of Stars (Cl*), and Star in Nebula (*iN)

\end{itemize}

For completeness we must also consider two other types of Galactic
sources, PN and {\bf Solar System Objects} (SSOs) such as comets, asteroids, and planets. The SSOs are identified by utilizing the \wise\ All-sky Known Solar System Object Possible Association List. We also identify two IR bright planets, Neptune and Uranus, which are included in the sample. 

\begin{figure*}[htp]
\begin{center}
\begin{tabular}{l}
\includegraphics[width=7in,trim=0 2in 0 0,clip]{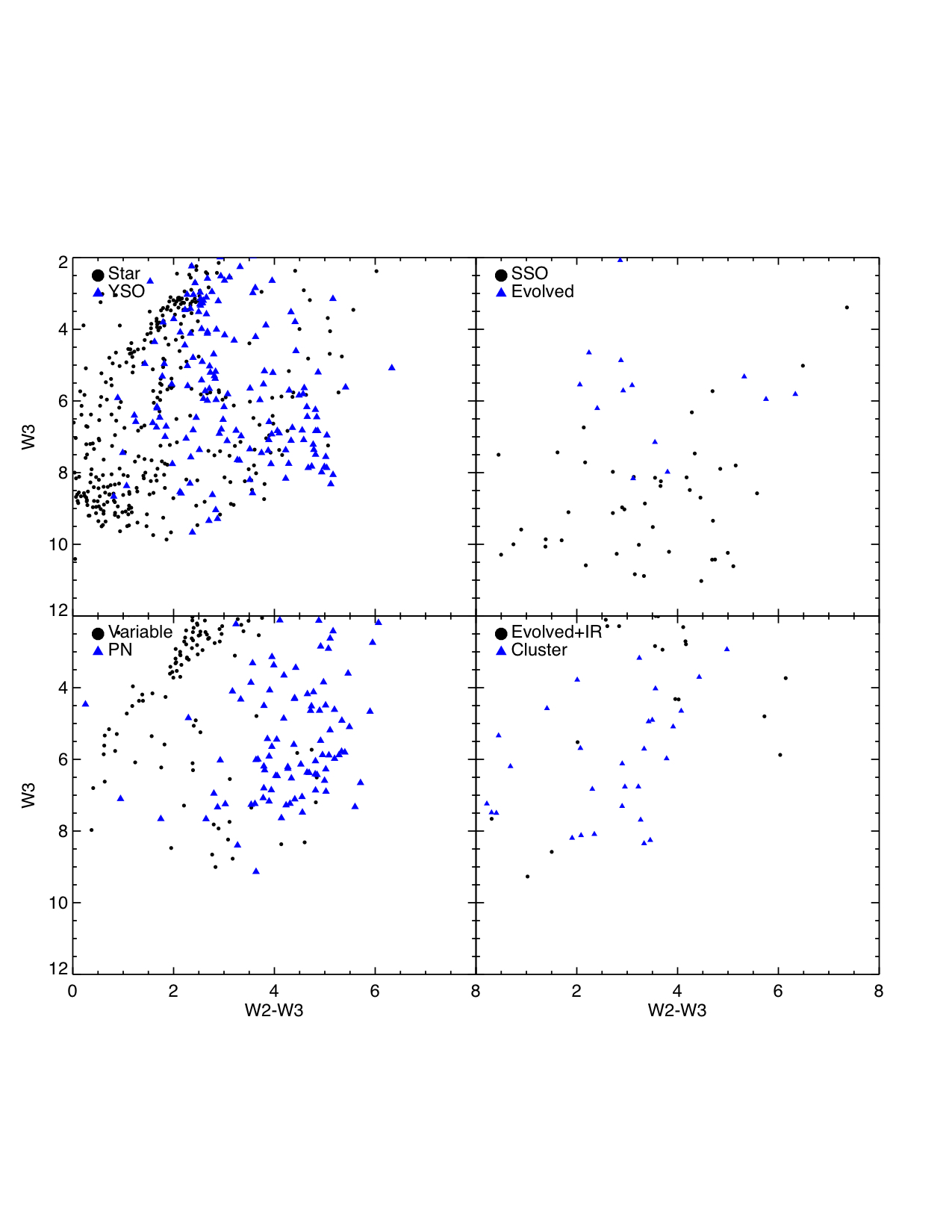}
\end{tabular}
\end{center}
\caption[CAPTION]{\label{GoldCMD2} Color-magnitude diagrams for a
  variety of galactic sources in the Extended Gold Sample.}
\end{figure*}

To characterize the \wise\ high latitude Galactic population we
present Figure~\ref{GoldCMD2} , which shows the color-magnitude
diagrams for the various Galactic classifications. As before we find
that the majority of the Galactic population are relatively rare at W3
$> 10$ and W2-W3 $\ge 4.5$, though we do find sources with very extreme W2-W3 colors, i.e.\ W2-W3 $\ge 6$. We also find, as was expected, that the two types of sources which confuse and contaminate our search the most are the YSO's and planetary nebulae (PNe). While we do find contaminating sources within the Galactic sample, the scarcity of astrophysical sources with W2-W3 $\ge 4.5$ should make our search relatively straightforward for high $\gamma$ K3's.  

\section{Visual Review and Grading of Sources in the W3 Extended Gold Sample}

\label{visualgrading}

A systematic search into the existence and numbers of possible
galaxy-spanning ETIs requires that we carefully inspect the most
interesting subsets of the Extended Gold Sample. 

\subsection{AGENT parameterization}

\label{agent}

For every source we apply the methodology described in
\citet{WrightDyson2} to estimate the AGENT parameters $\gamma$ (the
fraction of starlight reradiated as alien waste heat)
and \twaste\ (the waste heat's characteristic temperature) for each
galaxy, assuming $\nu=0$ (no nonthermal alien power disposal/emission)
and $\alpha=\gamma$ (that is, all of the waste heat originated as
stellar power).  We fit all four \wise\ bands to the three parameters
$(L/4 \pi d^2)$ (the bolometric flux of the source), $\gamma$, and \twaste.  

For consistency, we modeled every galaxy as having an intrinsic SED
represented by the old elliptical SED model {\it Ell13} of
\citet{Silva98}, of which a fraction $\alpha$ of the starlight is
absorbed and re-emitted as a blackbody at temperature \twaste.  Of
course, many galaxies have a significant blue stellar population and
dust.  Because the {\it Ell13} template has the least MIR emission of
all of the \citet{Silva98} galaxy templates, using it will allow us to measure the {\it maximum}
amount of MIR emission from each galaxy that {\it could} be attributed
to ETI's from \wise\ broadband photometry alone.  Its is thus
consistent with our desire to set upper limits to K3 waste heat emission.

\subsection{Visual inspection}

{\bf Using $\gamma$ as the primary parameter for prioritizing sources we
inspected, reviewed and graded every source in the W3 Extended Gold Sample with $\gamma 
\ge 0.25$, ${\sim}4000$ in all.}
 
\begin{figure*}[htp]
 \epsscale{0.9}
 \plotone{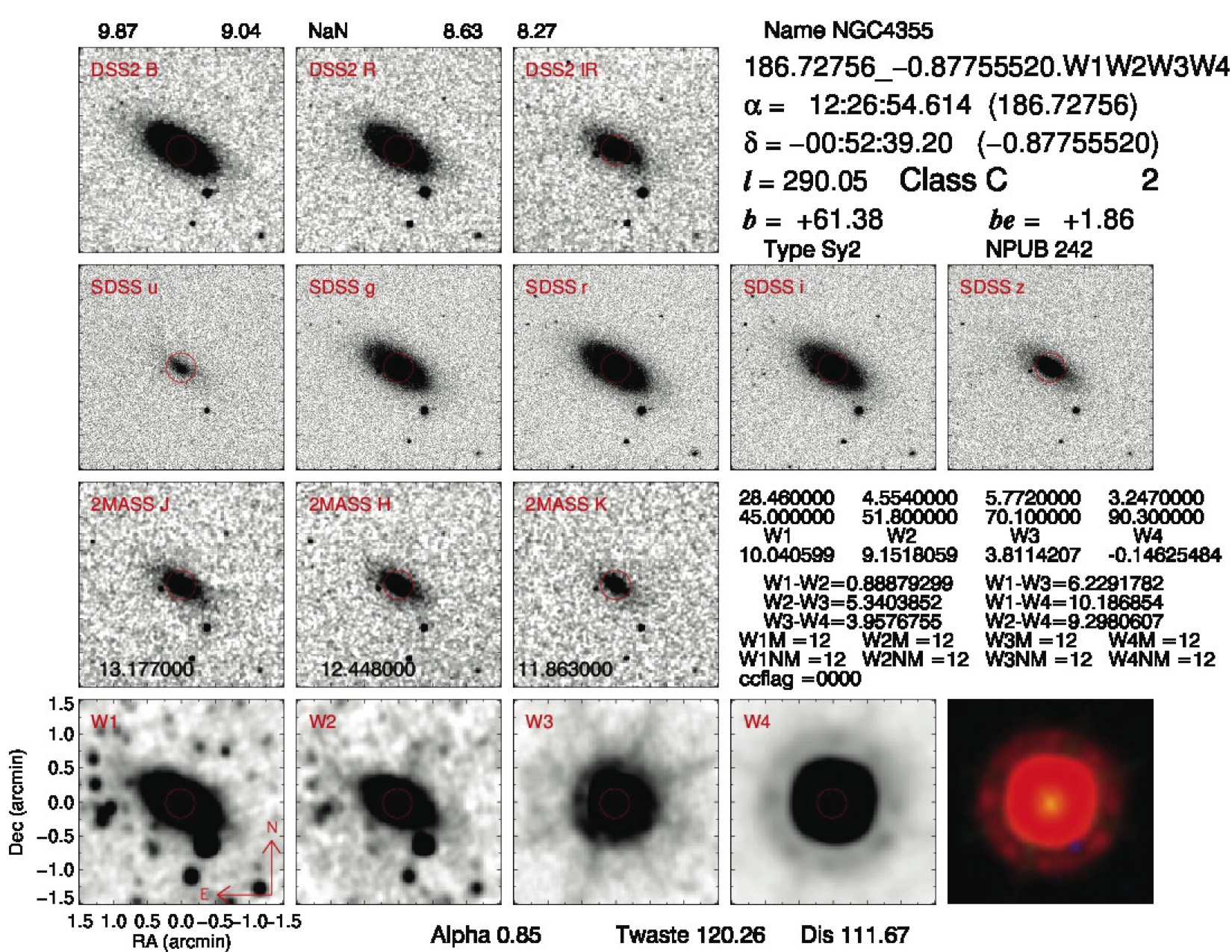}
 \caption{
Example of an ``at-a-glance'' chart used for detailed visual review
and classification of sources in the Extended Gold Sample.  The red
labels of the images shows the data sources; the scale is the same in
each image and given in the bottom-left-hand image.  The magnitudes of
the 2MASS imagery are given in the corresponding images.  The
bottom-right-hand image is a color composite from the four \wise\ bands
in the bottom row.  The SIMBAD name is in the upper right, along with
the file name, J2000 coordinates in sexagesimal and decimal, the
galactic coordinates and ecliptic latitude, the SIMBAD type, and the
number of publications reported by SIMBAD.  ``Class'' refers to the
grade of the source (C in this case, reflecting its large number of
associated publications, see Section~\ref{grading}).  Our \wise\ extended source photometry
appears to the right of the third row, including, for each band, the
RCHI2 parameter (large values indicate the source is well resolved), the signal-to-noise
ratio of the detection, and the magnitude.  Also reported are the
various color combinations, the number of visits to the field, and the
number of visits in which the source was detected.  Finally, the
contamination and confusion flags (CC{\tt \_}FLAG) are given for the four bands.
This particular source, W122654.61-005239.2,
illustrates what might pass for a high-$\gamma$\ KIII galaxy in
multiwavelength images; it is actually a dusty (and well-studied)
Seyfert 2 galaxy.  
 } 
 \label{fig:finder}
 \end{figure*}

To facilitate this visual review, we constructed an interactive GUI
system. This system provided an ``at-a-glance chart'' for each source
(Figure~\ref{fig:finder}) consisting of $2\arcmin \times 2\arcmin$
multiwavelength imaging cutouts (DSS, SDSS, 2MASS, and \wise) centered
on the \wise\ source position, a variety of hyperlinks to
supplementary imaging data and database queries centered on the
source position (e.g.\ \wise\ (L1B + Atlas images, SIMBAD, NED),
and printed information including source coordinates, \wise\ catalog colors and
magnitudes, number of
  publications listed in SIMBAD, number of times \wise\ observed the
  source position in each filter (W\#M) and number of times \wise\
  detected the source in each filter (W\#NM).

\subsection{Grading and Literature Search}
\label{grading}

The sample of red, extended objects in \wise\ is a combination
of sources, galactic and extragalactic; real and instrumental;
well-studied and new-to-science.  We developed a sorting scheme to help
us understand it better and prioritize sources for further
investigation.  

We defined a simple alphabetical grading scheme (A-F):

\begin{itemize}
\item A -- Astrophysical red source with no previous publications in SIMBAD, little or no
  ancillary survey data. {\em Highest priority candidate for
    observational follow-up.}
\item B -- Astrophysical red source with some ancillary data or few
  publications, which do {\em not} provide convincing evidence that
  its nature is understood. {\em Good candidate for follow-up.}
\item C -- Visual review of source and/or publication list provides
  convincing evidence that its nature is understood (although in some cases,
  the SIMBAD classification may be incorrect.)
\item D -- Astrophysical source that creates artifacts detected as red
  extended sources by \wise. Most
  frequently associated with a bright star or large region of bright nebulosity.
\item F -- Fake/false source or artifacts, such as a latent image. {\em
    Not} a real astrophysical red source. Does not belong in any astronomical catalog.
\end{itemize}

We next provide some details of the decision process leading to the
assignment of the various grades above, working backwards from F to A.

{\em Grade F.} 
Persistence artifacts, or ``latents,'' in the W3 band
constitute the most common class of false, extended red source meriting an F grade. Most
latents are culled by the initial visual classification (see Section~\ref{visual}), but some did make it into the Extended Gold Sample. The most
pernicious cases involve W3 latents that happen to fall on the
positions of legitimate W1+W2 sources; these closely resemble our
expectations for a high-$\gamma$ KIII galaxy in the \wise\
images. The visual review process readily discards such
sources. They have very low W3NM/W3M, failing to appear in the majority of the
individual \wise\ Level 1b frames (planets, comets, and asteroids also
exhibit this behavior, but do so in all bands and appear in the Known
Solar System Object Possible Association List, see Section~\ref{galactic}).
The wider-field \wise\ Atlas images typically reveal the bright
star responsible for producing a W3 latent.

{\em Grade D.}
Unlike F-grade sources, sources assigned D grades are real astrophysical
objects, but these are not valid, red extended \wise\ sources. There are two
main classes: (1) Saturated stars  (W3$\le 3.8$~mag) which would otherwise be point sources, and (2)
bright ``knots'' within larger regions of mid-IR nebulosity (for example, associated with
Galactic foreground emission). Our visual review of the \wise\ images
readily identifies such cases.

{\em Grade C.} 
For sources that pass the quality control checks associated with F and
D grades above, we next evaluate the citations listed for the
associated SIMBAD source(s). 

Sources that receive C grades typically have ${\ga}4$ citations,
which, taken together, convince
us that its astrophysical nature is understood. 

If our review of the literature for a given source point toward original object type
listed in SIMBAD, we accept that object type. 
In many cases we found either (1) that the preponderance of literature
pointed to a different object type or (2) the nearest SIMBAD source
returned by our automated matching was not the appropriate counterpart
for the \wise\ source. If necessary, we manually
overrode the SIMBAD object type and/or matching source for our catalog.

An example of a grade C source is shown in Figure~\ref{fig:finder}.

{\em Grades B and A.}
Our final two grades, B and A, represent real astrophysical extended
red sources with scant (${\la}4$) or zero existing literature citations, respectively;
these sources should be given high priority for further observational
followup to determine their nature.

Grade B sources are cited only in large, survey catalogs containing
minimal interpretation of individual sources. Commonly encountered
catalog papers containing Grade B (and also C) extended red \wise\
sources are listed in Table~\ref{tab:refs}.

\begin{deluxetable*}{ll}
\tabletypesize{\footnotesize}
 \tablewidth{0 pt}
 \tablecaption{ \label{tab:refs}
  Frequently Encountered Catalogs Matching Red Extended \wise\
  Sources}
\tablehead{Reference & Catalog Description} 
\startdata
\citet{UGC} & Arcsecond Positions of Uppsala General Catalog (UGC) Galaxies \\
\citet{UZC} & The Updated Zwicky Catalog (UZC) \\
\citet{ESO-Uppsala} & The surface photometry catalogue of the
ESO-Uppsala galaxies \\
\citet{2M++} & The 2M++ galaxy redshift catalogue (69,160 galaxies) \\
\citet{QDOT} &  The QDOT all-sky IRAS galaxy redshift survey \\
\citet{FlatGal} & The 2MASS-selected Flat Galaxy Catalog (18,020
disc-like galaxies) \\
\citet{positions} & Positions for 17,124 galaxies including 3301
new companions of UGC galaxies \\
\citet{UZC-SSRS2} & The UZC-SSRS2 Group Catalog (1168 galaxy groups) \\
\citet{PSCz} & The PSCz catalogue (15,411 IRAS galaxies) \\
\citet{KISO} & KISO survey for ultraviolet-excess galaxies.\\
\citet{Imperial} & The Imperial IRAS-FSC Redshift Catalogue (60,303 galaxies) \\
\enddata
\end{deluxetable*}

\begin{figure*}[htp]
\centering
\epsscale{0.9}
\plottwo{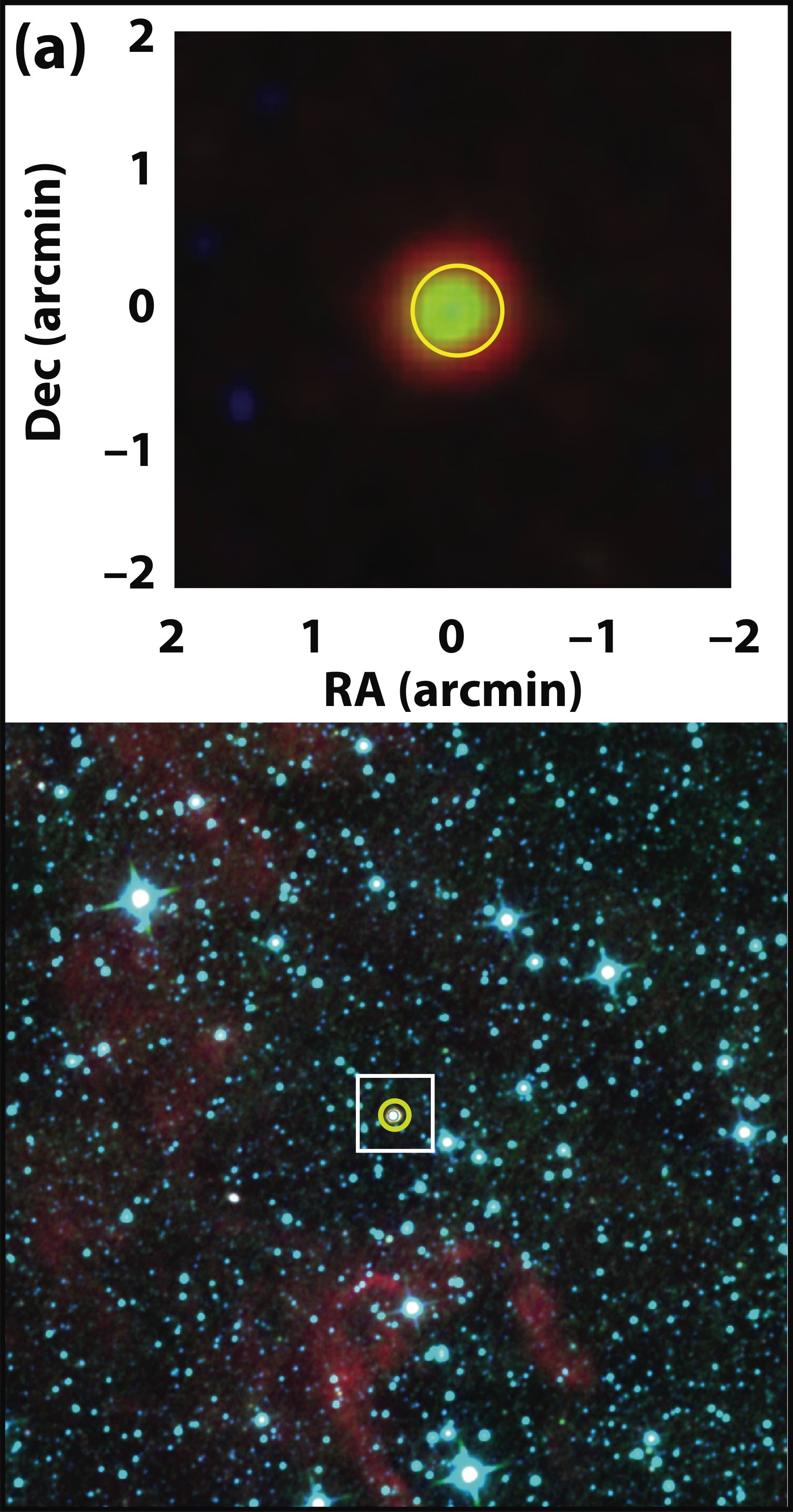}{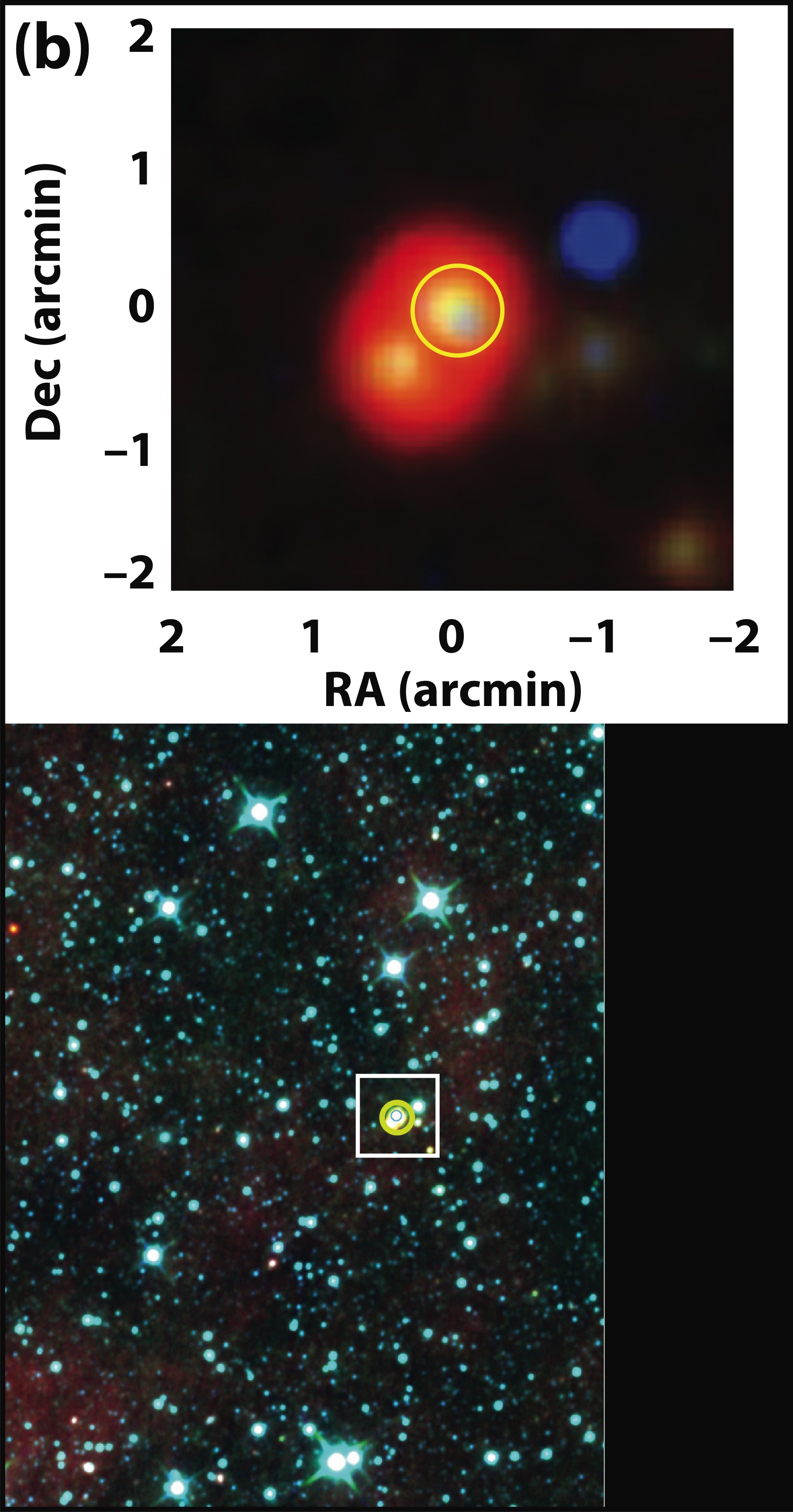}
\caption{Left (a): Source W224436.13+372533.7, an example of an isolated red
  source given an A grade. Right (b): Source W043329.56+645106.5, an unusual
  cluster of sources discovered by our search, also given
  an A grade.  Top
  panels show the color-composite \wise\ $2\arcmin \times 2\arcmin$
  ``postage stamp'' images from our ``at-a-glance'' charts (red = W4, green =
  W3,  blue = W1+W2), while bottom panels
  show the same sources in the wider $30\arcmin \times 30\arcmin$
    \wise\ atlas images (red = W3, green = W2, blue = W1).
} 
\label{fig:isolated}
\end{figure*}

The most promising candidates for observational follow-up are grade A
sources that are isolated, meaning they are neither part of a cluster of
red objects (for example, a young embedded star cluster or a galaxy
cluster hosting many mergers) nor associated with diffuse mid-IR
nebulosity (a hallmark of embedded star clusters in \hii~regions or an
indication that they may be associated with a large, extended
galaxy). An example of an isolated ``A'' source and a cluster of red sources is presented
in Figure~\ref{fig:isolated}.

Note that this grading scheme only loosely tracks ``interest'' from a
SETI perspective:  a well-studied galaxy might have an anomalously
high MIR luminosity and thus be an outstanding SETI target, while a
new-to-science protostar might be manifestly ordinary and so very low
priority.  The ``followup'' of class A sources is thus primarily to
determine their true nature and determine if they warrant further
study from either a natural astrophysics or SETI perspective.  

\begin{figure*}[htp]
\begin{center}
%\begin{tabular}{l}
\includegraphics*[width=7in,trim=0 4in 0 5.5in,clip]{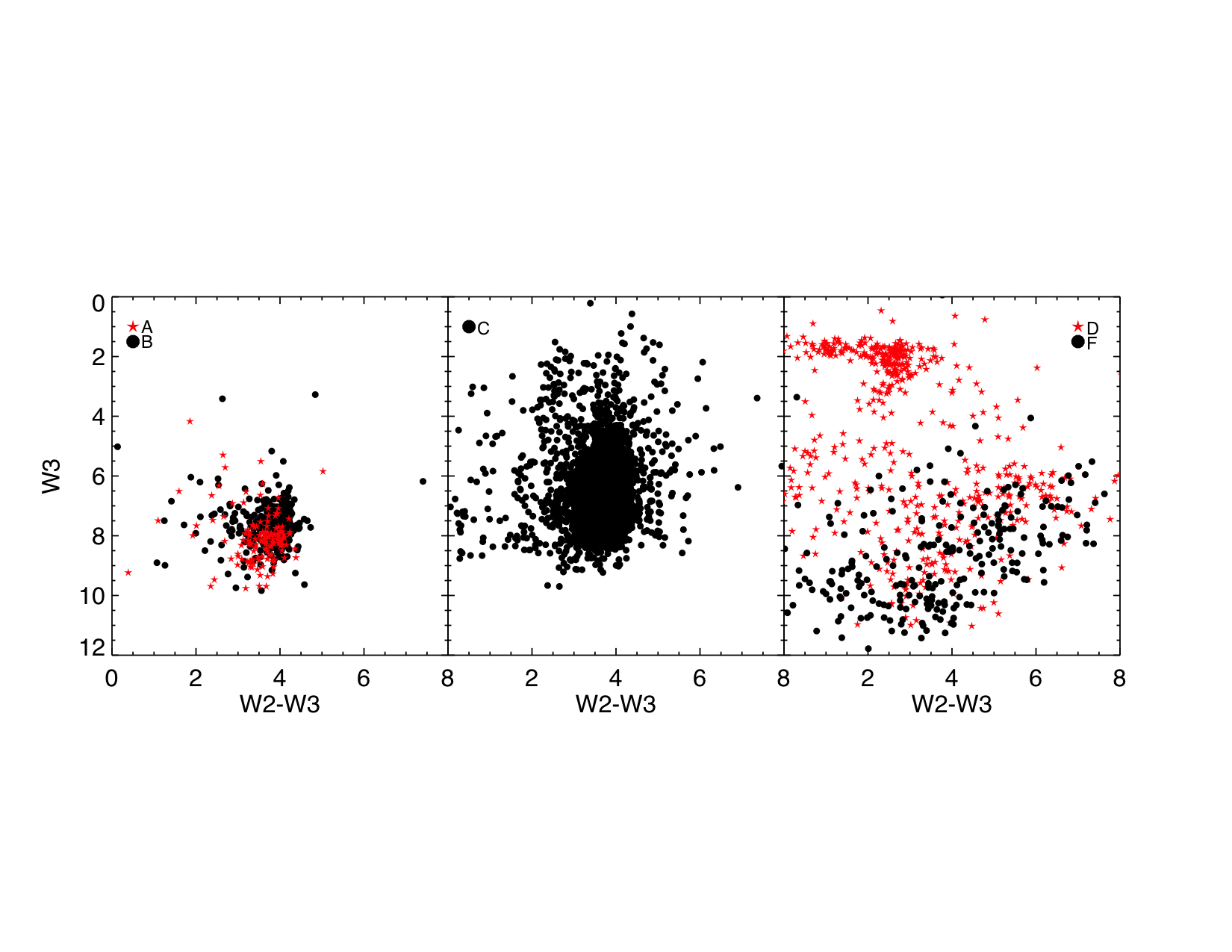}
\includegraphics*[width=7in,trim=0 4.5in 0 4.5in,clip]{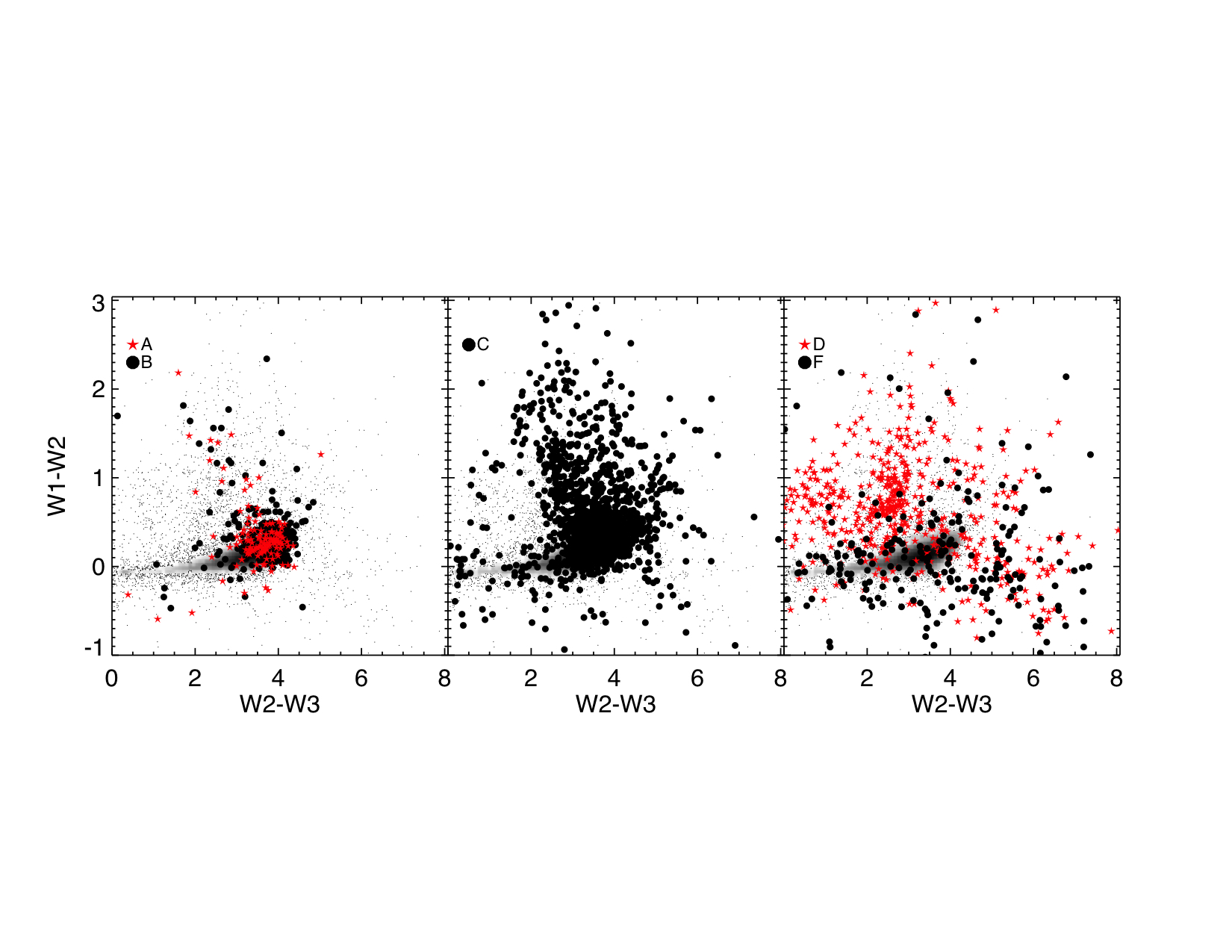}
%\end{tabular}
\end{center}
\caption[CAPTION]{\label{Classified} Color-magnitude
  diagrams (top) and color-color diagrams (bottom) for W3 Extended
  Gold sample sources which
  have been carefully vetted and classified as described in
  Section~\ref{grading}.  Sources with no or little literature presence (grades A and B, respectively) are in the left panels; well-studied sources (grade C) are in the middle; saturated stars and artifacts (grades D and F, respectively) are on the right. Unsurprisingly, grade A and B sources (i.e.\
  sources with little presence in the literature) are
  typically faint, and grade D and F sources have extreme colors and
  magnitudes (indicating that they are artifacts of the instrument or
  the analysis pipeline strongly affecting some bands but not others).  The density of all W3 Extended Gold Sample sources is
  indicated in grayscale in the bottom panels.  We use plots such as
  this to ensure that we understand the effects of making cuts based
  on our grading scheme.}
\end{figure*}

Figure~\ref{Classified} illustrates the colors and magnitudes of the
sources we have graded, and shows where these sources reside in this parameter space.

\section{W3 Extended Platinum Sample}

\label{platinum}

\subsection{Construction of the Sample}

The Extended Gold Sample was constructed to be relatively free from
non-extended or non-astrophysical sources, but our curation was
deliberately liberal so that borderline cases could be handled more
carefully on a case-by-case basis.  Having constructed the Gold Sample, we
performed this more careful analysis on those objects whose calibrated
photometry indicate that they are especially red.

As discussed in Section~\ref{calibrate}, we have performed magnitude corrections
using the photometry provided
by the \wise\ pipeline. For every source in the Extended Gold Sample we
now have photometry using four
different photometric systems (i.e.\ MPRO (profile), MAG
(aperture corrected magnitude), CMAG (Corrected Magnitude, see Section~\ref{calibrate}), and GMAG (elliptical aperture)).  While the corrected magnitudes are
reliable for most sources, there are special cases
where a different photometric system may yield more reliable results.
Given this, we have measured $\gamma$, $\twaste$, and $L/(4\pi d^2)$, (see
Section 4.0, and Paper II)
in each of the four different photometric systems.

To ensure that our choice of photometry does not cause us to miss any
red sources, we selected from the Extended Gold Sample all sources with 
$\gamma \ge 0.25$ in any of the photometric systems. This identified a total of 3,145
sources. We used our collected imaging, grading, and literature search
results (Section~\ref{grading}) to carefully and visually vet this sample to identify and remove the
most obvious contaminants, mostly nebulosity and saturated stars that
survived our very liberal curation of the Extended Gold Sample.  We identified and removed 366 obvious contaminants, reducing our list to 2779 objects.

We found that in a small percentage of cases, especially with
blended sources, an extended source is broken up into two or more
components in the All-sky catalog.  In these cases we choose the entry
coordinates closest to the extended source center and the photometric
system that best captures the true magnitudes of the extended source.
In some cases, the W3RCHI2 value of the catalog entry centered on an
extended source is below 3 because of the details of how the All-sky
catalog breaks up composite sources.

We segregated this sample into four different categories and determined
which photometric system best fits the source in question:

\begin{itemize}

\item Extended Source: Source is truly extended in the \wise\ images
\item Point Source Galaxy: Source appears to be a galaxy from
  O/IR imaging or other sources, but appears unresolved in the \wise\
  imagery, despite having a high W3RCHI2 value.
\item Point Source Star: Source appears to be stellar in nature.
\item Junk: Source is a contaminant, should not be in any catalog
  of real astrophysical sources.

\end{itemize}

In addition to classifying these sources as described above, we also
determine which photometric system best estimates the true photometry
of the given source. We use the following convention to indicate the
best possible \wise\ photometry in the source tables:

\label{photometry}

\begin{itemize}

\item 0: MPRO (Profile fit photometry)
\item 1: MAG (Aperture-corrected magnitude)
\item 2: CMAG (Corrected magnitude)
\item 3: GMAG (2MASS XSC elliptical aperture).

\end{itemize}

Using these conventions we identify 1296(46.7\%) classified as Extended,
974(35\%) classified as unresolved galaxies, 263(9.4\%) classified as stellar, and
246(8.9\%) classified as junk. Of the Extended sources, 563 have
$\gamma > 0.25$ in the photometric system most appropriate for that
source.  These compose our W3 Extended Platinum Sample of real,
extended, red sources in \wise. 

\subsection{Platinum Sample Published Catalog}
We present a catalog of 563 sources deemed to be extended and real in
the \wise\ 12$\mu$m filter and identified as the Platinum Sample as a
FITS file associated with this manuscript. This
list of sources have also been selected as those having $\gamma \ge
0.25$ in the preferred photometric system. Table~\ref{FITS} gives the detailed
parameter description for this electronic catalog. For brevity, we
present a minimal number of parameters representing the most important measurements for these
sources, and the \wise\ identifier so that users may cross-reference our
catalog to the \wise\ All-sky or ALL-WISE catalogs. 
Presented in this parameter list are the coordinates, optimal
\wise\ photometry and the photometric system used (see Section~\ref{photometry}), the
AGENT parameters, and pertinent information derived from the SIMBAD
database. Tables \ref{reddest}--\ref{extreme} have been derived using this catalog.

\begin{center}
\begin{deluxetable*}{lccl}
\tablewidth{0pt}
\tablecaption{Platinum Sample Published Catalog \label{FITS}}
\tablehead{\colhead{Parameter} & Type & Entry & Description}
\startdata
   DESIGNATION   &  STRING  &  J190101.24-181215.0 & \wise\ Designation \\
   NAME     &       STRING  &  PNA6651                              &
SIMBAD Identifier \\
   GRADE    &       STRING &   C & Grade (see Section~\ref{grading}) \\
   SIMBAD\_TYPE  &   STRING &   PN    & Object Type  \\
   R.A.      &        DOUBLE     &      285.25520& Right Ascension [degrees]\\
   Decl.       &      DOUBLE    &      -18.204188& Declination [degrees]\\
   W1BEST   &       FLOAT    &       15.4980 & W1 optimal photometry [mag] \\
   W2BEST     &     FLOAT    &       13.889&W2 optimal photometry [mag] \\
   W3BEST   &       FLOAT    &       9.0770&W3 optimal photometry [mag] \\
   W4BEST    &      FLOAT   &        3.0690& W4 optimal photometry [mag]\\
   PHOT\_SOURCE &    INT     &         1& Photometric System, see Section 5.1 \\
   TWASTE    &      FLOAT    &       72.923& Waste heat temperature [K]\\
   GAMMA      &     FLOAT    &      0.99294&   AGENT parameter $\gamma$\\
   BOL\_FLUX    &    FLOAT    &       255.83& Bolometric flux [\Lsun/pc$^2$]
\enddata
\end{deluxetable*}
\end{center}

\section{Extreme Extended Objects in the Platinum Sample}

\label{sec:extreme}

\subsection{Extreme \wise\ Colors}

The \wise\ filter system allows for the measurement of 6 different
photometric colors, i.e W1-W2, W2-W3, etc. In Table~\ref{reddest}, we present a
list of the 10 reddest (per color) objects contained within the
Extended Platinum Sample. These objects have been ordered by decreasing ``redness'', so
that the 10th object in the list represents the 10th reddest object
within the color explored. 

The most extreme MIR objects in the high latitude sky ($|b|
\ge 10$) appear to be dominated by PNs.  Other
objects with extreme colors include a pair of comets, three YSOs, a handful of Type {\sc ii} Seyfert galaxies, two
uncataloged sources in the LMC that escaped our mask of that region,
and some objects we discuss in Section~\ref{extremegamma}.
 
Since this study is primarily interested in finding ETIs of
extra-galactic origin we present Table~\ref{redgal}, which is a list
of the 10 reddest (per color) high latitude extragalactic objects
seen in the Extended Platinum Sample. We find that the colors of
galaxies are in general not as extreme as Galactic sources.  Most of the
galaxies on this list are Grade C, meaning that their nature has been
identified in the literature.  

Table~\ref{redgal} is dominated by galaxies classified in SIMBAD as
AGN of various stripes, which is unsurprising since those galaxies
are often characterized by their extreme MIR emission.  AGN are thus,
as expected, our primary confounders in our waste heat search for K3
civilizations.

\subsection{Extreme $\gamma$}

\label{extremegamma}

In the AGENT formalism (Section~\ref{agent}, \citet{WrightDyson2}) the
parameter $\gamma$ represents the fraction of starlight reemitted in
the MIR, at temperature \twaste.  We have measured maximal values for
this parameter assuming that there is no dust in any of our sources,
and that their underlying stellar population is that of an old
elliptical galaxy (so, virtually dust-free).  Since this is the
parameter of interest in searches for alien waste heat, we have sorted
the Extended Platinum Sample by this parameter.  We present the top 50
such galaxies in Table~\ref{highgamma}.

The galaxy with the highest measured $\gamma (= 0.85)$ is NGC 4355
(=NGC 4418), which is also has the most extreme colors in four of the six
\wise\ color combinations.  It is a well-studied Type {\sc ii} Seyfert galaxy in
the Virgo cluster with a (W1-W4) color of 10.19, with the extreme MIR
emission being due to the AGN. 

 The second source on our list is IRAS
04259-0440, a marginally resolved galaxy with modest presence in the
literature.  It has been studied in the context of being a Seyfert
galaxy or LINER \citet{Wu2011}, so we are convinced that the MIR
emission is understood. Nonetheless, given the extreme nature of this galaxy's infrared
emission, this galaxy would appear to warrant more attention than it
has received to date. 

The third galaxy in our list is the well-studied Arp 220, the quintessential local starburst
galaxy and a known active galaxy.  Fourth and fifth are UGCA 116 and IRAS F20550+1655-SE, both
pairs of interacting galaxies.  In all three cases, the extreme MIR colors
are clearly due to star formation triggered by galaxy interactions.

\subsection{Extreme Objects New to Science}

\label{new}

One of the primary objectives of this investigation is to search and
identify the most rare and extreme sources in the high latitude
infrared sky. The majority of the sources in our tables have already been
discovered and discussed in various articles, but there are still a
small number of objects not previously discovered or discussed in the
literature. 

In Table~\ref{extreme} we present 3 objects ($\gamma \ge 0.25$)
classified as As, meaning that they are essentially new to science.
In this section we also discuss an extreme and apparently anomalous
object that we gave a B grade, IRAS 15553-1409, and a particularly
interesting lower-$\gamma$ grade A source.

\subsubsection{IRAS 04287+6444: An Unusual Cluster of MIR Sources With
  no Optical Counterparts}

\begin{figure*}
\includegraphics[width=6.5in,trim=0 0.75in 0 0,clip]{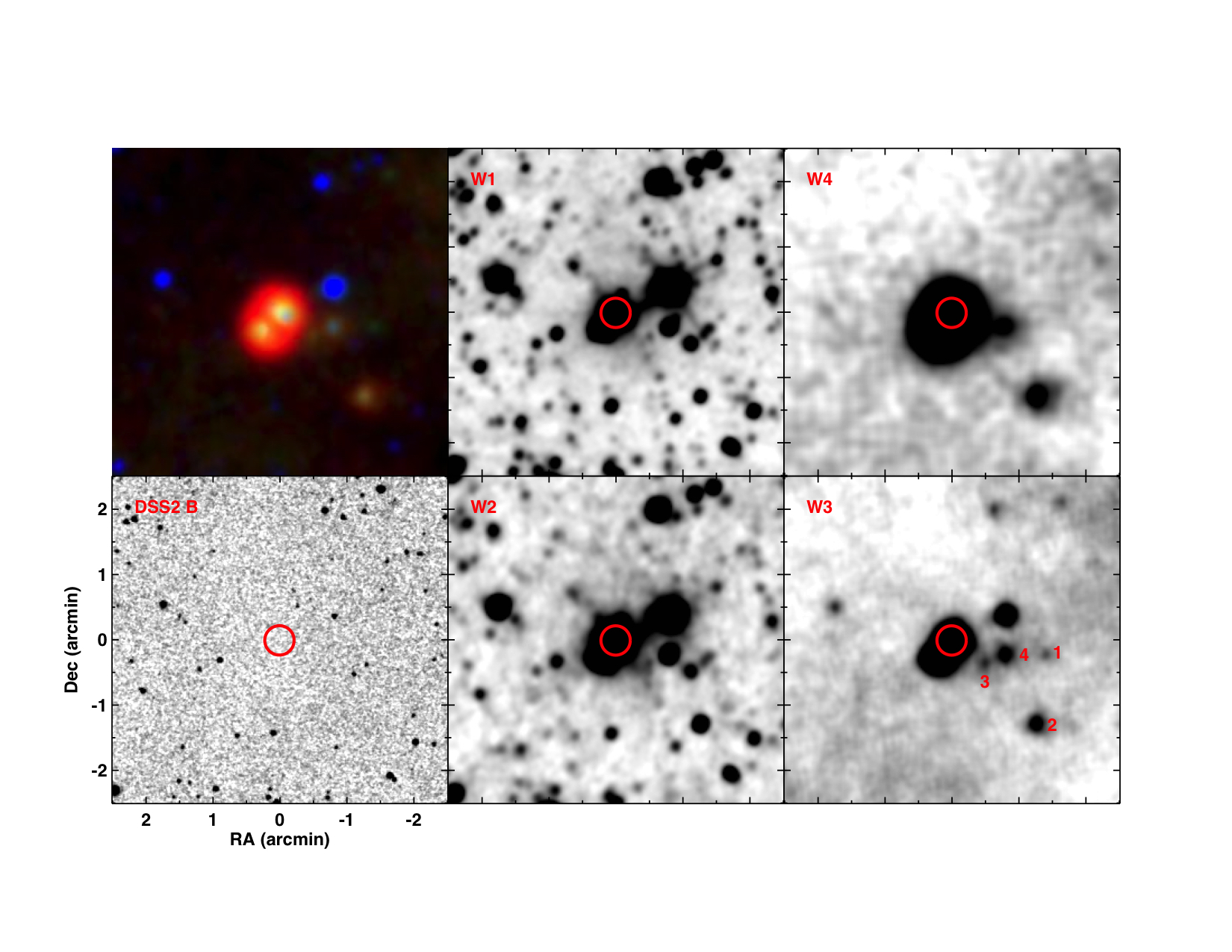}
\caption{Six views of the extremely red source IRAS 04287+6444 (WISE
 J043329.55+645106.5).  A red circle indicates the common position of the WISE
  emission peak in all six panels.  There is no hint of any emission
  in the optical (B band, lower left).  \wise\ reveals a large number of sources
  in the region in W1 and W2 (top and bottom middle), and a pair of extremely bright, blended
  sources in W4 (upper right).  The color composite image (upper left)
  shows that there are also four, fainter but also very red objects
  to the southwest of the primary pair.  We label these four sources
  in the W3 image (lower right).
   \label{04287a}}
\end{figure*}

Our most unusual objects are associated with IRAS 04287+6444.  The
brightest of these sources is slightly blended, which complicated our magnitude
corrections, giving our source an erroneously high $\gamma$ value
before our quality checks corrected the error.  This blending also triggered the high
 W3RCHI2 value that suggested this was extended source.  There appear to be at
least seven very red point sources clustered in this region in all.  We
 identify and number four fainter sources in Figure~\ref{04287a}
 (Bottom right); the other three or more all contribute to the
 brighter blend in the NE.  
 
NED reports five entries within 2$^\prime$ of these sources'
positions.  The two nearest detections are from
the 2MASS Extended Source Catalog (see below), and the third nearest 
entry is the IRAS counterpart to this source.  The other two NED
entries are a ROSAT detection \citep[1WGA J0433.4+6451][]{RASS} centered $\sim 33^{\prime\prime}$ 
away (outside the $20^{\prime\prime}$ astrometric precision of ROSAT), and the 1.4 GHz radio
source NVSS J043322+645120 \citep{NRAOVLA} centered $\sim 0.8^\prime$
away (consistent with being associated with our source).

\citet{IRAS_redshift} included the IRAS center in their optical spectroscopic
survey, and identified it as ``cirrus or dark cloud,'' although they
note that a significant number of such entries may in fact be
galaxies.  This non-detection is not surprising given the lack of
optical counterpart to these objects.

\begin{figure}
\plotone{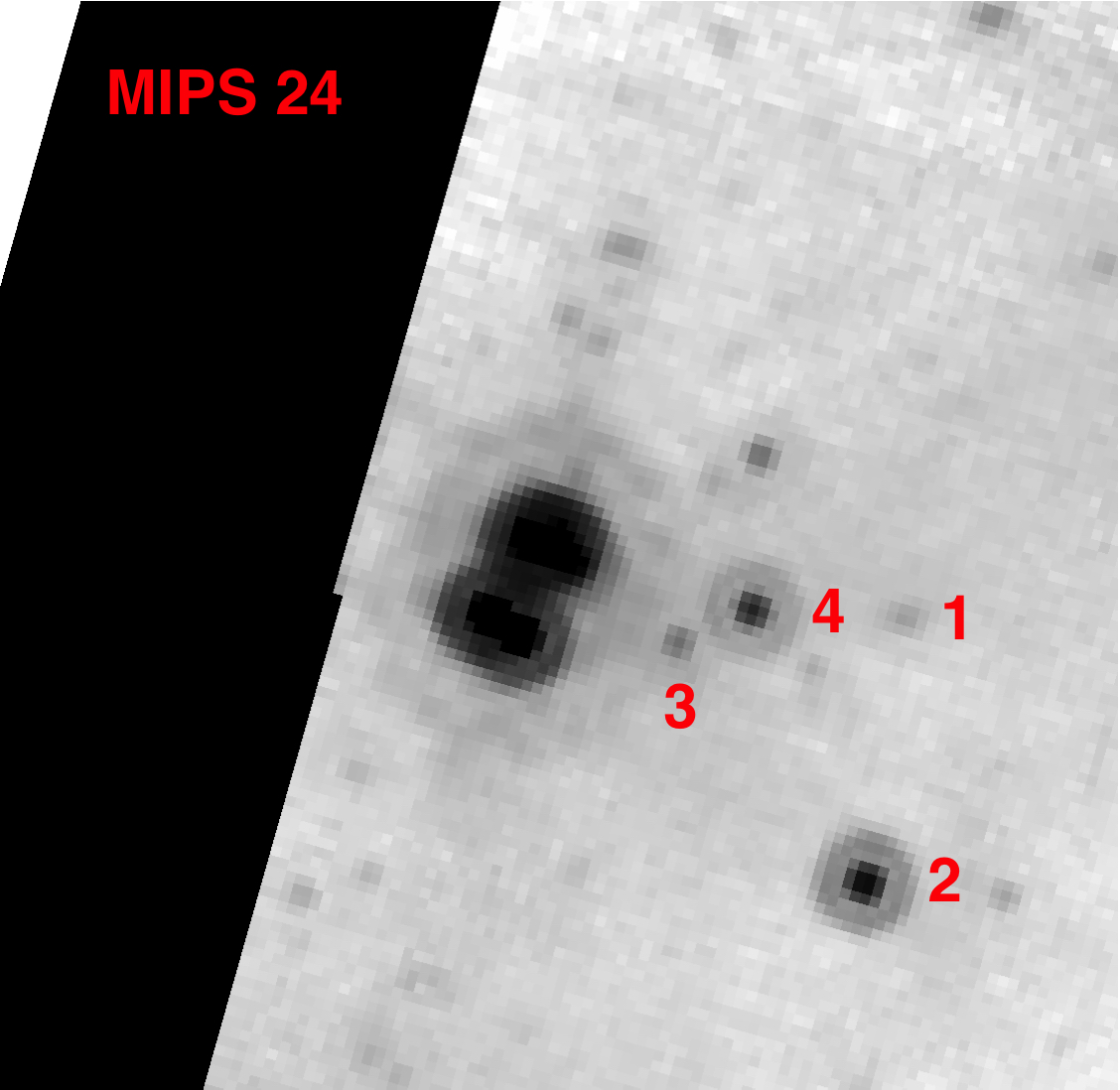}
\caption{Serendipitous archival MIPS imagery of IRAS 04287+6444.  The
  bright \wise\ source (which shows substantial substructure here) and
  the four fainter \wise\ sources are all detected in this 22$\mu$ image.\label{Spitzer}}
\end{figure}

We have found a serendipitous archival {\it Spitzer} MIPS image of
these sources, taken because they are within 15$^\prime$ of HD 28495,
a target observed with IRAC as part of the FEPS program \citep{FEPS}.
Figure~\ref{Spitzer} shows how the superior resolution of this 22$\mu$ imagery reveals substructure
to the SE component of the bright blend, and many sources undetected
by \wise\ (without color in formation, it is unclear whether these
sources are associated with IRAS 04287+6444).

Objects \#1 and \#3 are barely detected in the WISE images (not
appearing in the W4 band at all)
and thus provide little clues as to their true nature, but appear to
be fainter versions of the other red objects, with similar
colors. Objects \#2 and \#4 appear
cleanly detected in all four bands.

The angular proximity to HD 28495 is intriguing, but these are likely
unassociated since a common distance at 25 pc \cite{Hipparcos2} would imply
a projected separation of $\sim 2\times 10^4$ AU.  Nonetheless, the lack of optical
counterpart complicates efforts to rule out this scenario from proper motion.

\begin{figure*}
\includegraphics[width=7in,trim=0 2in 0 2.5in,clip]{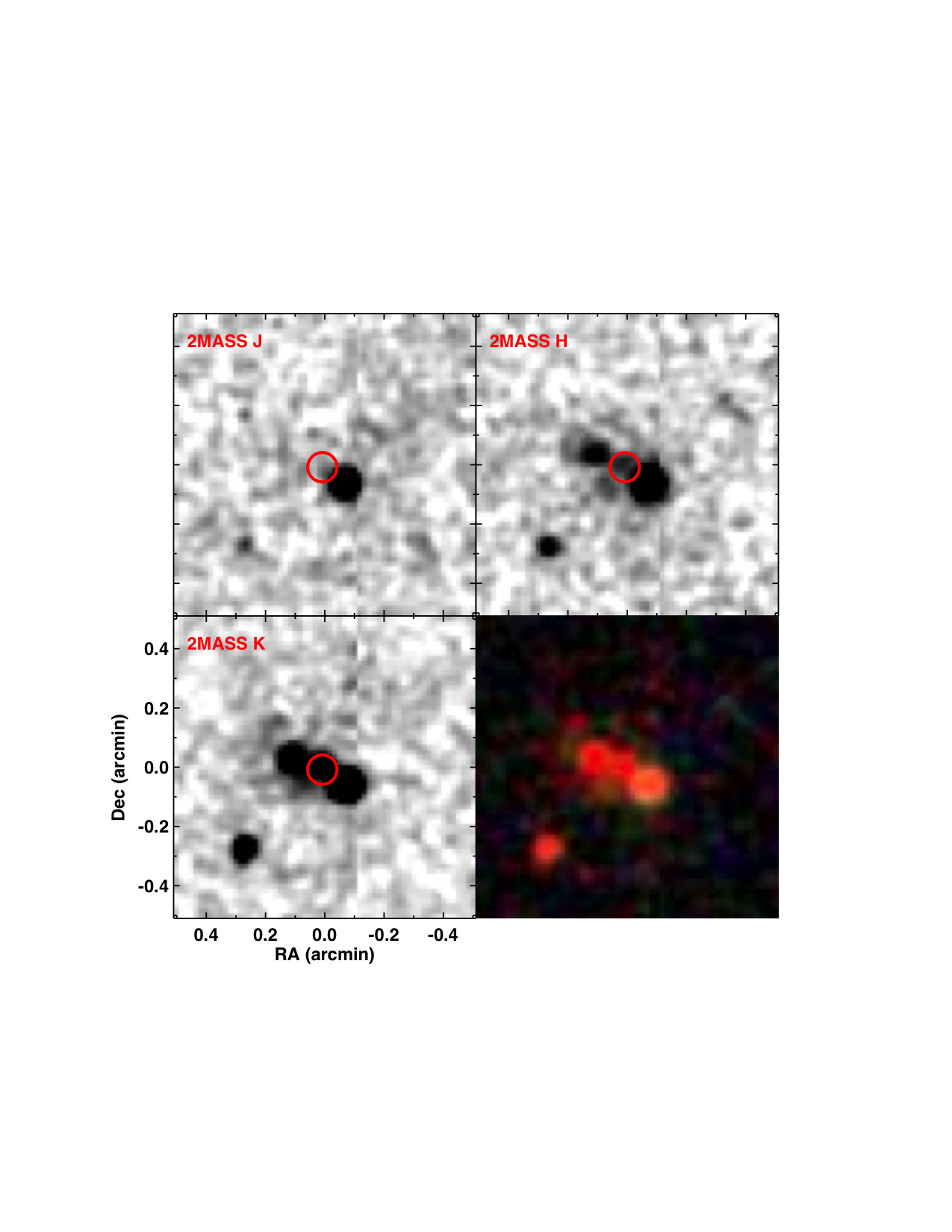}
\caption{Four closer-in, NIR views of the extremely red source IRAS 04287+6444 (WISE
 J043329.55+645106.5) from 2MASS. A red circle indicates the common position of the WISE
  emission peak in the three single-band panels.  Note the change in angular scale with respect
 to Figure~\ref{04287a}.  The K band image (lower left)
  reveals that the bright \wise\ source comprises at least three
  sources.  Of these, only the NE and SW sources are apparent in H
  band (upper right), and only the SW shows a J band counterpart
  (upper left).  The more distant SE object, responsible for making
  the \wise\ source appear extended, is detected in all three bands. The color composite (lower right) shows that all four
  sources have extremely red NIR colors. 
   \label{04287b}}
\end{figure*}

One possibility is that this is a previously uncataloged moderate-latitude ($b=11.5^\circ$)
dark cloud, and that these are an embedded cluster of young stellar
objects or protostars.  Figure~\ref{04287b} shows the 2MASS imagery
for this region, which has significantly better angular resolution.
This NIR imagery reveals that the brightest source in the \wise\
imagery comprises at least three sources, only one of which is evident
in J band.   Supporting this interpretation, \citet{YangCO} detected CO (J=1$\rightarrow$0) emission in the
direction of the IRAS source\footnote{The ``Association'' field from
  Table 2 of \citet{YangCO} for IRAS 04287+6444 confusingly
  reads ``HL Tau''.  HL Tau itself appears in the table two 
entries prior, where the ``Association'' field reads ``04288+6444,''
apparently a typo for 04287+6444.  We presume that \citeauthor{YangCO}
erroneously transposed the ``Association'' values for these entries.
If, instead, it is the target names that are transposed, then the
appropriate LSR radial velocity for IRAS 04287+6444 is 6.91 km/s, the FWHM is
4.3 km/s, and no kinematic distance is available.} with LSR radial velocity -13.27 km/s and
FWHM 3.4 km/s, implying a molecular cloud exists in this direction at
a kinematic distance of 840 pc (and thus a height of $\sim 170$ pc
below the Sun's position in the plane). 

If the sources are extragalactic, the most natural explanation is
that they are members of a galaxy cluster.  \citet{Edelson12}
identified the IRAS source as having a modest chance ($\sim 50\%$) of
being an AGN of some flavor based on the \wise , 2MASS, and X-ray
fluxes, however the X-ray detection may be unassociated, and it
appears this probability does not incorporate the fact that the source
has no optical counterpart or that it is not isolated.    

The lack of optical counterpart could be due to redshift and internal
extinction.  A significant population of high redshift ($z\
> 2$) and more luminous ($L_{IR} > 10^{13}$\Lsun ) Dust Obscured
galaxies (called Hot Dust-obscured Galaxies, or ``Hot DOGs") have recently been
identified by the WISE survey, \citep[see][for
a detailed discussion of these types of objects]{HyperLIRGs,
HyperLIRGsubmm,Bridge13,Stern2014}.  Indeed, the strong 22 $\mu$m
emission for these objects are reminiscent of Dust Obscured Galaxies (DOGS), either local ($z
\sim 0$) \citep{LocalDOGs} or at high redshift ($z \ge 2$)
\citep{DeyDOGs}.  

However, our objects are inconsistent with hot DOGs
since hot DOGs tend to have very little or no emission in the shorter
WISE bands.  And such extreme examples of dusty galaxies are not
typically highly clustered as our sources are. 

We tentatively favor the interpretation that this is a cluster of
young stellar objects embedded in and heavily extinguished by their
parent molecular cloud.  We are intrigued by these objects, and we hope that
spectroscopic observations can and will reveal their true nature in the future.  

\subsubsection{WISE J224436.12+372533.6: A new MIR-bright galaxy}

\begin{figure*}
\includegraphics[width=6.5in,trim=0 0.75in 0 0,clip]{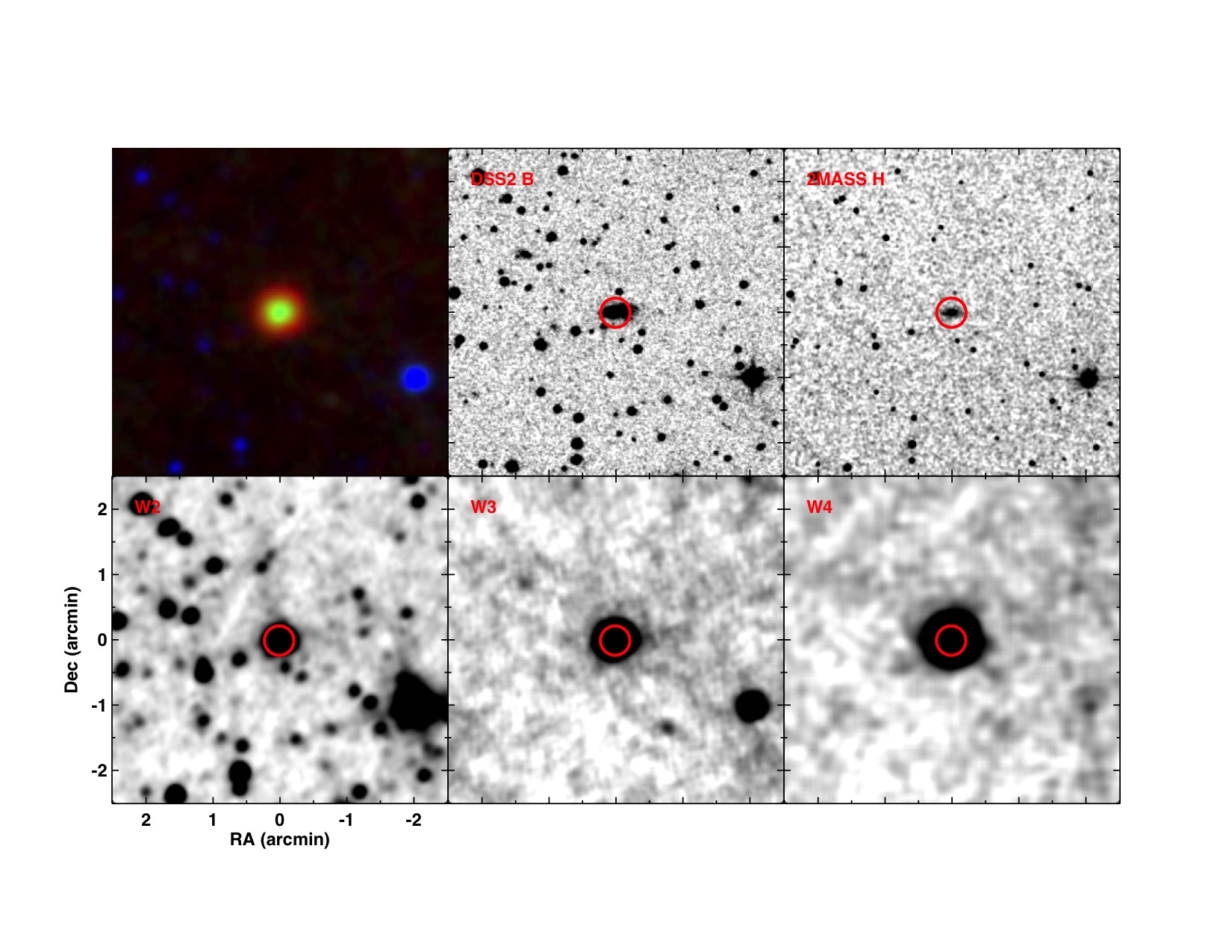}
\caption{The previously unstudied but MIR-bright galaxy
  WISE J224436.12+372533.6.  It was only barely detected by IRAS, but
  is easily detected by \wise\ as an extremely red MIR source.\label{W224436}
}
\end{figure*}

We gave the object WISE J224436.12+372533.6 (shown in
Figure~\ref{W224436}) an A grade because it has
no presence in the astronomical literature beyond having been noted in
the 2MASS Extended Source Catalog.  It is MIR bright and red, and
DSS and 2MASS imaging shows what appear to be a galaxy. It is just
barely detected by IRAS (it appears in only two bands in the IRAS
Faint Source catalog \citep{1992iras}), and so may have evaded
prior notice for that reason.  It also appears as a 1.4 GHz source in the NRAO VLA radio
survey \citet{NRAOVLA}.   This source deserves further study to
understand its superlative nature.

\subsubsection{IRAS 16329+8252: An MIR-bright galaxy at z=0.04?}

\begin{figure*}
\includegraphics[width=6.5in,trim=0 0.75in 0 0,clip]{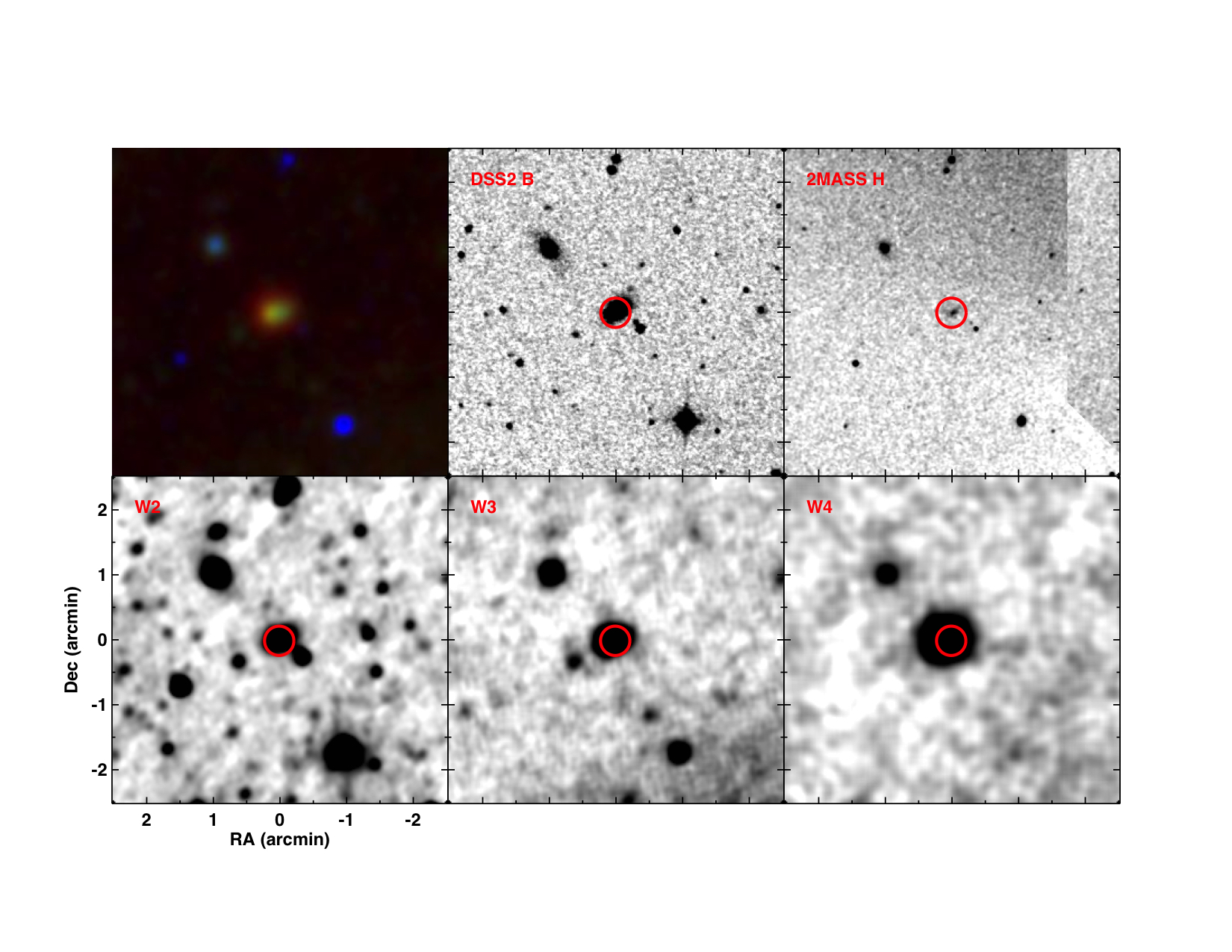}
\caption{Six views of the extremely red source IRAS 16329+8252.   A red circle indicates the common position of the WISE
  emission peak in all six panels.  The MIR
  morphology (upper left, where red = W4, green = W3, blue = W1+W2) is
  consistent with the optical B band (upper middle), although the
  shorter \wise\ bands (W2, lower right) appear slightly offset to the west, consistent with
  the H band imagery (upper right).   The W3 and W4 band imagery
  appear very slightly extended.  This is apparently a very MIR-bright
  z=0.039 galaxy, possibly disturbed by a neighbor to the northeast,
  or along the line of sight. \label{1627}}
\end{figure*}

We give this source an A grade because it has virtually no presence in
the literature beyond a single low resolution spectrum by
\citet{Chen11}.  If their identification of the emission lines in this
spectrum is correct, then it is a galaxy at $z=0.039$.  Its strong MIR
emission suggests large amounts of star formation, possibly triggered
by the disturbance of a nearby neighbor (see Figure~\ref{1627}).

\subsubsection{WISE J073504.83-594612.4: a pair of merging galaxies?}

We graded the source WISE J073504.83-594612.4 (Entry 2 in
Table~\ref{extreme}) A because it has virtually no presence in the
literature, appearing only in a handful of photometric catalogs,
including the 2MASS extended source catalog. It has no IRAS
counterpart.  It appears to be extragalactic.

\begin{figure*}
\includegraphics[width=6.5in,trim=0 0.75in 0 0,clip]{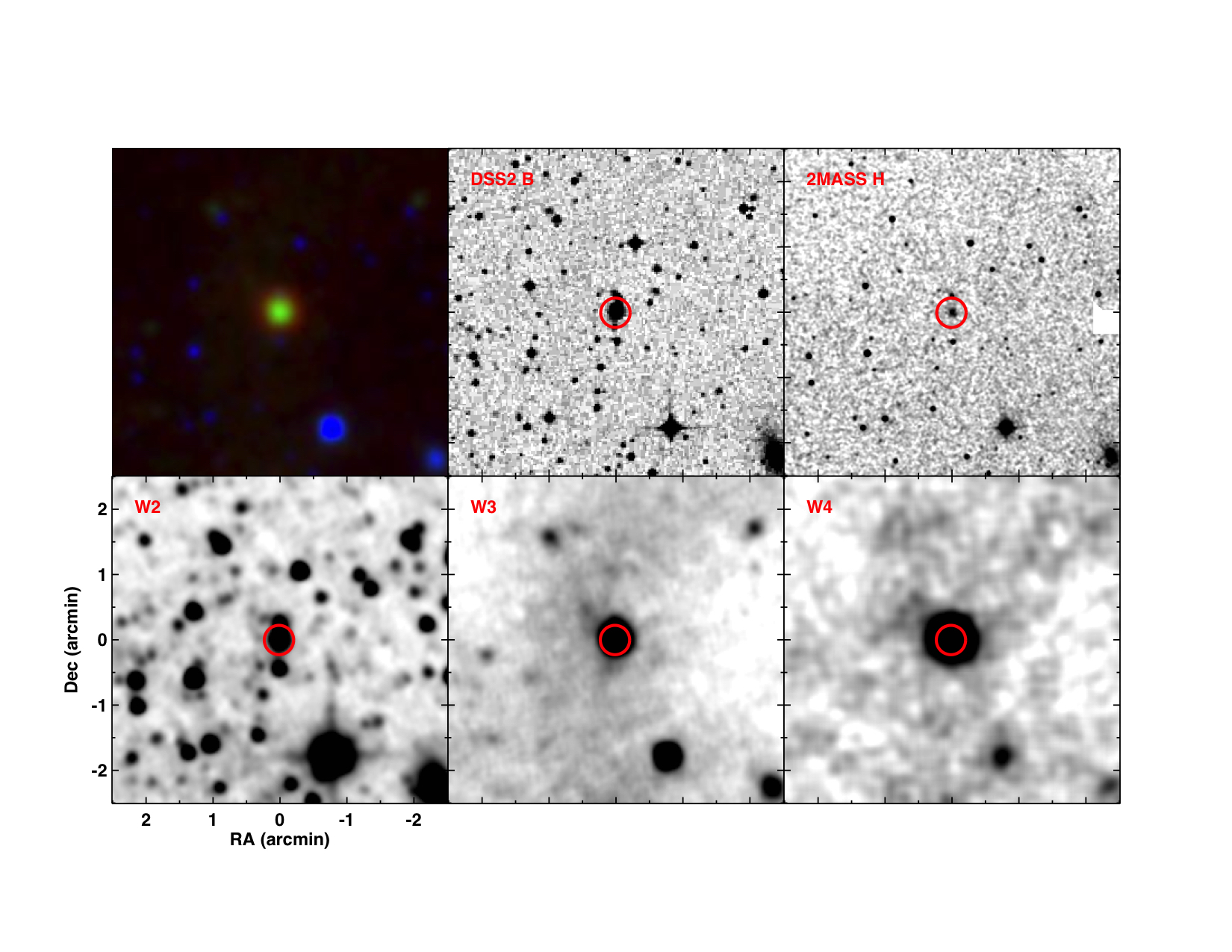}
\caption{Six views of the extremely red source WISE
  J073504.83-594612.4.  This source has
  no known IRAS counterpart.  The \wise\ color image  (upper left, where
  red = W4, green = W3, blue = W1+W2) shows little structure; it is
  barely extended in W3 and W4 (lower middle and right).  The H band image
(upper right) and W2 image (lower left) reveals two sources, one at the position of
  the \wise\ source and one $\sim 20{^{\prime\prime}}$ to the north, which are
  perhaps the bulges of a pair of galaxies.  B band shows a possible bridge of
  material between the two H band sources, indicating that this may
  be a pair of merging galaxies.  \label{0735}}
\end{figure*}

The \wise\ color image (Figure~\ref{0735}) shows little structure; it is
  barely extended in W3 and W4.  The H band and W2 imagery, which
  presumably trace the stellar populations of the bulges of the
  galaxies in this imagery, reveals two sources, one at the position of
  the \wise\ source and one $\sim 20^{\prime\prime}$ to the north.  The DSS B
  band image shows significant structure between these positions,
  suggesting that this is a disturbed or merging galaxy pair.  The
  bright W4 emission suggests that this disruption his generating
  significant star formation in the southern galaxy.

\subsubsection{IRAS 15553-1409: a large nebula of dust from a Be shell star?}
\label{sec:15553}

\begin{figure*}
\includegraphics[width=6.5in,trim=0 0.75in 0 0,clip]{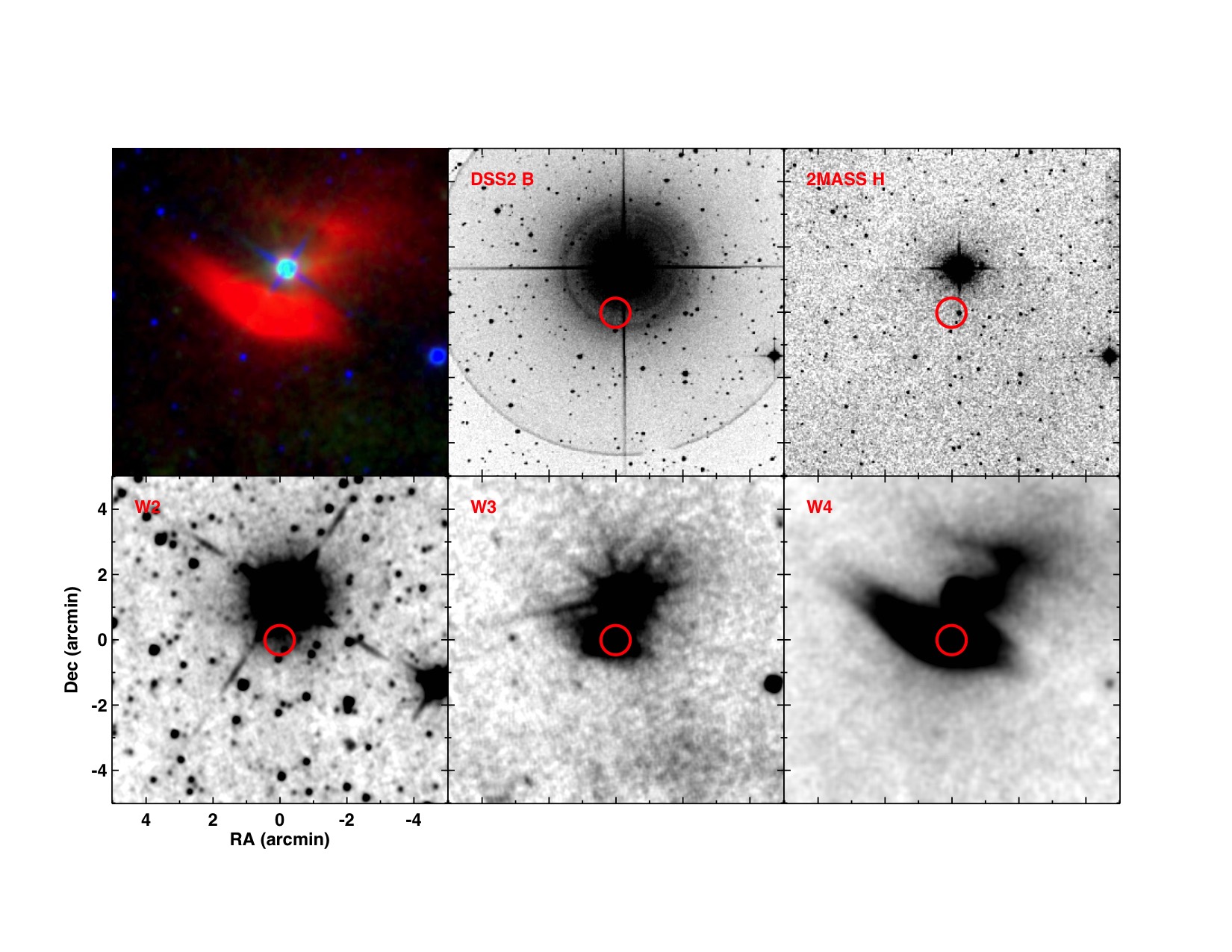}
\caption{Six views of the extremely red source IRAS 15553-1409 (WISE
  1558-1418).  A red circle indicates the common position of the WISE
  emission peak in all six panels.  The bright point source is the Be star 48 Librae.  The MIR
  morphology (upper left, where red = W4, green = W3, blue = W1+W2) is
  typical of a reflection nebula, but there is no emission obvious in
  the optical (B band, upper middle). The NIR (H band, upper right)
  and W2 bands (lower left) appear unremarkable.  The brightest part
  of the cloud is apparent in W3 (lower middle) and large
  amount of diffuse emission are obvious around the star in W4 (lower right).  \label{15553}}
\end{figure*}

The infrared source IRAS 15553-1409 is the reddest or second reddest
source in four of the six \wise\ colors.  We give this source a B grade because it has
virtually no presence it the literature, beyond \citet{Carballo92}, who identify
it as a potentially ``evolved Galactic object.''  \wise\ 
imagery reveals it to be a MIR-bright nebula associated with the
classical Be star 48 Librae.  Figure~\ref{15553} shows that the MIR
morphology and colors are similar to that 
of reflection nebulae, but in this case there is no apparent emission in the
optical.   

48 Librae is known to be a classical Be shell star
\citep[e.g.][]{48Lib_shell} and a member of the Sco-Cen Assocation.  
It has a distance of $\sim 140$ pc \citep{Hipparcos2}, expansion
velocities of 25 km/s \citep{BSCNotes}, and age $< 20$
Myr (Eric Mamajek, private
communication\footnote{\url{https://www.facebook.com/jason.wright.18062/posts/10204670796112674}}
and references therein.)  It is sometimes listed in the literature as
a giant star, but this is likely because its rapid rotation gives it an
anomalously low surface gravity, complicating its spectrally derived
luminosity classification.

This source is superlative because of its extreme MIR
colors and extent, although this is a byproduct of its proximity;
similar Be shell stars at greater distances would not be resolved.

\citet{AdamsIR} notes that the infrared excesses of main sequence
stars typically have one of four origins: circumstellar debris disks,
protoplanetary and protostellar disks around very young stars,
``cirrus hot spots'' caused by the illumination of ambient
interstellar dust, and the excretion disks of classical Be stars.  

\begin{figure}
\begin{center}
\includegraphics[width=3in]{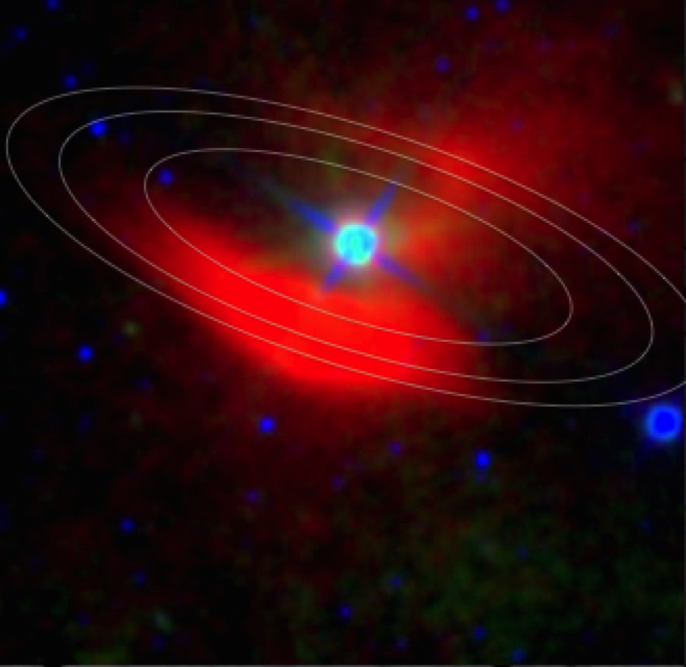}
\caption{A \wise\ color composite of the 48 Librae nebula, with
  ellipses overlain to illustrate the common symmetries of the arc
  structures in the nebula.  The ellipses describe concentric rings centered
  on the star inclined 19$^\circ$ from edge-on.  The physical size of
  the nebula is $\sim$ 15,000 AU.\label{symmetry}}
\end{center}
\end{figure}

Cirrus hot spots \citep{AdamsIR,vanBuren} can be simple reflection
nebulae \citep[the ``Pleiades phenomenon,''][]{Kalas} or the result of
bow-shocking by the star's wind or radiation as it moves through the
ISM \citep[e.g.][]{vanBuren,Povich08,Everett10}.  Inspection of the morphology
of typical reflection nebulae (including that of the Pleiades themselves and
those discovered by \citet{Kalas}) shows that 48 Librae does not
appear to be a typical example.  The 48 Librae nebula has no optical
counterpart, suggesting that the 22 micron emission is thermal, and
appears to have symmetries about the position of the star, suggesting
that it has some connection to the star beyond being 
illuminated by it (see Figure~\ref{symmetry}).

The most likely explanation of this nebula is that it is therefore a
ring or bow shock.  The bow shock and ring interpretations are complicated by the fact
that, despite being a Sco-Cen Association star, 48 Librae does not
appear to be embedded in a region of high density gas and dust, and that
its space motion is not large.

The proper motion of 48 Librae is SW \citep[$(\mu_\alpha, \mu_\delta) = (-12.44,
-16.73)$ mas/yr][]{Hipparcos2}, which is not in the direction of the brightest
sources of emission, but it appears that most of this
apparent motion is actually due to motion of the Sun with respect to
the local standard of rest.  Correcting for the solar motion (using radial
velocity $7.5\pm$ 1.8 km/s \citep{48LibRV} and the LSR of \citet{LSR}), the
proper motion is still roughly SW ($(\mu_\alpha, \mu_\delta) = (-2.4,
-0.9)$ mas/yr), corresponding to an LSR space velocity of 4 km/s, with
an error of a few km/s from our LSR correction.

Further complicating the bow shock interpretation is that both sides of the 48 Librae
nebula have similar arcs consistent with being portions of rings
inclined $\sim 20^\circ$ from edge-on, and centered on the star.  

Given these difficulties with the typical bow shock and ring models, we suggest that
the nebula have originated with the star itself.  Classical and shell
Be stars are not typically considered to be significant sites of dust
formation or sources of MIR excess \citep{Be_review}, although NIR
excesses, presumably due to circumstellar dust, are common, and many unclassified Be
stars are known to have MIR
excesses\citep[e.g.][]{Be_FIR,Be_dust,BE_NIR}.  48 Librae has one of
the larger and stronger disks known among classical Be stars (T.\
Rivinius, private communication).

Since 48 Librae is known to have shells, a time variable disk
\citep{48Librae}, and significant mass loss, it is reasonable that
some of this shell material would condense into dust near the star
before being lost \citep{Be_dust}.  The nebula is not too
large for this: the nebula has an angular size of order a few arcminutes, which at the
distance of 48 Librae ($\sim 150$ pc) corresponds to $\sim 15,000$ AU.  The expansion
velocity of the shells is 25 km/s, yielding a characteristic timescale
of $\sim 30,000$ yr, significantly shorter than the Be phase of a
star.   

So, we may be seeing shocked dust where the shells (really rings) of
excreted material collide, either with the ambient ISM or with
previously ejected material.  If so, then we expect the rings we observe
in the nebula to have a rough correspondence to the geometry of the
excretion disk.  The rings (which we fit by eye) have an
inclination of 70$^\circ$ (i.e.\ 20$^\circ$ from edge-on), and a
position angle of 72$^\circ$.  The actual inclination of the
excretion disk must be $\gtrsim 65^\circ$ (because it shows absorption
lines from the disk, T.\ Rivinius, private communication), and the
actual position angle is known from interferometry to be 50$^\circ
\pm$ 9$^\circ$ \citep{48Librae}, consistent with our rings at 1--2 $\sigma$, within the rough precision
with which we can define them.

Alternatively, we may be seeing excreted dust being illuminated by UV
radiation from 48 Librae itself, although this interpretation is
complicated by to the lack of apparent optical scattered light, and by
the patchiness of the emission.

It is unclear if the nebula contains more dust than could be plausibly
explained by the currently observed mass loss rate, but of course 48 Librae
could have had episodes of higher mass loss rates in the past. 

This object may be revealing to us that Be shell stars are, in fact, common sites of
dust generation.  This phenomenon warrants further study.

\section{Upper Limits on the Energy Supplies of Type {\sc iii}
  Kardashev Civilizations}

\subsection{Limits on energy supplies as a fraction of stellar luminosity}

We can use those sources in our study with the largest amounts of thermal emission
to set an upper limit for waste heat emission among the galaxies we
have surveyed.  We have found no sources with $\gamma > 0.85$, and the
50 galaxies we have found with $\gamma > 0.5$ appear to have 
natural origins to most of their MIR emission, although we
have not rigorously verified this.

These values for $\gamma$ were calculated under the assumption that our target
galaxies are composed of only two components: an old stellar
population, and alien waste heat originating entirely from
reprocessed starlight.  Since real galaxies have other sources of
MIR emission, these numbers are upper limits on the
integrated alien waste heat emitted by these galaxies.  

Our assumption that the origin of the alien waste heat is intercepted
starlight (i.e.\ $\gamma = \alpha$, where $\alpha$ is the fraction of
starlight intercepted, see Section~\ref{agent}) means that we have
assumed that the flux in the W1 and W2 bands, which in our model
constrain the total stellar luminosity, does not include the starlight
lost to alien factories.  If, instead, we assume that all alien waste
heat is generated by other means and that only a negligible fraction
of starlight is occulted ($\alpha\sim 0$), then we would infer a
higher value for the fraction of starlight emitted as waste heat.
Specifically, our limiting values of $\gamma = 0.85$ and 0.5
correspond to $\gamma_{\alpha=0} = 5.7$ and 1 \citep[see Section
3.2][]{WrightDyson2}.  

In other words, we shave shown that there are no galaxies resolved by
\wise\ with MIR luminosities consistent with alien energy
supplies in excess of 5.7 times the starlight in their galaxies (i.e.\
we have ruled out $\gamma = \epsilon > 5.7$).  If all of 50 galaxies
in our list turn out to have purely natural origins to their emission,
then this upper limit drops to $\gamma = \epsilon < 1$.

If we assume that any large alien energy supply will be based on starlight (that is, $\gamma
\sim \alpha$ and $\epsilon \sim 0$), then our upper limit is much
tighter:  no resolved galaxies exist in our search area with more than
85\% of their starlight reprocessed by alien factories, a limit which
will drop to 50\% when our 50 high-$\gamma$ galaxies are more carefully vetted.

Translating these numbers into physical units (erg s$^{-1}$) will
require a more detailed modeling of the stellar and nonstellar
components of the galaxies we have surveyed, a project which is beyond
the scope of this paper.  We hope to pursue this in a future paper.

\subsection{Number of galaxies surveyed}
\label{number}

Translating our upper limits into an upper limit on the frequency of
K3's requires knowledge of the number of galaxies we have effectively
surveyed.  We cannot use our Gold or Platinum Samples to estimate this
number because they included color cuts to remove stars that also
removed elliptical and other dust-free galaxies.  Even if we had
imposed no such cuts, there are many galaxies that would be resolved
in the W3 band if they were MIR bright, but are unresolved
--- or in some cases undetected --- in \wise\ because their MIR surface
brightness is below the \wise\ detection limits.

To estimate the number of galaxies that {\it would have been included}
in our sample {\it if} they had $\gamma > 0.5$, we can use the number
of sources in the W1 and W2 bands.  The W1 band, in particular, has
better angular resolution that W3, and is primarily sensitive to
stellar photospheres, so is in many ways a clean band for estimating
the angular extent of galaxies around which alien factories
might reside.

One concern with using \wise\ data for this purpose is that, as we
have seen, the source counts include many non-galaxies (including artifacts), and many point
sources have erroneously large values of RCHI2.  To mitigate this, we have used the 2MASS
Extended Source Catalog (XSC), which is relatively clean of point
sources and is composed almost entirely of  galaxies.

We first cross-matched \wise\ to the 2MASS XSC, and selected only those
sources with $|b| \ge 10$.  We then examined the NIR properties of MIR-red galaxies in the 2MASS
XSC by examining the relationship between the WXRCHI2 values for the
W1 and W2 bands and the W3RCHI2 parameter, which we used for the
Platinum sample.  We restricted our analysis to 9,589 matched sources
with (W1-W3 $\ge$ 3.8) for the W1RCHI2 analysis, and 14,927
sources with (W2-W3 $\ge$ 3.5) for the W2RCHI2 analysis.

\begin{figure*}
\includegraphics*[width=6in]{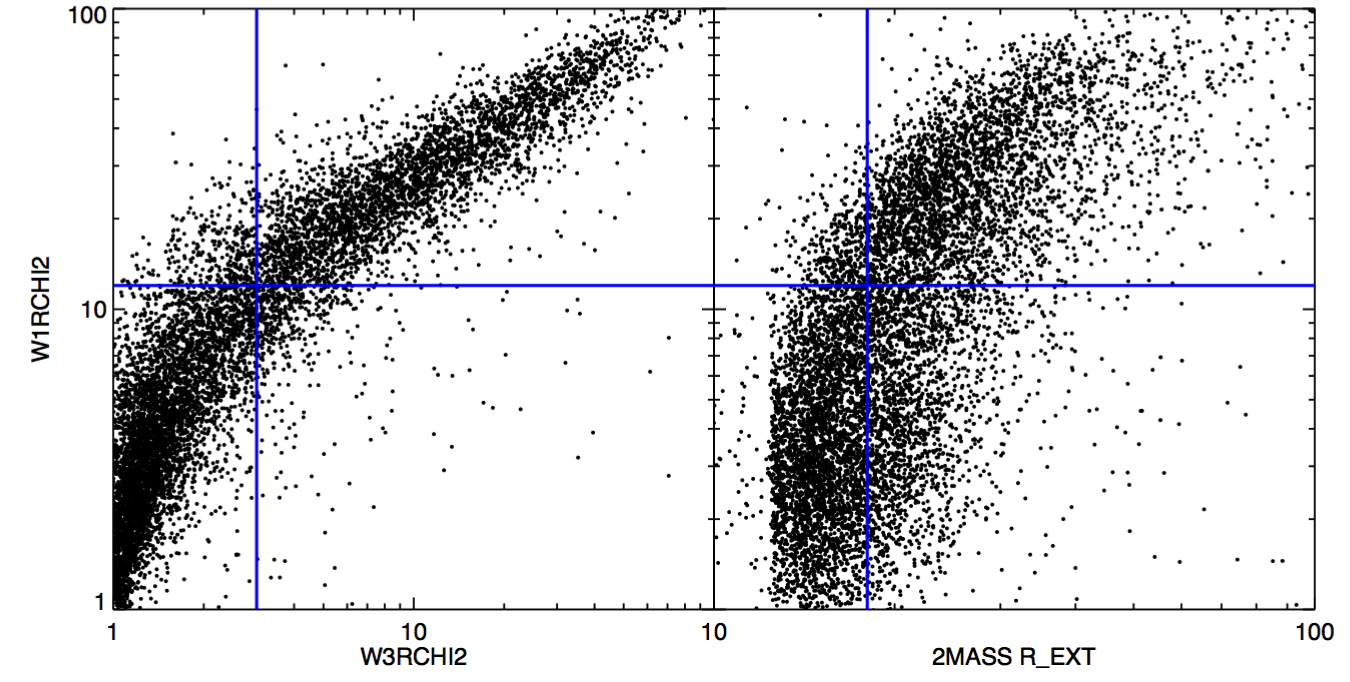}
\includegraphics*[width=6in]{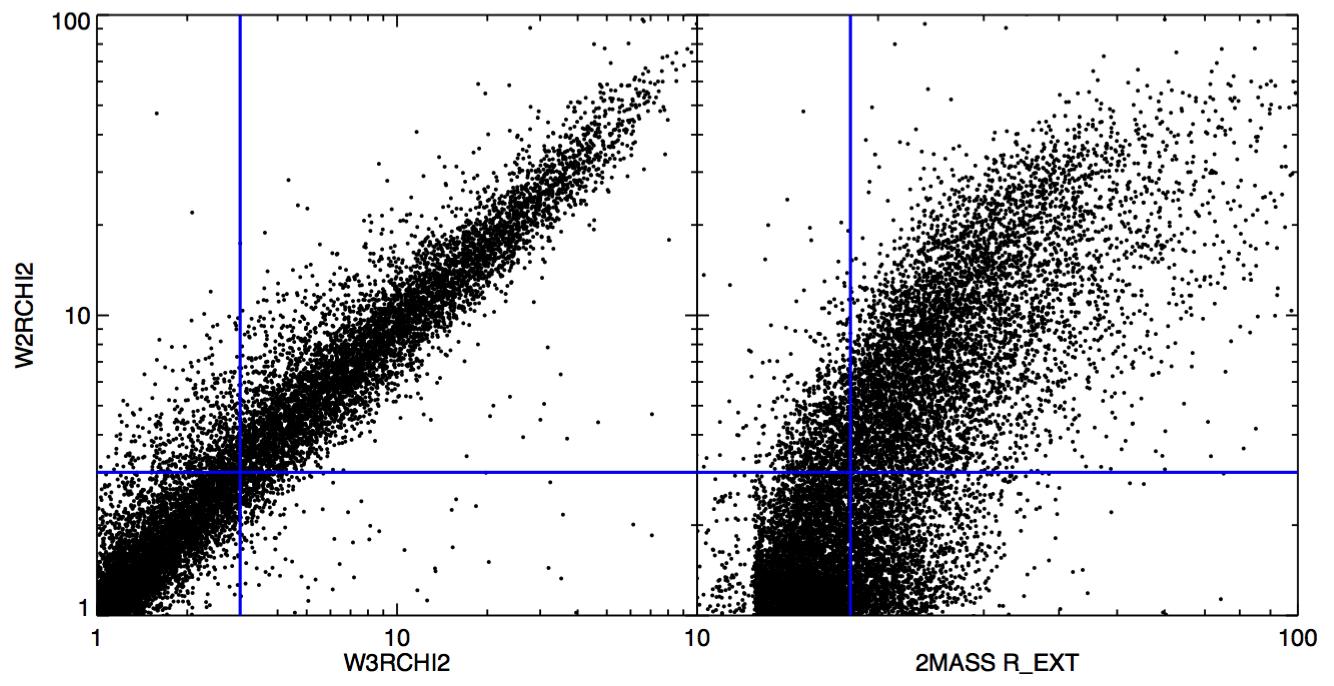}
\caption{Correlations among size parameters for red extended sources
 in \wise\ with 2MASS Extended Source Catalog (XSC) counterparts.  The
  y-axis in all plots corresponds to the RCHI2 parameter, which
  measures the goodness-of-fit of a source to a model for a point
  source (large values indicate a poor fit, so an extended source).
  The x-axis of the left panels is the RCHI2 parameter in the W3 band,
  (we used W3RCHI2$=3$, indicated by the vertical lines, to identify
  extended sources in our catalogs).  The horizontal lines mark the
  approximate RCHI2 value in the W1 and W2 bands for typical red
  sources that meet this criterion.  The
  x-axis of the right panels is the R\_EXT parameter of the XSC, in
  units of arcseconds, describing the NIR angular size of the source.
 The vertical line shows the value of this parameter 
 18$^{\prime\prime}$, that best corresponds to the W1 and W2 RCHI2
 values typical of barely extended sources in W3. {\it
   Top:} correlations between the W1 RCHI2 parameter.  {\it Bottom:}
  correlations with the W2 RCHI2 parameter.   \label{count1}}
\end{figure*}

The left hand panels of Figure~\ref{count1}
show the relationship between the W1RCHI2 and W2CHI2 parameters (which
describe the degree to which galaxies are resolved in those bands, see
Section~\ref{rchi2}) and the vs.\ W3RCHI2 parameter we used to define a
source as ``extended'' in our survey.

These panels show that, for the red sources we used to construct this
figure, we can use W1RCHI2 $> 12$, and W2RCHI2 $>3$ as
proxies for our actual criterion W3RCHI2$>3$. Using the full
\wise-2MASS XSC cross-matching (that is, not imposing any color cuts), we find
1,589,099 sources common to both catalogs.  Of these, 1,463,781 have
$|b| \ge 10$.  Of these, 111,617 have W1RCHI2 $\ge 12$ and 104,039 have
W2RCHI2 $\ge 3$.  These figures are consistent, suggesting that we
have surveyed $1 \times 10^5$ galaxies.  The only previous search for
K3's in the refereed literature, that of \citet{annis99a}, surveyed
163 galaxies.

As a check, we also used the R\_EXT parameter in the XSC, which
corresponds to a measure of the NIR angular size of these galaxies.
The right hand panels in Figure~\ref{count1} show that
R\_EXT corresponds to the angular sizes we 
are interested in with good sensitivity (i.e., the test  R\_EXT$>18^{\prime\prime}$ has a
low false negative rate), although it is not very specific (i.e.\ it has
roughly a 50\% false positive rate) for these red sources.   

A query of the entire XSC with $|b| \ge 10$ and
R\_EXT$\ge 18$  yields 229,813 sources.  A random sampling of 100 of
these sources reveals that all are present in the \wise\ All-sky
catalog, 45 have W1RCHI2$>12$ and 43 have
W2RCHI2$>3$.  These numbers are consistent with the specificity we estimated among the
MIR-red sources in Figure~\ref{count1}.  We also tested 100 random
\wise\ sources satisfying our extended source criteria in W1 and W2, and
find that 86 and 87 of them, respectively, have
R\_EXT$>18^{\prime\prime}$, also consistent with the sensitivity suggested
in Figure~\ref{count1}.  

We conclude that there are $\sim 1\times 10^5$ galaxies with sufficient angular
size that they would have been included in our platinum sample if they
had had significant W3 emission.  In our survey of these $\sim 1\times 10^5$ galaxies, we have found
that there are no alien, non-stellar energy supplies in excess of 5.7 times the
stellar luminosity of their host galaxy, and no alien
supercivilizations reprocessing as much as 85\% of their starlight into the
MIR.  We have found 50 galaxies consistent with 50\%
reprocessing, all of which are presumably extraordinary, but entirely
natural, star-forming galaxies.  Verification of the natural origin of
the MIR flux in these 50 galaxies will thus lower our upper limit to 50\%.

% \subsection{other stuff}

% The two highest $\gamma$ sources in this list, W0621-6856 and
% W0621-6749, with $\gamma$ of 0.99 and 0.66, respectively, appear to be
% associated with the LMC and are most likely of galactic origin (AGB,
% PN, or Y*O). It should be noted that the types presented in this list
% are strictly tentative since very little is known about them. The
% waste heat temperature is a very powerful diagnostic to distinguish
% natural from artificial emission and it is important to consider when
% searching for signs of galaxy-spanning ETIs. By segregating the
% sources in this list into three temperature categories, too cold
% $T_{\text{waste}} \le 180$ K, moderate $180 < T_{\text{waste}} \le
% 370$ K, and too hot $T_{\text{waste}} > 370$ K we are able to more
% easily identify our best KIII candidates. Using this table we find 12
% sources to be too cold, 2 sources to be too hot, and the remaining 29
% sources appear to have moderate waste heat temperatures and make up
% the best KIII candidates in our sample.  

% In Table~\ref{extreme} we also present a list of 95 objects ($\gamma \ge 0.25$)
% classified as type Bs, again sorted by $\gamma$. Using the waste heat
% temperature we find 74 being too cold, 2 appearing to be too hot, and
% the remaining 71 being of moderate waste heat temperatures. In total
% we present 95 very extreme infrared sources of moderate waste heat
% temperatures which make excellent candidates for follow-up
% observations.  

%\subsection{Best KIII candidates}

\section{Conclusions}
\label{conclusions}

We have produced a clean catalog of the reddest extended sources in
outside the Zone of Avoidance using in the \wise\ All-sky catalog, and
corrected that catalog's photometry of extended sources to be consistent with careful
aperture photometry at the 3--5\% level.   We used the point
source goodness-of-fit parameter W\#RCHI2 to identify extended sources,
and various tests (including visual inspection and interrogation of
the Level 1b \wise\ data) to clean this sample of instrumental and
data pipeline artifacts and point sources.

We have graded each of our sources in terms of its presence in the
published literature, to determine whether the nature of its MIR
emission is well understood.  

Our motivation is to use this catalog to perform the first
extragalactic search for waste heat from galaxy-spanning alien
supercivilizations.  To that end, we have used the AGENT formalism of
\citet{WrightDyson2} to interpret the \wise\ SEDs of these sources as
ordinary elliptical galaxies with alien waste heat luminosities equal
to a fraction $\gamma$ of the starlight and characteristic temperature
\twaste.  This is an inappropriate model for natural sources,
especially spirals and star-forming galaxies, but it provides a
conservative upper limit on the true $\gamma$ paramater for the
galaxy.  

We find that there are no galaxies in our sample of $1\times 10^5$ galaxies with fit values of
$\gamma > 0.85$, meaning that no galaxies resolved by \wise\ contain
galaxy-spanning supercivilizations with energy supplies greater than
85\% of the starlight in the galaxy (unless this energy is not
primarily expelled as light in the \wise\ bandpasses).  We have
further identified all 50 resolved galaxies in our sample with fit
values of $\gamma > 0.5$.  More detailed SED modeling of these galaxies, including the
use of other bands, will allow for more stringent upper limits, and we
will perform such modeling in the future. 

We also identify 93 sources with $\gamma > 0.25$ but very little study in
the scientific literature.  Three of these sources are MIR-bright and red galaxies
that are essentially new to science, having little or no literature
presence beyond bare mentions of a detection by IRAS or other surveys.    

Verification that the MIR flux in all of these galaxies is
predominantly from natural sources (e.g., through 
SED modeling across many more bands than \wise\ offers or
spectroscopy) will push our upper limit on galaxy-spanning alien
energy supplies in our sample of $1\times 10^5$ galaxies down to 50\% of the available starlight.  In the meantime,
these are the best candidates in the Local Universe for Type {\sc iii}
Kardashev civilizations.  This limit will improve upon the limit of $\alpha < 75\%$ found by
\citet{annis99a} in 57 spiral and 106 elliptical galaxies.  

We find that the Be shell star 48 Librae has a large extended MIR
nebula.  If the source of this dust is 48 Librae itself, it would suggest,
surprisingly, that dwarf Be shell stars can be sites of significant
dust production.  

We have also found a previously unstudied cluster of MIR-bright sources with no
optical counterparts and very red colors.  They appear to be Galactic sources
associated with a cloud, and so are likely part of a previously
unstudied star forming region.

In the appendices, we have also illustrated how WISE can be used to rule out broad
classes of K3 civilizations as being responsible for the lack of
emission in so-called \ion{H}{1} dark galaxies and the anomalous
colors and morphologies of ``red'' (or ``passive'') spirals.  We find
a sample of five ``red'' spirals with red MIR
and $(NUV-r)$ colors, which are inconsistent with high levels of star
formation but consistent with high levels of alien waste heat. 
Significant internal extinction would be a satisfactory natural
explanation for these colors, but until that is ruled out these
galaxies are some of the best candidates for K3's in our search to date.

\acknowledgements

This research is supported entirely by the John Templeton Foundation
through its New Frontiers in Astronomy and Cosmology, administered by
Don York of the University of Chicago.  We are grateful for the
opportunity provided by this grant to perform this research. 

We thank Jason Young and Sharon X.\ Wang for discussions on the
typical sizes and surface brightnesses of galaxies, especially LSBs.
We thank Caryl Gronwall, and Lea Hagan for their assistance with NED.
We thank Tom Jarrett for sharing a preliminary version of his extended
source catalog.  We thank the \wise\ team for making the \wise\ All-sky survey, and especially Davy Kirkpatrick for useful
discussions and assistance navigating the \wise\ data products.
We thank Richard Wade for illuminating the nature of 48 Librae, and
Leisa Townsley, Stan Owocki, Thomas Rivinius, Howard Bond, and Eric Fiegelson for
discussions about the nature of dust around Be stars.  

We thank Eric Mamajek for his inordinate efforts in hunting down
photometry and kinematics for our strange no-optical-counterpart
cluster and 48 Librae.

This publication makes use of data products from the Wide-field
Infrared Survey Explorer, which is a joint project of the University
of California, Los Angeles, and the Jet Propulsion
Laboratory (JPL)/California Institute of Technology (Caltech), funded by the National
Aeronautics and Space Administration (NASA).  This work is based in part on observations made with the Spitzer
Space Telescope, which is operated by JPL/Caltech under a contract with NASA. This research has made use of the NASA/IPAC Extragalactic Database (NED) which is operated byJPL/Caltech, under contract with NASA.

The Digitized Sky Surveys were produced at the Space Telescope Science
Institute under U.S. Government grant NAG W-2166. The images of these
surveys are based on photographic data obtained using the Oschin
Schmidt Telescope on Palomar Mountain and the UK Schmidt
Telescope. The plates were processed into the present compressed
digital form with the permission of these institutions.  The Second
Palomar Observatory Sky Survey (POSS-II) was made by the California
Institute of Technology with funds from the National Science
Foundation, the National Geographic Society, the Sloan Foundation, the
Samuel Oschin Foundation, and the Eastman Kodak Corporation. 

Funding for the Sloan Digital Sky Survey (SDSS) has been provided by
the Alfred P. Sloan Foundation, the Participating Institutions, NASA, the National Science
Foundation, the U.S. Department of Energy, the Japanese
Monbukagakusho, and the Max Planck Society. The SDSS Web site is
\url{http://www.sdss.org/}.

The SDSS is managed by the Astrophysical Research Consortium (ARC) for
the Participating Institutions. The Participating Institutions are The
University of Chicago, Fermilab, the Institute for Advanced Study, the
Japan Participation Group, The Johns Hopkins University, Los Alamos
National Laboratory, the Max-Planck-Institute for Astronomy (MPIA),
the Max-Planck-Institute for Astrophysics (MPA), New Mexico State
University, University of Pittsburgh, Princeton University, the United
States Naval Observatory, and the University of Washington.

Based on observations made with the NASA Galaxy Evolution
Explorer. GALEX is operated for NASA by the California Institute
of Technology under NASA contract NAS5-98034.

The Center for Exoplanets and Habitable Worlds is supported by the Pennsylvania State University, the Eberly College of Science, and the Pennsylvania Space Grant Consortium.

\appendix

\section{\ion{H}{1} dark Galaxies}

So-called ``dark galaxies'' are galaxies with no detectable stellar
component (i.e.\ composed entirely of dark matter, and perhaps gas).
The HIPASS \citep[e.g.][]{NOIRCAT} and ALFALFA \citep{ALFALFA} 
surveys of \ion{H}{1} are sensitive to neutral hydrogen in galaxies,
and those detections with little no detectable optical emission are
candidate ``dark galaxies'' or ``almost dark galaxies''
\citep{AlmostDark}.    

Such galaxies are consistent with $\alpha\sim 1$ --- that is,
galaxies in which all or nearly all of the starlight is absorbed to
power a galaxy-wide civilization, or K3.   If $\gamma \sim \alpha$ (i.e.\
the energy were all radiated as waste heat) and \twaste$>\gtrsim 150$
K, such a galaxy would be very bright in the MIR, and the \^G survey
would find them easily.  

As a check, and to ensure that we include even those known \ion{H}{1}
galaxies that might not be resolved by \wise, we have examined the
ALFALFA and NOIRCAT (Northern HIPASS Optical/IR Catalog) survey
galaxies using \wise\ imaging.

To within 5$^{\prime\prime}$, only 67 of the 730 sources in the ALFALFA
sample and 78 of the 1002 sources from the NOIRCAT (northern HIPASS)
sample matched to a \wise\ source.  We constructed ``at-a-glance''
charts for all 1732 sources from the two surveys.  In no case
is there a bright MIR source coincident with the \ion{H}{1} and little
or no optical emission.\footnote{We point out that
at least one object in Platinum
sample (entry 50 in Table~\ref{extreme}, HIPASS J1403-50) is a HIPASS
galaxy, and was given a grade of B. This object is not, however,
optically faint.}  It is safe to say that these dark galaxies
are probably not examples of K3s.   

We note that a similar analysis was performed by
\citet{DarkGalaxyWISE}, though probably for a different purpose.

\section{Passive Spirals}

One possible application of energy for a galaxy-spanning civilization
might be to alter the underlying stellar population in some way, for
instance suppressing high-mass star formation to prevent supernova
explosions.  Such a galaxy might appear morphologically similar to a
spiral, but lack supernova progenitors and other signatures
of high mass star formation.  This is consistent with ``passive
spirals'' or ``red spirals'' \citep[see][and references
therein]{Masters10}, and so we explore here if such galaxies are
anomalously bright in the MIR.

\citet{Cortese12} argued that optically selected ``red spirals'' are
not, in fact, passive, but simply have old populations in addition to
active current star formation.  That is, in very high mass systems ($> 10^{10}
$\Msun) optical colors are more sensitive to star formation history
than to the instantaneous star formation rate.  \citeauthor{Cortese12}
argued that $NUV - r$ colors were a better proxy for
activity/quiescence than optical colors, and showed that ``red
spirals'' typically have $(NUV - r) < 4$, consistent with star
formation rates typical of ordinary spirals.  \citeauthor{Cortese12}
selected only face-on spirals to minimize the effects of extinction in
the $NUV$ (which would ruin this band's diagnostic power).

It follows, then that a ``red spiral'' with an anomalously red
$(NUV-r)$ color might be a truly passive spiral, which would make a
high MIR flux anomalous.  

We investigated the sample of red spiral galaxies presented in
\citet{Masters10}.  We matched their list of 5433 sources to \wise\
using a 3$^{\prime\prime}$ search radius and recovered \wise\ photometry for $\sim$ 99.5\% of the  
sample.  We recover \wise\ photometry for 27 more sources by opening up
the search radius to 20$^{\prime\prime}$. For our NUV photometry we matched
this list to GALEX and recovered photometry for $\sim$ 88\% of the
sources.   All of these Galaxy Zoo galaxies have good $r$-band
photometry from the Sloan Digital Sky Survey.

\begin{figure*}[htp]
\begin{center}
\begin{tabular}{l}
\includegraphics[width=7in,trim=0 3in 0 2in,clip]{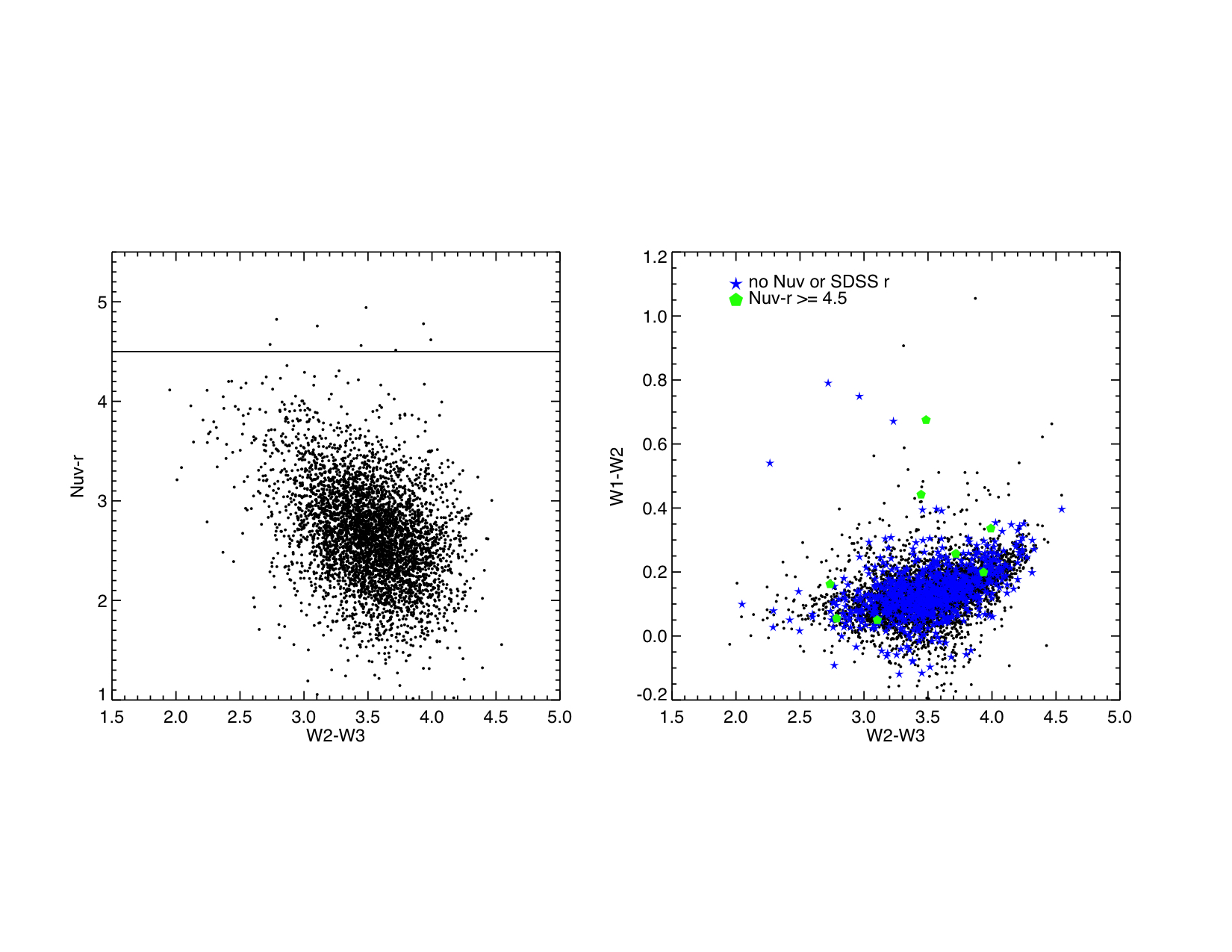}
\end{tabular}
\end{center}
\caption[CAPTION]{\label{Galex} (Left) We plot GALEX NUV-r versus W2-W3 for a large sample of red spiral galaxies. The solid horizontal line represents NUV-r $\ge 4.5$. (Right) We plot W1-W2 versus W2-W3 
for approximately 100\%  of the red spiral galaxy sample. Green points indicate the extreme NUV-r sources, while the blue points represent sources which could not be included in the left hand plot due to the lack of photometric measurement.\label{galex} }
\end{figure*}

The left panel of Figure~\ref{galex} shows $NUV-r$ colors for our sample.  In
agreement with \citeauthor{Cortese12} we find that most red spirals
have $(NUV-r) < 4$, with some scattering to higher values (consistent
with some having extreme extinction).  Eight galaxies in our sample
have $(NUV-r) > 4.5$ which would put them firmly in the ``red sequence'' of passive
galaxies, if they were unextinguished.  

The right panel of Figure~\ref{galex} shows a \wise\ color-color plot of the
entire sample (irrespective of the availability of GALEX NUV data),
with the anomalously red galaxies highlighted.  Five of these
sources have moderately red $(W2-W3)$ colors, consistent with high
rates of star formation.  Either these five galaxies suffer from
significant extinction, or they are examples of anomalous MIR-bright, passive
spiral galaxies, perhaps consistent with a K3.  Visual examination
reveals that none are edge-on; all are at modest inclination.

We list these five potentially anomalous sources with their AGENT
parameters in Table~\ref{tab:galex}.

\begin{deluxetable*}{rrrrl}
\tablewidth{0pt}
\tablecaption{Five spirals red in optical, $NUV-r$, and MIR colors \label{tab:galex}}
\tablehead{\colhead{\wise\ designation} & $(NUV-r)$ & \twaste &$\gamma$ &SIMBAD name}
\startdata
% J082006.62+550920.9   &4.57&  196 &0.071 & SDSS J082006.62+550920.9 \\
% J082959.07+304340.2   &4.82&  250 &0.055 & 2MASX J08295908+3043398  \\
 J085428.94+084444.4   &4.62&  179 &0.258 & 2MASX J08542894+08444430 \\
% J100445.51+530650.7   &4.76&  248 &0.075 & SDSS J100445.48+530650.5 \\
 J113325.93+141618.9   &4.94&  180 &0.198 & 2MASX J11332590+1416186  \\ 
 J142619.17+473357.8   &4.51&  180 &0.202 & SDSS J142619.17+473357.7 \\
 J142859.55+605000.5   &4.56&  163 &0.212 & 2MASX J14285959+6050005  \\
 J230616.43+135856.4   &4.78&  239 &0.172 & 2MASXJ23061640+1358560   
\enddata   
\end{deluxetable*}

     % ra                     dec            WISE Designation  SIMBAD Designation
     %   125.02760       55.155832  
     %   127.49615       30.727836  
     %   133.62060       8.7456941  
     %   151.18965       53.114101  
     %   173.35805       14.271943  
     %   216.57990       47.566066  
     %   217.24816       60.833497  
     %   346.56847       13.982355  

% \begin{figure*}[htp]
% \begin{center}
% \begin{tabular}{l}
% \includegraphics[scale=0.71,angle=90]{WISE-2MASS_paper_plots.ps}
% \end{tabular}
% \end{center}
% \caption[CAPTION]{\label{fig:7} We show W1 versus W1-W2 (Left) and W1 versus W1RCHI2 (Right) for sources in the Control field which satisfied our search criteria, simply W1MPRO $>$ 8.1 and W1RCHI2 $\ge$ 3.0. The black points 
% represent all sources in the field, while the sources matched to the 2MASS XSC are presented as red triangles. We use the left hand plot to derive a more robust set of selection criteria to get a more accurate measurement of the reliability. 
% The additional selection criteria are shown by the blue solid lines.}
% \end{figure*}

% \begin{figure*}[htp]
% \begin{center}
% \begin{tabular}{l}
% \includegraphics[scale=0.75]{Mock_paper_plots_v2.ps}
% \end{tabular}
% \end{center}
% \vspace{-1.2in}
% \caption[CAPTION]{\label{Mock2} We plot our predicted mock catalog \wise\ colors, W1-W2 (Top) and W2-W3 (Bottom) versus $T_\text{waste}$ (Left) and $\gamma$ (Right). The red points in the left panel are
% for sources with $\gamma \ge 0.25$ while the red points on the right are for those with $250 \le T_\text{waste} \le 350$ K. }
% \end{figure*}

\clearpage
\LongTables
%\begin{landscape}
\begin{deluxetable}{lccccccccccccl}
\tabletypesize{\scriptsize}
\tablecolumns{12} 
\tablewidth{0pt}
\tablecaption{Extreme WISE Colors \label{reddest}}
\tablehead{
\colhead{RA} &
\colhead{DEC} &
\colhead{SIMBAD} &
\colhead{Grade} &
\colhead{W1} &
\colhead{W2} &
\colhead{W3} &
\colhead{W4} &
\colhead{Color} &
\colhead{$\gamma$} &
\colhead{$T_{\rm waste}$} &
\colhead{Photometry} &
\colhead{Note}\\
\colhead{hhmmss.ss}&
\colhead{ddmmss.s}&
\colhead{type}& &
\colhead{mag.}&
\colhead{mag.}&
\colhead{mag.}&
\colhead{mag.}&
\colhead{mag.}& &
\colhead{K}& &}
\startdata
\cutinhead{W1-W2 Most Extreme Sources}
072533.95&+292919.9 &PN     &C &     11.20 &      8.89 &      4.33 &      0.44 &      2.31 &      0.83 &         123 &           2 &NGC 2371                              \\
213652.95&+124719.2 &PN     &C &     11.68 &      9.49 &      5.25 &      0.53 &      2.19 &      0.92 &          98 &           2 &NGC 7094                              \\
180600.64&+002235.5 &PN     &C &     15.12 &     13.01 &      8.52 &      3.56 &      2.11 &      0.90 &         248 &           1 &PN-G028.0+10.2                       \\
054206.16&+090511.6 &PN     &C &     10.06 &      7.98 &      4.07 &     -0.33 &      2.09 &      0.83 &         106 &           2 &NGC 2022                             \\
060253.97&-710310.0 &LIN    &C &     11.59 &      9.52 &      6.14 &      3.25 &      2.07 &      0.66 &         349 &           0 &2MASX J06025406-7103104              \\
122430.65&-184700.7 &PN     &C &      9.61 &      7.66 &      4.33 &      0.00 &      1.95 &      0.71 &         108 &           2 &NGC 4361                             \\
214035.92&+663525.8 &Y*O    &B &     10.05 &      8.25 &      5.58 &      3.05 &      1.80 &      0.45 &         403 &           1 &2MASS J21403852+6635017               \\
194357.75&-140913.5 &PN     &C &      8.85 &      7.09 &      3.14 &     -0.88 &      1.75 &      0.72 &         118 &           2 &NGC 6818                              \\
193121.41&-723921.3 &Sy2    &C &     10.29 &      8.54 &      5.55 &      2.06 &      1.75 &      0.36 &         137 &           2 &Antennae                    \\
003443.48&-000226.6 &Sy2    &C &     11.72 &      9.98 &      6.57 &      3.84 &      1.74 &      0.54 &         347 &           2 &2MFGC-403                            \\
\cutinhead{W2-W3 Most Extreme Sources}
091742.98&+131214.8 &Comet      &C &     12.76 &     11.50 &      5.02 &      0.99 &      6.49 &      0.93 &         204&           2 &29P/Schwassmann-Wachmann 1 1927 V1   \\
155812.25&-141806.2 &IR     &B &     16.48 &     15.39 &      9.13 &      4.03 &      6.26 &      0.99 &          89&           0 &IRAS-15553-1409                      \\
235821.58&-320748.6 &Comet      &C &     16.56 &     15.71 &      9.92 &      4.97 &      5.79 &      0.97 &          93&           1 &Garradd 2009 P1  \\
045608.93&-674947.5 &       &C &     14.39 &     14.51 &      8.79 &      6.52 &      5.71 &      0.51 &         211&           1 &   Object in  LMC    \\
122654.61&-005239.1 &Sy2    &C &     10.04 &      9.15 &      3.81 &     -0.15 &      5.34 &      0.85 &         120&           2 &NGC 4355                              \\
052703.47&-664959.2 &       &C &     15.10 &     14.63 &      9.32 &      7.90 &      5.31 &      0.45 &         282&           0 &   Object in LMC                     \\
100918.85&-805128.1 &PN     &C &     10.81 &      9.80 &      4.61 &      2.88 &      5.19 &      0.59 &         290&           2 &NGC 3195                             \\
055542.61&+032331.8 &H2G    &C &     10.99 &     10.12 &      4.94 &      1.52 &      5.18 &      0.72 &         142&           2 &UGCA 116                             \\
061452.97&-062244.4 &Y*O    &B &      8.80 &      8.31 &      3.15 &      1.78 &      5.16 &      0.42 &         289&           2 &IRAS 06124-0621                       \\
004036.06&+410117.9 &HII    &C &     14.43 &     13.76 &      8.62 &      6.45 &      5.13 &      0.52 &         263&           0 &BA1-441                              \\
\cutinhead{W1-W3 Most Extreme Sources}
091742.98&+131214.8 &Comet      &C &     12.76 &     11.50 &      5.02 &      0.99 &      7.74 &      0.93 &         204 &           2 &29P/Schwassmann-Wachmann 1 1927 V1   \\
155812.25&-141806.2 &IR     &B &     16.48 &     15.39 &      9.13 &      4.03 &      7.34 &      0.99 &          89 &           0 &IRAS-15553-1409                      \\
072533.95&+292919.9 &PN     &C &     11.20 &      8.89 &      4.33 &      0.44 &      6.87 &      0.83 &         123 &           2 &NGC 2371                              \\
235821.58&-320748.6 &Comet      &C &     16.56 &     15.71 &      9.92 &      4.97 &      6.64 &      0.97 &          93 &           1 &Garradd 2009 P1  \\
180600.64&+002235.5 &PN     &C &     15.12 &     13.01 &      8.52 &      3.56 &      6.60 &      0.90 &         248 &           1 &PN-G028.0+10.2                       \\
213652.95&+124719.2 &PN     &C &     11.68 &      9.49 &      5.25 &      0.53 &      6.43 &      0.92 &          98 &           2 &NGC 7094                              \\
190101.24&-181215.0 &PN     &C &     15.50 &     13.89 &      9.08 &      3.07 &      6.42 &      0.99 &          73 &           1 &PN A6651                              \\
185334.98&+330152.2 &PN     &C &     11.79 &     10.33 &      5.48 &      0.85 &      6.30 &      0.93 &         100 &           1 &Ring Nebula                          \\
122654.61&-005239.1 &Sy2    &C &     10.04 &      9.15 &      3.81 &     -0.15 &      6.23 &      0.85 &         120 &           2 &NGC 4355                              \\
041415.79&-124420.7 &PN     &C &      9.12 &      7.98 &      2.91 &     -0.05 &      6.21 &      0.69 &         258 &           2 &NGC 1535                             \\
\cutinhead{W3-W4 Most Extreme Sources}
190101.24&-181215.0 &PN     &C &     15.50 &     13.89 &      9.08 &      3.07 &      6.01 &      0.99 &          73 &           1 &PNA6651                              \\
095551.65&+694046.6 &Em*    &C &      6.38 &      5.40 &      3.73 &     -1.79 &      5.53 &      0.75 &          77 &           1 &[DOB2000]-19                         \\
032910.31&+312158.0 &Y*O    &C &      6.76 &      5.85 &      3.62 &     -1.63 &      5.25 &      0.73 &          83 &           1 &[LAL96]-213                          \\
155812.25&-141806.2 &IR     &B &     16.48 &     15.39 &      9.13 &      4.03 &      5.10 &      0.99 &          89 &           0 &IRAS-15553-1409                      \\
215935.03&-392308.7 &PN     &C &     16.42 &     15.54 &     10.75 &      5.70 &      5.05 &      0.95 &          90 &           1 &IC 5148                               \\
180600.64&+002235.5 &PN     &C &     15.12 &     13.01 &      8.52 &      3.56 &      4.96 &      0.90 &         248 &           1 &PN-G028.0+10.2                       \\
235821.58&-320748.6 &Comet      &C &     16.56 &     15.71 &      9.92 &      4.97 &      4.94 &      0.97 &          93 &           1 &Garradd 2009 P1  \\
071448.80&-465733.6 &PN     &C &     11.68 &     10.09 &      6.39 &      1.59 &      4.79 &      0.86 &          95 &           2 &ESO 256-19                            \\
213652.95&+124719.2 &PN     &C &     11.68 &      9.49 &      5.25 &      0.53 &      4.72 &      0.92 &          98 &           2 &NGC 7094                              \\
160426.50&+404059.1 &PN     &C &     12.54 &     12.04 &      7.48 &      2.78 &      4.70 &      0.87 &          98 &           2 &NGC 6058                             \\
\cutinhead{W1-W4 Most Extreme Sources}
155812.25&-141806.2 &IR     &B &     16.48 &     15.39 &      9.13 &      4.03 &     12.45 &      0.99 &          89 &           0 &IRAS-15553-1409                      \\
190101.24&-181215.0 &PN     &C &     15.50 &     13.89 &      9.08 &      3.07 &     12.43 &      0.99 &          73 &           1 &PN A6651                              \\
091742.98&+131214.8 &Comet      &C &     12.76 &     11.50 &      5.02 &      0.99 &     11.77 &      0.93 &         204 &           2 &29P/Schwassmann-Wachmann 1 1927 V1   \\
235821.58&-320748.6 &Comet      &C &     16.56 &     15.71 &      9.92 &      4.97 &     11.58 &      0.97 &          93 &           1 &Garradd 2009 P1  \\
180600.64&+002235.5 &PN     &C &     15.12 &     13.01 &      8.52 &      3.56 &     11.56 &      0.90 &         248 &           1 &PN-G028.0+10.2                       \\
213652.95&+124719.2 &PN     &C &     11.68 &      9.49 &      5.25 &      0.53 &     11.15 &      0.92 &          98 &           2 &NGC 7094                              \\
185334.98&+330152.2 &PN     &C &     11.79 &     10.33 &      5.48 &      0.85 &     10.94 &      0.93 &         100 &           1 &Ring Nebula                          \\
072533.95&+292919.9 &PN     &C &     11.20 &      8.89 &      4.33 &      0.44 &     10.76 &      0.83 &         123 &           2 &NGC 2371                              \\
215935.03&-392308.7 &PN     &C &     16.42 &     15.54 &     10.75 &      5.70 &     10.71 &      0.95 &          90 &           1 &IC 5148                               \\
160628.19&-354520.9 &PN     &C &     10.61 &      9.25 &      4.52 &      0.08 &     10.53 &      0.89 &         105 &           2 &PN-G341.5+12.1                       \\
\cutinhead{W2-W4 Most Extreme Sources}
155812.25&-141806.2 &IR     &B &     16.48 &     15.39 &      9.13 &      4.03 &     11.36 &      0.99 &          89 &           0 &IRAS-15553-1409                      \\
190101.24&-181215.0 &PN     &C &     15.50 &     13.89 &      9.08 &      3.07 &     10.82 &      0.99 &          73 &           1 &PN A6651                              \\
235821.58&-320748.6 &Comet      &C &     16.56 &     15.71 &      9.92 &      4.97 &     10.73 &      0.97 &          93 &           1 &Garradd 2009 P1  \\
091742.98&+131214.8 &Comet      &C &     12.76 &     11.50 &      5.02 &      0.99 &     10.52 &      0.93 &         204 &           2 &29P/Schwassmann-Wachmann 1 1927 V1   \\
215935.03&-392308.7 &PN     &C &     16.42 &     15.54 &     10.75 &      5.70 &      9.83 &      0.95 &          90 &           1 &IC 5148                               \\
185334.98&+330152.2 &PN     &C &     11.79 &     10.33 &      5.48 &      0.85 &      9.48 &      0.93 &         100 &           1 &Ring Nebula                          \\
180600.64&+002235.5 &PN     &C &     15.12 &     13.01 &      8.52 &      3.56 &      9.45 &      0.90 &         248 &           1 &PN-G028.0+10.2                       \\
122654.61&-005239.1 &Sy2    &C &     10.04 &      9.15 &      3.81 &     -0.15 &      9.30 &      0.85 &         120 &           2 &NGC 4355                              \\
160426.50&+404059.1 &PN     &C &     12.54 &     12.04 &      7.48 &      2.78 &      9.25 &      0.87 &          98 &           2 &NGC 6058                             \\
160628.19&-354520.9 &PN     &C &     10.61 &      9.25 &      4.52 &      0.08 &      9.17 &      0.89 &         105 &           2 &PN-G341.5+12.1                       \\
\enddata

\end{deluxetable}

\clearpage
%\end{landscape}

\clearpage
\LongTables
%\begin{landscape}
\begin{deluxetable}{lcccccccccccc}
\tabletypesize{\scriptsize}
\tablecolumns{12} 
\tablewidth{0pt}
\tablecaption{Galaxies With Extreme \wise\ Colors \label{redgal}}
\tablehead{
\colhead{RA} &
\colhead{DEC} &
\colhead{SIMBAD} &
\colhead{Grade\tablenotemark{1}} &
\colhead{W1} &
\colhead{W2} &
\colhead{W3} &
\colhead{W4} &
\colhead{Color} &
\colhead{$\gamma$} &
\colhead{$T_{\rm waste}$} &
\colhead{Photometry\tablenotemark{2}} &
\colhead{Note} \\
\colhead{hhmmss.ss}&\colhead{ddmmss.s}&\colhead{type}&&\colhead{mag.}&\colhead{mag.}&\colhead{mag.}&\colhead{mag.}&\colhead{mag.}&&\colhead{K}&&}
\startdata
\cutinhead{W1-W2 Most Extreme Galaxies}
060253.97&-710310.0 &LIN    &C &     11.59 &      9.52 &      6.14 &      3.25 &      2.07 &      0.66 &         349 &           0 &2MASX J06025406-7103104              \\
193121.41&-723921.3 &Sy2    &C &     10.29 &      8.54 &      5.55 &      2.06 &      1.75 &      0.36 &         137 &           2 &SUPER ANTENNAE                    \\
003443.48&-000226.6 &Sy2    &C &     11.72 &      9.98 &      6.57 &      3.84 &      1.74 &      0.54 &         347 &           2 &2MFGC-403                            \\
080106.63&-660914.9 &G      &C &     10.20 &      8.51 &      5.24 &      2.17 &      1.69 &      0.52 &         338 &           2 &6dFGSg J080106.6-660915               \\
002153.60&-791007.8 &Sy2    &C &     10.06 &      8.39 &      5.22 &      2.14 &      1.67 &      0.49 &         341 &           2 &2MASX J00215355-7910077              \\
100125.94&+154612.2 &G      &C &      7.99 &      6.33 &      4.06 &      1.14 &      1.65 &      0.34 &         401 &           2 &NGC 3094                             \\
033639.05&-205406.8 &GiG    &C &      9.58 &      7.98 &      4.69 &      1.50 &      1.60 &      0.50 &         328 &           2 &NGC 1377                             \\
183820.32&-652539.0 &Sy2    &C &      9.37 &      7.80 &      4.32 &      1.21 &      1.57 &      0.52 &         322 &           2 &ESO 103-35                           \\
173801.51&+561325.9 &Sy2    &C &     11.08 &      9.57 &      6.19 &      3.51 &      1.51 &      0.45 &         343 &           2 &2MASX J17380143+5613257              \\
112402.72&-282315.4 &Sy2    &C &     10.81 &      9.31 &      5.98 &      3.60 &      1.50 &      0.43 &         359 &           2 &IRAS 11215-2806                       \\
\cutinhead{W2-W3 Most Extreme Galaxies}
122654.61&-005239.1 &Sy2    &C &     10.04 &      9.15 &      3.81 &     -0.15 &      5.34 &      0.85 &         120&           2 &NGC 4355                              \\
055542.61&+032331.8 &H2G    &C &     10.99 &     10.12 &      4.94 &      1.52 &      5.18 &      0.72 &         142&           2 &UGCA 116                             \\
042826.03&-043349.2 &LIN    &C &     11.62 &     10.99 &      5.92 &      2.00 &      5.06 &      0.79 &         121&           2 &IRAS 04259-0440                      \\
103833.62&-071014.4 &G      &C &      9.49 &      9.10 &      4.14 &      0.80 &      4.97 &      0.61 &         146&           1 &IC 630                               \\
001850.88&-102236.6 &EmG    &C &     10.76 &     10.29 &      5.41 &      2.37 &      4.89 &      0.52 &         162&           0 &MCG 02-01-051                        \\
181338.76&-574356.9 &G      &C &     11.26 &     10.84 &      5.99 &      2.57 &      4.85 &      0.60 &         142&           0 &IC 4686                              \\
225234.71&+244349.4 &AGN    &C &     11.44 &     10.87 &      6.06 &      2.67 &      4.81 &      0.60 &         143&           2 &Mrk 309                              \\
010426.95&-640712.2 &H2G    &C &     11.23 &     10.81 &      6.03 &      2.78 &      4.78 &      0.54 &         150&           0 &ESO 79-16                            \\
195405.20&+495647.0 &G      &B &     12.57 &     12.16 &      7.39 &      4.81 &      4.76 &      0.39 &         195&           0 &LEDA 200363                          \\
043400.03&-083444.9 &AGN    &C &      8.74 &      8.21 &      3.54 &     -0.07 &      4.67 &      0.63 &         133&           2 &NGC 1614                             \\
\cutinhead{W1-W3 Most Extreme Galaxies}
122654.61&-005239.1 &Sy2    &C &     10.04 &      9.15 &      3.81 &     -0.15 &      6.23 &      0.85 &         120 &           2 &NGC 4355                              \\
055542.61&+032331.8 &H2G    &C &     10.99 &     10.12 &      4.94 &      1.52 &      6.06 &      0.72 &         142 &           2 &UGCA 116                             \\
205826.80&-423900.3 &LIN    &C &     11.08 &      9.69 &      5.33 &      1.75 &      5.75 &      0.65 &         135 &           2 &ESO 286-19                           \\
042826.03&-043349.2 &LIN    &C &     11.62 &     10.99 &      5.92 &      2.00 &      5.69 &      0.79 &         121 &           2 &IRAS 04259-0440                      \\
003652.44&-333316.8 &AGN    &C &     10.52 &      9.09 &      4.84 &      1.17 &      5.69 &      0.66 &         131 &           2 &ESO 350-38                           \\
124930.16&-112403.4 &Sy2    &C &     11.46 &     10.10 &      5.90 &      2.79 &      5.55 &      0.59 &         286 &           1 &IRAS 12468-1107                       \\
060253.97&-710310.0 &LIN    &C &     11.59 &      9.52 &      6.14 &      3.25 &      5.45 &      0.66 &         349 &           0 &2MASX J06025406-7103104              \\
012002.63&+142142.5 &LIN    &C &     10.84 &      9.67 &      5.41 &      2.24 &      5.42 &      0.49 &         154 &           0 &MCG+02-04-025                        \\
011607.20&+330521.7 &Sy2    &C &     11.09 &      9.78 &      5.70 &      2.57 &      5.38 &      0.54 &         286 &           0 &NGC 449                               \\
225234.71&+244349.4 &AGN    &C &     11.44 &     10.87 &      6.06 &      2.67 &      5.38 &      0.60 &         143 &           2 &Mrk 309                              \\
\cutinhead{W3-W4 Most Extreme Galaxies}
131503.51&+243707.8 &Q?     &C &     10.44 &     10.32 &      7.05 &      2.45 &      4.60 &      0.61 &          99 &           2 &IC 860                               \\
153457.25&+233011.5 &SyG    &C &      9.52 &      8.96 &      4.55 &      0.19 &      4.35 &      0.79 &         107 &           2 &Arp 220                               \\
125145.54&+254628.5 &IG     &C &      9.74 &      9.58 &      6.24 &      1.92 &      4.32 &      0.53 &         107 &           2 &NGC 4747                             \\
000820.57&+403755.9 &Sy2    &C &     12.41 &     12.27 &      8.58 &      4.44 &      4.14 &      0.54 &         113 &           3 &2MASX J00082041+4037560              \\
042759.96&-475445.8 &LIN    &C &      9.52 &      9.44 &      6.79 &      2.81 &      3.98 &      0.27 &         115 &           2 &CARAFE NEBULA                   \\
122654.61&-005239.1 &Sy2    &C &     10.04 &      9.15 &      3.81 &     -0.15 &      3.96 &      0.85 &         120 &           2 &NGC 4355                              \\
065558.96&-404912.2 &G      &B &     11.12 &     10.72 &      7.34 &      3.41 &      3.93 &      0.43 &         119 &           2 &6dFGSg J065559.0-404912              \\
042826.03&-043349.2 &LIN    &C &     11.62 &     10.99 &      5.92 &      2.00 &      3.92 &      0.79 &         121 &           2 &IRAS 04259-0440                      \\
102508.18&+170914.1 &IG     &C &     12.56 &     12.15 &      8.06 &      4.15 &      3.92 &      0.58 &         121 &           0 &NGC 3239                              \\
202825.49&-330420.5 &G      &B &     10.33 &     10.32 &      7.15 &      3.29 &      3.86 &      0.32 &         121 &           2 &ESO 400-28                           \\
\cutinhead{W1-W4 Most Extreme Galaxies}
122654.61&-005239.1 &Sy2    &C &     10.04 &      9.15 &      3.81 &     -0.15 &     10.19 &      0.85 &         120 &           2 &NGC 4355                              \\
042826.03&-043349.2 &LIN    &C &     11.62 &     10.99 &      5.92 &      2.00 &      9.61 &      0.79 &         121 &           2 &IRAS 04259-0440                      \\
055542.61&+032331.8 &H2G    &C &     10.99 &     10.12 &      4.94 &      1.52 &      9.47 &      0.72 &         142 &           2 &UGCA 116                             \\
003652.44&-333316.8 &AGN    &C &     10.52 &      9.09 &      4.84 &      1.17 &      9.35 &      0.66 &         131 &           2 &ESO 350-38                           \\
153457.25&+233011.5 &SyG    &C &      9.52 &      8.96 &      4.55 &      0.19 &      9.33 &      0.79 &         107 &           2 &Arp 220                               \\
205826.80&-423900.3 &LIN    &C &     11.08 &      9.69 &      5.33 &      1.75 &      9.33 &      0.65 &         135 &           2 &ESO 286-19                           \\
205724.32&+170738.5 &G      &C &     10.57 &      9.88 &      5.27 &      1.43 &      9.15 &      0.70 &         124 &           2 &IRAS F20550+1655-SE             \\
133955.96&-313824.4 &AGN    &C &      8.27 &      7.00 &      2.95 &     -0.73 &      9.01 &      0.60 &         130 &           2 &NGC 5253                              \\
150029.00&-262649.2 &H2G    &C &     11.96 &     11.36 &      6.82 &      3.08 &      8.88 &      0.65 &         128 &           2 &2MASX J15002897-2626487              \\
231546.75&-590314.5 &Sy2    &C &     10.69 &      9.67 &      5.34 &      1.87 &      8.83 &      0.57 &         139 &           2 &ESO 148-2                            \\
\cutinhead{W2-W4 Most Extreme Galaxies}
122654.61&-005239.1 &Sy2    &C &     10.04 &      9.15 &      3.81 &     -0.15 &      9.30 &      0.85 &         120 &           2 &NGC 4355                              \\
042826.03&-043349.2 &LIN    &C &     11.62 &     10.99 &      5.92 &      2.00 &      8.98 &      0.79 &         121 &           2 &IRAS 04259-0440                      \\
153457.25&+233011.5 &SyG    &C &      9.52 &      8.96 &      4.55 &      0.19 &      8.77 &      0.79 &         107 &           2 &Arp 220                               \\
055542.61&+032331.8 &H2G    &C &     10.99 &     10.12 &      4.94 &      1.52 &      8.60 &      0.72 &         142 &           2 &UGCA 116                             \\
205724.32&+170738.5 &G      &C &     10.57 &      9.88 &      5.27 &      1.43 &      8.45 &      0.70 &         124 &           2 &IRAS F20550+1655-SE             \\
234709.20&+153548.3 &Sy2    &C &     11.02 &     10.56 &      5.91 &      2.20 &      8.36 &      0.65 &         129 &           2 &MCG+02-60-017                        \\
103833.62&-071014.4 &G      &C &      9.49 &      9.10 &      4.14 &      0.80 &      8.31 &      0.61 &         146 &           1 &IC 630                               \\
150029.00&-262649.2 &H2G    &C &     11.96 &     11.36 &      6.82 &      3.08 &      8.29 &      0.65 &         128 &           2 &2MASX J15002897-2626487              \\
043400.03&-083444.9 &AGN    &C &      8.74 &      8.21 &      3.54 &     -0.07 &      8.28 &      0.63 &         133 &           2 &NGC 1614                             \\
181338.76&-574356.9 &G      &C &     11.26 &     10.84 &      5.99 &      2.57 &      8.27 &      0.60 &         142 &           0 &IC 4686                              \\
\enddata

\tablenotetext{1}{{\it Grade} refers to how well understood an object 
  is (see Section~\ref{grading}.)  All of these objects have grade C indicating that they are understood, having been discussed in the refereed literature.}
\tablenotetext{2}{{\it Photometry} refers to the source of the
  photometry used in the table: 0:Profile fit; 1:Aperture-corrected; 2:Calibrated, 3:Elliptical aperture. See Section~\ref{photometry}}

\end{deluxetable}

\clearpage
%\end{landscape}

\clearpage
\LongTables
%\begin{landscape}
\begin{deluxetable}{lcccccccccccc}
\tabletypesize{\scriptsize}
\tablecolumns{11} 
\tablewidth{0pt}
\tablecaption{Galaxies With Extreme \wise\ $\gamma$ Values \label{highgamma}}
\tablehead{
\colhead{RA} &
\colhead{DEC} &
\colhead{SIMBAD} &
\colhead{Grade\tablenotemark{1}} &
\colhead{W1} &
\colhead{W2} &
\colhead{W3} &
\colhead{W4} &
\colhead{$\gamma$} &
\colhead{$T_{\rm waste}$} &
\colhead{Photometry\tablenotemark{2}} &
\colhead{Note} \\
\colhead{hhmmss.ss}&\colhead{ddmmss.s}&\colhead{type}&&\colhead{mag.}&\colhead{mag.}&\colhead{mag.}&\colhead{mag.}&&\colhead{K}&&}
\startdata
122654.61&-005239.1 &Sy2    &C &     10.04 &      9.15 &      3.81 &     -0.15 &      0.85&         120 &           2 &NGC 4355                              \\
042826.03&-043349.2 &LIN    &C &     11.62 &     10.99 &      5.92 &      2.00 &      0.79&         121 &           2 &IRAS 04259-0440                      \\
153457.25&+233011.5 &SyG    &C &      9.52 &      8.96 &      4.55 &      0.19 &      0.79&         107 &           2 & Arp 220 \\
055542.61&+032331.8 &H2G    &C &     10.99 &     10.12 &      4.94 &      1.52 &      0.72&         142 &           2 &UGCA 116                             \\
205724.32&+170738.5 &G      &C &     10.57 &      9.88 &      5.27 &      1.43 &      0.70&         124 &           2 &IRAS F20550+1655-SE             \\
060253.97&-710310.0 &LIN    &C &     11.59 &      9.52 &      6.14 &      3.25 &      0.66&         349 &           0 &2MASX J06025406-7103104              \\
003652.44&-333316.8 &AGN    &C &     10.52 &      9.09 &      4.84 &      1.17 &      0.66&         131 &           2 &ESO 350-38                           \\
150029.00&-262649.2 &H2G    &C &     11.96 &     11.36 &      6.82 &      3.08 &      0.65&         128 &           2 &2MASX J15002897-2626487              \\
234709.20&+153548.3 &Sy2    &C &     11.02 &     10.56 &      5.91 &      2.20 &      0.65&         129 &           2 &MCG +02-60-017                        \\
205826.80&-423900.3 &LIN    &C &     11.08 &      9.69 &      5.33 &      1.75 &      0.65&         135 &           2 &ESO 286-19                           \\
043400.03&-083444.9 &AGN    &C &      8.74 &      8.21 &      3.54 &     -0.07 &      0.63&         133 &           2 &NGC 1614                             \\
131503.51&+243707.8 &Q?     &C &     10.44 &     10.32 &      7.05 &      2.45 &      0.61&          99 &           2 &IC 860                               \\
103833.62&-071014.4 &G      &C &      9.49 &      9.10 &      4.14 &      0.80 &      0.61&         146 &           1 &IC 630                               \\
181338.76&-574356.9 &G      &C &     11.26 &     10.84 &      5.99 &      2.57 &      0.60&         142 &           0 &IC 4686                              \\
225234.71&+244349.4 &AGN    &C &     11.44 &     10.87 &      6.06 &      2.67 &      0.60&         143 &           2 &Mrk 309                              \\
151806.13&+424444.8 &LIN    &C &     10.69 &     10.05 &      5.60 &      1.96 &      0.60&         132 &           2 &IRAS F15163+4255-NW             \\
133955.96&-313824.4 &AGN    &C &      8.27 &      7.00 &      2.95 &     -0.73 &      0.60&         130 &           2 &NGC 5253                              \\
121539.36&+361935.1 &SBG    &C &     11.13 &     10.70 &      6.31 &      2.59 &      0.59&         128 &           1 &NGC 4228                              \\
124930.16&-112403.4 &Sy2    &C &     11.46 &     10.10 &      5.90 &      2.79 &      0.59&         286 &           1 &IRAS 12468-1107                       \\
102508.18&+170914.1 &IG     &C &     12.56 &     12.15 &      8.06 &      4.15 &      0.58&         121 &           0 &NGC 3239                              \\
231546.75&-590314.5 &Sy2    &C &     10.69 &      9.67 &      5.34 &      1.87 &      0.57&         139 &           2 &ESO 148-2                            \\
134442.10&+555313.3 &Sy2    &C &     10.22 &      9.00 &      5.36 &      1.51 &      0.55&         123 &           2 &Mrk 273                               \\
121346.00&+024840.3 &LIN    &C &     12.11 &     11.39 &      7.02 &      3.54 &      0.55&         139 &           2 &LEDA 39024                            \\
062722.52&-471046.7 &IG     &C &     11.52 &     11.06 &      6.66 &      3.11 &      0.55&         135 &           3 &ESO 255-7                            \\
025941.29&+251415.0 &G      &C &     11.89 &     11.46 &      6.99 &      3.49 &      0.54&         138 &           1 &NGC 1156                              \\
011607.20&+330521.7 &Sy2    &C &     11.09 &      9.78 &      5.70 &      2.57 &      0.54&         286 &           0 &NGC 449                               \\
010426.95&-640712.2 &H2G    &C &     11.23 &     10.81 &      6.03 &      2.78 &      0.54&         150 &           0 &ESO 79-16                            \\
003443.48&-000226.6 &Sy2    &C &     11.72 &      9.98 &      6.57 &      3.84 &      0.54&         347 &           2 &2MFGC-403                            \\
000820.57&+403755.9 &Sy2    &C &     12.41 &     12.27 &      8.58 &      4.44 &      0.54&         113 &           3 &2MASX J00082041+4037560              \\
021037.63&-154624.2 &G      &C &     10.95 &     10.49 &      6.01 &      2.54 &      0.54&         140 &           2 &NGC 814                              \\
005404.02&+730505.7 &G      &C &      9.31 &      8.75 &      4.23 &      0.84 &      0.54&         143 &           2 &MCG+12-02-001                        \\
130220.39&-154559.0 &GiG    &C &      9.86 &      9.43 &      5.14 &      1.55 &      0.53&         134 &           2 &MCG-02-33-099                        \\
125145.54&+254628.5 &IG     &C &      9.74 &      9.58 &      6.24 &      1.92 &      0.53&         107 &           2 &NGC 4747                             \\
134818.91&-505838.8 &G      &B &     11.08 &     10.65 &      6.00 &      2.73 &      0.52&         149 &           0 &2MASX J13481892-5058391              \\
001850.88&-102236.6 &EmG    &C &     10.76 &     10.29 &      5.41 &      2.37 &      0.52&         162 &           0 &MCG-02-01-051                        \\
080106.63&-660914.9 &G      &C &     10.20 &      8.51 &      5.24 &      2.17 &      0.52&         338 &           2 &6dFGSg J080106.6-660915               \\
183820.32&-652539.0 &Sy2    &C &      9.37 &      7.80 &      4.32 &      1.21 &      0.52&         322 &           2 &ESO 103-35                           \\
043548.45&+021529.6 &G      &C &     11.17 &     10.03 &      6.03 &      2.54 &      0.52&         138 &           2 &UGC 3097                             \\
002131.65&-483728.7 &IG     &C &      8.71 &      8.24 &      3.89 &      0.44 &      0.51&         140 &           2 &NGC 92                               \\
102751.28&-435414.5 &GiG    &C &      7.48 &      6.83 &      2.58 &     -0.88 &      0.51&         139 &           2 &6dFGSg J102751.3-435414              \\
054323.63&+540044.2 &G      &B &     11.54 &     11.07 &      6.52 &      3.24 &      0.51&         149 &           2 &2MASX J05432362+5400439              \\
064651.13&-645727.7 &IG     &C &     12.10 &     11.78 &      7.30 &      3.90 &      0.50&         142 &           0 &ESO 87-41                             \\
105918.14&+243234.6 &LIN    &C &     10.90 &     10.12 &      5.63 &      2.44 &      0.50&         153 &           0 &2XMM J105918.1+243234                \\
130842.02&-242257.8 &Sy2    &C &     11.29 &     10.22 &      5.91 &      2.72 &      0.50&         153 &           2 &PKS 1306-241                         \\
033639.05&-205406.8 &GiG    &C &      9.58 &      7.98 &      4.69 &      1.50 &      0.50&         328 &           2 &NGC 1377                             \\
002153.60&-791007.8 &Sy2    &C &     10.06 &      8.39 &      5.22 &      2.14 &      0.49&         341 &           2 &2MASX J00215355-7910077              \\
012002.63&+142142.5 &LIN    &C &     10.84 &      9.67 &      5.41 &      2.24 &      0.49&         154 &           0 &MCG+02-04-025                        \\
034648.35&+680546.5 &GiG    &C &      6.95 &      6.61 &      2.54 &     -1.12 &      0.49&         131 &           2 &IC 342                                \\
022557.05&-244240.8 &IG     &C &     11.80 &     11.57 &      7.55 &      3.82 &      0.49&         127 &           1 &AM 0223-245                           \\
233614.11&+020917.9 &AGN    &C &      9.31 &      9.04 &      4.73 &      1.24 &      0.49&         138 &           2 &NGC 7714                             \\
\enddata
\tablenotetext{1}{{\it Grade} refers to how well understood an object 
  is (see Section~\ref{grading}.)  Grade B is reseverd for objects with very little presence in the refereed literature, such that we are not convinced the object's true nature has been carefully verified.  Grade A is given to objects that are effectively new to science having, at most, been detected in surveys but never examined.}
\tablenotetext{2}{{\it Photometry} refers to the source of the
  photometry used in the table: 0:Profile fit; 1:Aperture-corrected; 2:Calibrated, 3:Elliptical aperture. See Section~\ref{photometry}}

\end{deluxetable}

\clearpage
%\end{landscape}

\clearpage
\LongTables
%\begin{landscape}
\begin{deluxetable}{ccccccccccccc}
\tabletypesize{\scriptsize}
\tablecolumns{12}
\tablewidth{0pt}
\tablecaption{Extreme WISE $\gamma$ sources (Grades A+B)\label{extreme}}
\tablehead{
\colhead{NLS \#} &
\colhead{RA} &
\colhead{DEC} &
\colhead{SIMBAD} &
\colhead{Grade\tablenotemark{1}} &
\colhead{W1} &
\colhead{W2} &
\colhead{W3} &
\colhead{W4} &
\colhead{$\gamma$} &
\colhead{$T_{\rm waste}$ (K)}&
\colhead{Photometry\tablenotemark{2}} &
\colhead{Note}\\
&\colhead{hhmmss.ss}&\colhead{ddmmss.s}&\colhead{type}&&\colhead{mag.}&\colhead{mag.}&\colhead{mag.}&\colhead{mag.}&&\colhead{K}&&}
\startdata
1&224436.12&+372533.6 &       &A &     11.27 &      9.79 &      6.92 &      4.44 &      0.35&         381 &           2 &                                     \\
2&073504.83&-594612.4 &       &A &     12.86 &     12.59 &      8.03 &      5.80 &      0.30&         229 &           0 &                                  \\
3&162721.02&+824538.7 &IR     &A &     13.39 &     13.18 &      9.41 &      6.31 &      0.27&         156 &           0 &IRAS 16329+8252                      \\
4&155812.25&-141806.2 &IR     &B &     16.48 &     15.39 &      9.13 &      4.03 &      0.99&          89 &           0 &IRAS 15553-1409                      \\
5&031805.43&-663024.0 &Cl*    &B &     12.12 &     11.69 &      6.92 &      3.42 &      0.61&         138 &           1 &[L2004]-n1313-341                    \\
6&005342.56&-723920.9 &EB*    &B &     13.17 &     12.78 &      8.43 &      4.77 &      0.56&         131 &           0 &OGLE J005342.56-723921.4             \\
7&125552.98&-765608.7 &HH     &B &     12.38 &     10.67 &      8.14 &      6.92 &      0.53&         516 &           3 &HH-54H2                              \\
8&134818.91&-505838.8 &G      &B &     11.08 &     10.65 &      6.00 &      2.73 &      0.52&         149 &           0 &2MASX J13481892-5058391              \\
9&054323.63&+540044.2 &G      &B &     11.54 &     11.07 &      6.52 &      3.24 &      0.51&         149 &           2 &2MASX J05432362+5400439              \\
10&214140.73&-454223.0 &IR     &B &     11.57 &     10.54 &      7.33 &      3.28 &      0.50&         115 &           2 &IRAS F21384-4556                     \\
11&012933.33&-733344.3 &Y*?    &B &     10.57 &      9.73 &      6.57 &      2.57 &      0.45&         117 &           2 &[GQH2007]-50                         \\
12&214035.92&+663525.8 &Y*O    &B &     10.05 &      8.25 &      5.58 &      3.05 &      0.45&         403 &           1 &2MASS J21403852+6635017               \\
13&031803.99&-663234.1 &WR*    &B &     13.30 &     12.93 &      8.61 &      5.32 &      0.45&         147 &           1 &[HC2007]-15                          \\
14&065558.96&-404912.2 &G      &B &     11.12 &     10.72 &      7.34 &      3.41 &      0.43&         119 &           2 &6dFGSg J065559.0-404912              \\
15&163600.57&+101340.4 &G      &B &     13.09 &     12.56 &      8.22 &      5.09 &      0.43&         156 &           1 &2MASX J16360060+1013396              \\
16&061452.97&-062244.4 &Y*O    &B &      8.80 &      8.31 &      3.15 &      1.78 &      0.42&         289 &           2 &IRAS 06124-0621                       \\
17&093016.68&+212150.5 &G      &B &     11.75 &     11.31 &      6.94 &      3.86 &      0.41&         159 &           2 &SDSSCGB-51158.1                      \\
18&020805.42&-291432.7 &G      &B &     11.64 &     10.16 &      6.98 &      4.19 &      0.41&         346 &           2 &2dFGRSTGS306Z043                     \\
19&224458.08&-014600.4 &G      &B &     11.10 &     10.55 &      6.18 &      3.16 &      0.41&         162 &           2 &2MASX J22445816-0145589              \\
20&195405.20&+495647.0 &G      &B &     12.57 &     12.16 &      7.39 &      4.81 &      0.39&         195 &           0 &LEDA 200363                          \\
21&044738.41&-172601.8 &G      &B &     11.74 &     11.39 &      7.38 &      4.02 &      0.39&         144 &           2 &ESO 552-5                            \\
22&113231.06&+742242.0 &G      &B &     11.59 &     11.30 &      7.06 &      3.89 &      0.39&         153 &           2 &2MASX J11323098+7422421              \\
23&154813.36&-245309.6 &EmG    &B &     11.29 &     11.00 &      6.74 &      3.59 &      0.39&         155 &           1 &ESO 515-7                            \\
24&181603.19&+473705.4 &G      &B &     11.46 &     11.13 &      6.79 &      3.74 &      0.39&         160 &           2 &2MASX J18160312+4737056              \\
25&024143.24&+454626.6 &G      &B &     11.71 &     11.35 &      7.22 &      3.99 &      0.38&         150 &           2 &2MASX J02414325+4546272              \\
26&045510.80&+053512.6 &GiC    &B &     12.52 &     12.12 &      8.00 &      4.81 &      0.38&         152 &           2 &2MASX J04551070+0535126              \\
27&181041.37&+250723.2 &G      &B &     11.88 &     11.64 &      7.19 &      4.23 &      0.38&         166 &           0 &2MASX J18104135+2507238              \\
28&061647.61&-090133.6 &IR     &B &      8.64 &      8.18 &      3.34 &      1.60 &      0.37&         280 &           2 &IRAS 06144-0900                       \\
29&185222.44&-293620.7 &EmG    &B &     10.77 &     10.49 &      6.52 &      3.18 &      0.37&         144 &           2 &6dFGSg J185222.4-293621              \\
30&134547.40&+700445.9 &G      &B &     11.24 &     10.87 &      6.59 &      3.58 &      0.37&         162 &           2 &2MASX J13454733+7004455              \\
31&025559.96&+474819.3 &G      &B &     10.74 &     10.49 &      6.39 &      3.16 &      0.37&         150 &           2 &MCG+08-06-022                        \\
32&052134.60&-660547.1 &Y*O    &B &     13.02 &     12.39 &      7.88 &      5.30 &      0.36&         197 &           1 &[GC2009]J052134.67m660548.3          \\
33&083538.40&-011407.1 &G      &B &     11.68 &     10.65 &      6.82 &      4.38 &      0.36&         310 &           2 &2MASX J08353838-0114072              \\
34&173429.01&-040541.7 &G      &B &     13.11 &     12.58 &      8.20 &      5.44 &      0.35&         180 &           0 &6dFGSg J173428.9-040542              \\
35&045542.30&-712034.9 &Y*?    &B &     13.52 &     13.18 &      8.30 &      6.55 &      0.35&         273 &           1 &GMP 19                                \\
36&195751.88&-322128.2 &EmG    &B &     10.70 &     10.40 &      6.28 &      3.15 &      0.35&         155 &           2 &6dFGSg J195751.9-322128              \\
37&191227.31&-290235.7 &EmG    &B &     11.66 &     11.33 &      6.97 &      4.10 &      0.35&         171 &           0 &ESO 459-7                            \\
38&132021.98&-233225.9 &EmG    &B &     11.51 &     10.67 &      6.71 &      3.71 &      0.35&         164 &           2 &2MASX J13202200-2332256              \\
39&152309.66&-393448.2 &G      &B &     11.74 &     11.28 &      7.13 &      4.17 &      0.34&         165 &           0 &2MASX J15230967-3934481              \\
40&035715.22&-261859.3 &G      &B &     13.96 &     13.51 &      9.37 &      6.39 &      0.34&         164 &           0 &APMBGC 483+082+074                   \\
41&054746.71&-444953.8 &IR     &B &     11.95 &     11.63 &      7.49 &      4.47 &      0.33&         162 &           2 &IRAS 05463-4450                       \\
42&064233.41&-645925.4 &G      &B &     13.41 &     13.12 &      9.01 &      5.96 &      0.33&         160 &           0 &AM 0642-645                           \\
43&025609.76&-153943.5 &G      &B &     12.60 &     12.01 &      8.33 &      5.04 &      0.33&         147 &           3 &6dFGSg J025609.8-153946              \\
44&042851.45&+693447.1 &G      &B &     11.72 &     11.32 &      6.89 &      4.26 &      0.33&         190 &           0 &2MASX J04285125+6934469              \\
45&150253.22&+165508.4 &GiG    &B &     10.85 &     10.51 &      6.25 &      3.40 &      0.33&         172 &           0 &MCG+03m38m076                        \\
46&025039.58&+414012.5 &IR     &B &     10.91 &     10.73 &      7.24 &      3.65 &      0.32&         133 &           1 &IRAS 02474+4127                      \\
47&102050.93&-171859.4 &EmG    &B &     11.95 &     11.63 &      7.38 &      4.52 &      0.32&         172 &           0 &MCG-03-27-005                        \\
48&182552.75&+375241.6 &IR     &B &     12.10 &     11.91 &      7.77 &      4.74 &      0.32&         161 &           0 &IRAS 18241+3750                      \\
49&202825.49&-330420.5 &G      &B &     10.33 &     10.32 &      7.15 &      3.29 &      0.32&         121 &           2 &ESO 400-28                           \\
50&140330.96&-504643.1 &HI     &B &     13.45 &     13.38 &      9.81 &      6.26 &      0.32&         134 &           0 &HIPASS J1403-50                       \\
51&092338.22&-251634.9 &EmG    &B &     11.11 &     10.86 &      6.69 &      3.74 &      0.32&         166 &           2 &6dFGSg J092338.2-251635              \\
52&013506.96&-412611.9 &G      &B &      9.64 &      9.55 &      5.76 &      2.40 &      0.32&         143 &           2 &6dFGSg J013506.9-412612               \\
53&093548.86&-291955.6 &EmG    &B &     11.84 &     10.62 &      7.34 &      5.08 &      0.32&         355 &           0 &ESO 434-13                           \\
54&194112.82&+630542.9 &G      &B &     11.76 &     11.42 &      7.30 &      4.37 &      0.31&         167 &           2 &2MASX J19411289+6305430              \\
55&105416.74&-394019.3 &G      &B &     11.28 &     11.07 &      7.12 &      3.98 &      0.31&         155 &           2 &ESO 318-23                           \\
56&125324.16&-234545.6 &EmG    &B &     12.69 &     12.43 &      8.40 &      5.37 &      0.31&         161 &           0 &6dFGSg J125324.2-234546              \\
57&044016.10&-444525.7 &G      &B &     11.77 &     11.46 &      7.38 &      4.43 &      0.31&         166 &           2 &6dFGSg J044016.1-444526              \\
58&054421.56&-135311.9 &G      &B &     12.02 &     11.65 &      7.29 &      4.70 &      0.31&         192 &           0 &2MASX J05442151-1353116              \\
59&174709.21&-643817.6 &WR?    &B &      9.54 &      9.18 &      6.11 &      2.37 &      0.30&         125 &           2 &[CB2009]-A1-C1                       \\
60&004250.05&-365243.2 &G      &B &     11.46 &     11.24 &      6.87 &      4.22 &      0.30&         187 &           0 &6dFGSg J004250.1-365241              \\
61&111859.15&-400013.2 &G      &B &     11.45 &     11.08 &      6.55 &      4.42 &      0.30&         248 &           0 &2MASX J11185912-4000135              \\
62&051521.42&-262817.1 &EmG    &B &     11.52 &     11.25 &      7.34 &      4.25 &      0.30&         157 &           2 &ESO 486-39                           \\
63&040819.05&-611605.7 &IG     &B &     11.11 &     10.89 &      6.91 &      3.87 &      0.30&         160 &           2 &ESO 118-4                            \\
64&111932.31&-471015.7 &G      &B &     12.05 &     11.83 &      7.94 &      4.82 &      0.30&         155 &           2 &2MASX J11193186-4710218              \\
65&050147.35&-181000.8 &IG     &B &     12.20 &     11.99 &      7.57 &      5.02 &      0.30&         195 &           0 &NGC 1739                              \\
66&063226.08&-243208.3 &G      &B &     12.50 &     12.32 &      8.13 &      5.29 &      0.30&         173 &           0 &ESO 490-11                           \\
67&070912.66&-440731.7 &G      &B &     12.58 &     12.40 &      8.17 &      5.38 &      0.29&         177 &           0 &ESO 256-16                           \\
68&045605.74&+015932.1 &       &B &     12.17 &     11.96 &      7.75 &      4.97 &      0.29&         177 &           0 &                                     \\
69&173431.80&+471301.7 &G      &B &     11.75 &     11.49 &      7.29 &      4.54 &      0.29&         179 &           2 &2MASX J17343177+4713010              \\
70&222149.97&+395024.0 &GiG    &B &     12.43 &     12.08 &      7.63 &      5.31 &      0.29&         219 &           1 &IRAS 22196+3935                       \\
71&105052.15&+010944.2 &GiG    &B &     12.09 &     11.75 &      7.36 &      4.94 &      0.29&         208 &           0 &IC 649S                              \\
72&114543.59&-114712.6 &EmG    &B &     11.30 &     11.00 &      6.55 &      4.21 &      0.29&         216 &           0 &6dFGSg J114543.6-114712              \\
73&024506.40&-020727.7 &G      &B &     12.01 &     11.83 &      7.93 &      4.86 &      0.29&         158 &           3 &2MFGC 2186                           \\
74&142837.03&-394844.1 &IG     &B &     11.93 &     11.79 &      7.60 &      4.81 &      0.28&         176 &           2 &ESO 326-24                           \\
75&232610.59&-303106.1 &EmG    &B &     11.45 &     11.14 &      6.91 &      4.29 &      0.28&         189 &           2 &6dFGSg J232610.6-303106              \\
76&161833.99&+132425.9 &G      &B &     13.08 &     12.63 &      9.02 &      5.85 &      0.28&         152 &           3 &2MASX J16183392+1324253              \\
77&075803.01&-642929.2 &G      &B &     11.89 &     11.54 &      7.11 &      4.93 &      0.28&         238 &           0 &6dFGSg J075803.0-642930              \\
78&225451.03&+374220.8 &IG     &B &     12.75 &     12.12 &      7.96 &      5.49 &      0.28&         206 &           1 &2MASX J22545096+3742205               \\
79&064540.95&+433407.5 &G      &B &     11.81 &     11.61 &      7.46 &      4.72 &      0.27&         180 &           2 &2MASX J06454097+4334069               \\
80&064339.31&-271217.6 &GiG    &B &     10.83 &     10.64 &      6.79 &      3.74 &      0.27&         159 &           2 &6dFGSg J064339.3-271218              \\
81&202933.28&-441815.2 &G      &B &     12.20 &     11.97 &      7.60 &      5.18 &      0.27&         206 &           0 &2MASX J20293311-4418140              \\
82&174949.84&-484826.9 &PN     &B &     12.20 &     12.19 &      9.16 &      5.40 &      0.27&         124 &           0 &PN-G343.2-10.8                       \\
83&030445.12&+074739.2 &G      &B &     12.47 &     12.30 &      8.34 &      5.42 &      0.27&         167 &           0 &2MASX J03044511+0747394              \\
84&055652.46&-052303.8 &LSB    &B &     11.58 &     11.40 &      7.78 &      4.55 &      0.26&         149 &           2 &6dFGSg J055652.3-052309              \\
85&051646.24&-122059.4 &G      &B &     10.53 &     10.33 &      6.44 &      3.47 &      0.26&         164 &           1 &6dFGSg J051646.2-122100              \\
86&165033.09&-672039.6 &G      &B &     11.89 &     11.53 &      7.28 &      4.88 &      0.26&         210 &           0 &LEDA 96592                           \\
87&194605.40&+640850.1 &G      &B &     10.96 &     10.55 &      6.97 &      3.84 &      0.26&         154 &           2 &2MASX J19460544+6408494              \\
88&120913.87&+265237.4 &G      &B &     12.00 &     11.82 &      8.05 &      5.00 &      0.26&         159 &           2 &LEDA 38612                           \\
89&235521.97&-563457.1 &G      &B &     13.02 &     12.71 &      8.42 &      6.07 &      0.26&         214 &           0 &6dFGSg J235522.0-563456              \\
90&054738.65&-103552.8 &G      &B &     12.79 &     12.50 &      8.88 &      5.76 &      0.25&         155 &           3 &6dFGSg J054738.7-103552              \\
91&140736.99&+160121.6 &G      &B &     12.03 &     11.68 &      7.70 &      4.98 &      0.25&         181 &           2 &2MASX J14073693+1601212              \\
92&093930.15&+062613.0 &G      &B &     11.98 &     11.72 &      7.49 &      5.04 &      0.25&         203 &           1 &2MASX J09393017+0626133              \\
93&095814.56&-381356.5 &       &B &     12.82 &     12.68 &      8.80 &      5.87 &      0.25&         166 &           0 &                                     \\

\enddata
\tablenotetext{1}{{\it Grade} refers to how well understood an object 
  is (see Section~\ref{grading}.)  Grade B is reseverd for objects with very little presence in the refereed literature, such that we are not convinced the object's true nature has been carefully verified.  Grade A is given to objects that are effectively new to science having, at most, been detected in surveys but never examined.}
\tablenotetext{2}{{\it Photometry} refers to the source of the
  photometry used in the table: 0:Profile fit; 1:Aperture-corrected; 2:Calibrated, 3:Elliptical aperture. See Section~\ref{photometry}}

\end{deluxetable}

\clearpage
%\end{landscape}

% \clearpage
% \LongTables
% \begin{landscape}
% \input{kiii_table7_favorites.txt}
% \clearpage
% \end{landscape}

\end{document}